%% file: main.tex
\documentclass[a4paper,11pt]{article}
\pdfoutput=1 % if your are submitting a pdflatex (i.e. if you have
\usepackage{jcappub} % for details on the use of the package, please
\usepackage[T1]{fontenc} % if needed
\usepackage{braket, xcolor, soul}

\definecolor{greennn}{rgb}{0.0, 0.56, 0.0}
\definecolor{palatinatepurple}{rgb}{0.41, 0.16, 0.38}

\def\CL{{\cal L}}
\newcommand{\be}{\begin{equation}}
\newcommand{\ee}{\end{equation}}
\newcommand{\bea}{\begin{eqnarray}}
\newcommand{\eea}{\end{eqnarray}}
\newcommand{\barr}{\begin{array}}
\newcommand{\earr}{\end{array}}
\def\bal#1\eal{\begin{align}#1\end{align}}

\title{Classical Cosmological Collider Physics and Primordial Features}

\include{rezadefs}

\author[a]{Xingang Chen,}
\author[b,c]{Reza Ebadi,}
\author[d,e]{Soubhik Kumar}

\affiliation[a]{Institute for Theory and Computation, Harvard-Smithsonian Center for Astrophysics, 60
Garden Street, Cambridge, MA 02138, USA}
\affiliation[b]{Department of Physics, University of Maryland, College Park, MD 20742, USA}
\affiliation[c]{Quantum Technology Center, University of Maryland, College Park, MD 20742, USA}
\affiliation[d]{Berkeley Center for Theoretical Physics, Department of Physics,
University of California, Berkeley, CA 94720, USA}
\affiliation[e]{Theoretical Physics Group, Lawrence Berkeley National Laboratory, Berkeley, CA 94720, USA}
\emailAdd{xingang.chen@cfa.harvard.edu}
\emailAdd{ebadi@umd.edu}
\emailAdd{soubhik@berkeley.edu}

\abstract{
Features in the inflationary landscape can inject extra energies to inflation models and produce on-shell particles with masses much larger than the Hubble scale of inflation. This possibility extends the energy reach of the program of cosmological collider physics, in which signals associated with these particles are generically Boltzmann-suppressed.
We study the mechanisms of this classical cosmological collider in two categories of primordial features. In the first category, the primordial feature is classical oscillation, which includes the case of coherent oscillation of a massive field and the case of oscillatory features in the inflationary potential. The second category includes any sharp feature in the inflation model.
All these classical features can excite unsuppressed quantum modes of other heavy fields which leave observational signatures in primordial non-Gaussianities, including the information about the particle spectra of these heavy degrees of freedom.
}

\begin{document} 
\maketitle
\flushbottom

\section{Introduction}

In the context of inflationary cosmology, the study of primordial non-Gaussianity (NG) offers an exciting way to investigate physics at very high energy scales. Given the Hubble scale $H$ during inflation can be as high as $5\times 10^{13}$~GeV~\cite{Planck:2018jri}, NG can probe (on-shell) interactions at scales far beyond the reach of any laboratory experiments in the foreseeable future. While theoretical and observational aspects of NG have been studied in the context of various inflation models (for reviews see~e.g. \cite{Bartolo:2004if, Liguori:2010hx, Chen:2010xka, Wang:2013zva}), relatively recently it has been emphasized that particles with masses $m\gtrsim H$ can give rise to non-analytic momentum dependence and angular dependence of cosmological correlators from which the \textit{on-shell} mass-spin information of the heavy particles can be extracted~\cite{Chen:2009zp,Arkani-Hamed:2015bza}. This program, being similar to how the mass-spin spectroscopy is done at particle colliders, has been dubbed as ``cosmological collider physics'' and it has been a focus of various recent works to study how particle physics may be probed at the Hubble energy scale or higher
\cite{Chen:2009we,Chen:2009zp,Baumann:2011nk,Assassi:2012zq,Chen:2012ge,Sefusatti:2012ye,Norena:2012yi, Pi:2012gf,Noumi:2012vr,Gong:2013sma,Emami:2013lma, Arkani-Hamed:2015bza,Chen:2015lza,Dimastrogiovanni:2015pla,Kehagias:2015jha,Chen:2016nrs,Lee:2016vti,Meerburg:2016zdz,Chen:2016uwp,Chen:2016hrz,Chen:2017ryl,Kehagias:2017cym,An:2017hlx,An:2017rwo,Iyer:2017qzw,Kumar:2017ecc,Chen:2018sce,Chen:2018xck,Chua:2018dqh,Kumar:2018jxz,Wu:2018lmx,MoradinezhadDizgah:2018ssw,Saito:2018omt,Tong:2018tqf,Alexander:2019vtb,Lu:2019tjj,Hook:2019zxa,Hook:2019vcn,Kumar:2019ebj,Wang:2019gbi,Liu:2019fag,Wang:2019gok,Wang:2020uic,Li:2020xwr,Wang:2020ioa,Fan:2020xgh,Kogai:2020vzz,Bodas:2020yho,Aoki:2020zbj,Arkani-Hamed:2018kmz,Baumann:2019oyu,Baumann:2020dch,Maru:2021ezc,Lu:2021gso, Lu:2021wxu,Wang:2021qez,Tong:2021wai,Pinol:2021aun, Cui:2021iie,Tong:2022cdz,Reece:2022soh,Pimentel:2022fsc,Qin:2022lva}.

In most models the amplitude of cosmological production of particles with $m\gtrsim H$ is proportional to $ e^{-\pi m/H}$~\cite{Birrell:1982ix} and the resulting NG signatures get exponentially suppressed as $m$ is increased. 
This sets an upper limit $\sim H$ of the sensitivity of the cosmological collider to heavy particles. 
On the other hand, if the mass of a mediating particle is much less than $\mathcal{O}(H)$, although the amplitude of the signal is not directly affected by the mass, it is increasingly difficult for the cosmological collider to distinguish this particle from a massless particle \cite{Chen:2009zp}.
Therefore, given the exciting prospect of doing on-shell heavy particle spectroscopy, it is important to ask if there are plausible and generic scenarios where particles with masses $m\gg H$ can also give rise to observably large NG signatures.

This question is also interesting from a particle physics standpoint since the renormalization group running of the Standard Model (SM) gauge couplings hints at a possible Grand Unification scale (see~\cite{ParticleDataGroup:2020ssz} for a review and references) around $10^{14}-10^{16}$~GeV depending on the absence or presence of supersymmetry. Furthermore, in the context of high-scale see-saw mechanisms (see also~\cite{ParticleDataGroup:2020ssz}  for a review and references), the right-handed neutrinos can have masses around similar scales. It might also be true that a beyond-Standard-Model (BSM) theory, living at scales around $H$, has a particle spectrum with masses spanning several orders of magnitude, similar to the SM flavor spectrum. Such a spread in particle masses can already happen during inflation due to SM itself~\cite{Chen:2016hrz, Chen:2016uwp,Chen:2016nrs,Kumar:2017ecc}. Any information about the physics at super-$H$ scales may also be very useful to understand the UV completion of low energy effective field theories (EFT) of inflation. All these aspects give a strong motivation to look for mechanisms through which super-$H$ particles can leave distinctive signatures in the cosmological collider.

Recently in Refs.~\cite{Chen:2018xck,Hook:2019zxa,Hook:2019vcn,Wang:2019gbi,Wang:2020ioa,Bodas:2020yho,Sou:2021juh} it has been pointed out that a current coupling $\propto\partial_\mu\phi \mathcal{J}^\mu$, where $\phi$ is the inflaton and $\mathcal{J}^\mu$ is a current made out of heavy fields, can indeed produce particles with masses much larger than $H$. The key idea in such setups is that a $\partial_\mu\phi \mathcal{J}^\mu$ coupling introduces a new scale $\dot{\phi}_0^{1/2}$ in the dynamics via the term $\dot{\phi}_0\mathcal{J}^0$ when $\phi$ is set to its background homogeneous value $\phi_0$.\footnote{One way to see why $\dot{\phi}_0^{1/2}$ is relevant instead of $\dot{\phi}_0$, is to note that for scalars, fermions or gauge bosons $\partial_\mu\phi \mathcal{J}^\mu/\Lambda$ is a dimension-5 operator characterized by an EFT cut off scale $\Lambda$. A controlled derivative expansion within the EFT requires $\Lambda >\dot{\phi}_0^{1/2}$~\cite{Creminelli:2003iq} and therefore, one is left with the term $\kappa \mathcal{J}^0$ where $\kappa< \dot{\phi}_0^{1/2}$ sets the limiting mass scale.} Given the normalization of the primordial scalar power spectrum, $\dot{\phi}_0^{1/2}\approx 60 H$~\cite{Planck:2018jri}, this allows for a production of super-$H$ particles up to masses $\lesssim \dot{\phi}_0^{1/2}$ for various spins. These examples assume the shift symmetry for the inflaton, $\phi\rightarrow \phi+\text{constant}$, and thus the scale-invariance of the density perturbation is preserved in these models.

In this work, we will take an entirely different route to excite the super-$H$ particles. From the UV-completion model-building point of view, inflaton is rolling in a landscape built by many fields. It is natural to expect the possible existence of some shift-symmetry violating features along the inflaton trajectory in the inflationary landscape, both within the scales probed by the cosmic microwave background (CMB) and the large-scale structure (LSS) observations and outside those scales. This possibility introduces the ``primordial features''. (For reviews of such primordial features see~\cite{Chen:2010xka, Chluba:2015bqa, Slosar:2019gvt}.) The energy scales corresponding to such features are determined by the nature of the feature (such as its sharpness or oscillation frequency), and is independent of the value of $\dot{\phi}_0^{1/2}$. Once present, the heavy particle production is quite independent of the details of the features, the properties of the particles and the details of couplings between them.
For example, such primordial features can excite heavy, \textit{real} scalar fields
for which a non-trivial current coupling of the form $\propto\partial_\mu\phi \mathcal{J}^\mu$ does not exist.

Primordial features are also well-motivated from the observational side. Currently the strongest constraints on primordial features come from the measurements of CMB anisotropies in temperature and polarization, the power spectra of which are consistent with a featureless power-law \cite{WMAP:2003syu,Planck:2018jri}. Nonetheless, there are some interesting feature-like anomalies \cite{WMAP:2003syu,Planck:2018jri} in both the large and short scales, which may be due to either sharp or resonant features. In particular, since primordial features have strictly correlated predictions for the statistical properties of the CMB temperature, polarization, and the LSS maps, future experiments on CMB polarization and LSS will soon be able to test these candidates and look for new features, with much improved precision. 
If any such features are discovered, they would not only reveal some vital information about the inflation scenario, but also establish the existence of the super-Hubble cosmological collider and provide a strong motivation to search for new particles excited by these energies.

In this work we investigate how such primordial features can play a role in the context of the cosmological collider in exciting super-$H$ particles and how such particles leave their mass spectra information in cosmological observables. 

The simplest possibility is that some sharp feature in the landscape excites classical oscillation of a massive field. This oscillation will induce some characteristic oscillatory {\em scale-dependent} features in the power spectrum and non-Gaussianities of the primordial density perturbations, which encode the information of the mass of the heavy particle.
This has been studied in the context of classical primordial standard clocks
\cite{Chen:2011zf, Chen:2011tu, Chen:2012ja, Battefeld:2013xka, Gao:2013ota, Noumi:2013cfa, Saito:2012pd, Saito:2013aqa, Chen:2014joa, Chen:2014cwa, Huang:2016quc, Domenech:2018bnf, Braglia:2021ckn, Braglia:2021sun, Braglia:2021rej}.
In this work, we will instead focus on the quantum, in contrast to classical, excitation of other massive fields by features in the landscape.
We will study two such scenarios. In both scenarios, primordial features in the model generate slow-roll violating, time-dependent evolution in the couplings to some massive fields. These features in the couplings inject extra energy to the model that excite the massive fields quantum-mechanically.

In the first scenario, the coupling contains a small component of background oscillation. 
Such an oscillation may be induced by a subdominant, but high frequency, oscillatory component in the slow-roll potential or other background parameters, such as in the resonant models  \cite{Chen:2008wn,Flauger:2009ab,Flauger:2010ja,Chen:2010bka}.
Such an oscillation may also be induced by a classically oscillating massive field that is excited due to a sharp feature on the inflationary landscape. Notice that the above-mentioned classical oscillation of a massive field studied in 
\cite{Chen:2011zf, Chen:2011tu, Chen:2012ja, Battefeld:2013xka, Gao:2013ota, Noumi:2013cfa, Saito:2012pd, Saito:2013aqa, Chen:2014joa, Chen:2014cwa, Huang:2016quc, Domenech:2018bnf, Braglia:2021ckn, Braglia:2021sun, Braglia:2021rej}
is now used as a background feature to excite a different massive field.

In the second scenario, a parameter that couples a massive field to the inflaton undergoes a sharp change. This can arise if such a coupling is realised as a (time-dependent) VEV of some other scalar field which undergoes a sharp transition due to its dynamics in the landscape. In this paper, we simply model such a sharp change as a step function, but in general it can be any kind of sharp changes, such as in \cite{Starobinsky:1992ts,Adams:2001vc,Bean:2008na,Achucarro:2010da, Bartolo:2013exa}. Since a temporal step function has a fourier support over a wide frequency range including frequencies larger than heavy particle mass, such a sharp feature can excite super-$H$ particles. 

In both scenarios, we will demonstrate that we can have unsuppressed NG from super-$H$ particles and how the spectra of particles get encoded in NG.

This mechanism can, for example, increase the currently known real-scalar reach of the cosmological collider from $\mathcal{O}(H)$ to a much higher scale determined by the frequency of an oscillatory background or the sharpness of a feature.
Keeping in mind the existing CMB and LSS constraints on such primordial features, we will describe parameter space benchmarks where cosmological collider signatures could arise in the near future surveys. While our focus will be on heavy, real scalar particles, both the mechanisms can also be applied to excite particles with non-zero spins/charge.

We emphasize that, while primordial feature models, including oscillatory and sharp features, can produce a variety of BSM signals in the primordial density perturbations,
the focus of this work is the signatures of the massive fields that are produced on shell. Such signatures are the central theme of the cosmological collider physics program. 
Because the excitation mechanisms studied in this paper rely on classical features in the background evolution of inflation models, in contrast to the excitation of massive fields by Hubble energy in previous works, we dub this subject as the ``classical cosmological collider physics''.

The organization of the paper is as follows. In Sec.~\ref{sec.generalities}, we discuss some general aspects of searches for heavy particles in cosmological correlation functions, especially in the context of primordial features. In Sec.~\ref{sec.setup} we describe how primordial features can be coupled to inflaton-heavy field system in a simplified parametrization without relying on a detailed model. In Sec.~\ref{sec.osc}, we focus on the case of oscillatory features. We first provide parametric estimates of the bispectrum and identify the dominant contribution to the bispectrum that contains the heavy field-mediated non-analytic signal.
We then calculate the full bispectrum numerically, and also provide an analytical computation in the squeezed limit. In Sec.~\ref{sec.sharp} we move on to the sharp feature scenario and numerically compute the full bispectrum. We then specialize to two specific classes of inflationary models manifesting classical background oscillations in Sec.~\ref{sec.applications}, and using our analytical results, evaluate the strength of the NG as a function of heavy field mass. Finally, we summarize the mechanisms considered in this work and conclude in Sec.~\ref{sec.summary}. Appendices contain some technical details that supplement the calculations in the main text.

\section{Primordial features as classical cosmological collider}
\label{sec.generalities}

\subsection{General requirements for heavy particle production}

The primary goal of the present work is to investigate the mechanisms by which quantum modes of massive fields (with mass $\gg H$) are excited by classical primordial features, and the signatures of these massive fields in cosmological correlation functions. 

Generally speaking, if there is any abrupt (characterized by a time scale much less than $1/H$) deviations from slow-roll in the background evolution of the inflation model, they inject extra energies and excite any massive fields coupled to this background. In principle, the interactions between the massive fields and the background can be quite arbitrary.
On the other hand, in order to capture main physics with simple examples, in this paper we will proceed with a specific example of such couplings, and we will model the time dependence of the background features with analytical ansatz. 

In our example, the Lagrangian can be schematically written as
\begin{align}
\mathcal{L}=\mathcal{L}_\phi+ \mathcal{L}_\chi + \mathcal{L}_{\rm int}(\partial\phi,\chi),  
\end{align}
where $\mathcal{L}_{\phi}$ and $\mathcal{L}_{\chi}$ denote the Lagrangian for the inflaton and a massive field $\chi$, respectively. 
$\mathcal{L}_\phi$ satisfies the slow-roll conditions at the leading order, but may contain some shift symmetry breaking terms in order to incorporate the phenomenology of primordial features.
For simplicity, in the interaction part, $\mathcal{L}_{\rm int}(\partial\phi,\chi)$, the inflaton couples only derivatively to the heavy sector, e.g.~$(\partial\phi)^2 \chi$, so that the slow-roll conditions can be more easily preserved at the leading order and when the features are absent. 
The slow-roll breaking features in $\mathcal{L}_\phi$ introduces feature-like time-dependence in the background evolution of $\phi$, which in turn introduces feature-like time-dependence in the coupling between $\phi$ and $\chi$. We will consider the following two types of features.

\begin{itemize}
    \item Oscillatory features, which can be due to (a) oscillations of another heavy (clock) field, excited by some sharp features on the inflationary landscape, or (b) oscillations of the inflaton itself, due to features in the inflaton potential.
    \item Sharp features, which can be due to a sharp change in the coupling of a heavy field and the inflaton. Physically, such a change can take place if the coupling is controlled by a time-dependent VEV of another field undergoing a transition.
\end{itemize}
To make the analysis more model-independent without losing the main physics, we will not derive these time-dependent features from explicit models of $\CL_{\phi}$. Instead, we will just simply model these time-dependence by some analytical ansatz.

\subsection{Observables encoding masses of heavy particles}
Following their production, the heavy particles can decay into inflaton fluctuations. It is therefore interesting to ask which cosmological observables encode the spectra information about the heavy particles.

Primordial features themselves introduce scale-dependent features in the power spectrum. Depending on the type of feature model, these scale-dependence can be sinusoidal, resonant, or characteristic of a classically oscillating massive field. In this work, we are interested in the signature of a massive field $\chi$ quantum-mechanically excited by these classical features. Since we will ignore the classical oscillation of the $\chi$ field, the correction it introduces to the power spectrum does not carry the information of its mass. We need to consider non-Gaussianities and study the effect not captured by low-energy effective field theories.

One way to get such signatures is to consider the bispectrum, i.e., the three-point correlation function involving momenta $(\vec{k}_1,\vec{k}_2,\vec{k}_3)$. In particular, the bispectrum can allow for the so-called squeezed-limit configurations $k_3\ll k_1,k_2$ for which the latter two modes exit the horizon at conformal time, say $\eta_1$, which is much later compared to say $\eta_3$, the conformal time of horizon exit of $k_3$. This gives us a chance to probe \textit{shape}-dependent oscillations in the bispectrum due to the excited field $\chi$ with mass $m_\chi$, $e^{im_\chi (t_3-t_1)}\sim (\eta_3/\eta_1)^{-im_\chi/H}\sim(k_3/k_1)^{im_\chi/H}$, in addition to the above-mentioned scale-dependent oscillations due to the primordial feature. Here, we have used the relation between physical time $t$ and conformal time $\eta$, $\eta = -e^{-H t}/H$. Both these features in the power spectrum and bispectrum will be explained in more detail in the following sections.\footnote{Here we add a short comment on several terminologies used in the context of the cosmological collider physics. There are two types of non-analytical shape-dependence in NG as signatures of the quantum fluctuations of a massive field, which are collectively called the cosmological collider (CC) signal. One of them is a non-analytical power-law dependence, which happens when the mass of the particle is on the lower side of $\CO(H)$. The other is a non-analytical oscillatory dependence (i.e.~power-law with a complex power), which happens when the mass of the particle is on the higher side of $\CO(H)$. The latter is often called the ``clock signal'' due to its usage in the context of primordial standard clocks. Mathematically, these two cases are connected through analytical continuation. In this paper, we work in the latter case.}
We would also like to emphasize that the primary observable signatures that we will consider in the following will only involve tree-level diagrams, and hence can be straightforwardly evaluated.

Before proceeding, we would like to point out some key differences between the scenario in this work and those in some other works studied in the past.
The scenarios of heavy particle production through shift-symmetric derivative couplings have been studied recently in the context of cosmological collider physics, including examples in which the inflaton is coupled to the current $\mathcal{J}^\mu$ of heavy fermions~\cite{Chen:2018xck, Hook:2019zxa}, gauge bosons~\cite{Wang:2020ioa}, or softly broken $U(1)$-charged scalars \cite{Bodas:2020yho}. In these models, the density perturbations are scale-invariant primarily due to the shift symmetry of the Lagrangians. Although, for each massive mode with any comoving wavenumber, there is some time-dependent mass \cite{Chen:2018xck, Hook:2019zxa, Wang:2020ioa} or coupling \cite{Bodas:2020yho} associated with it (which introduces an energy up to $\propto \dot{\phi}_0^{1/2}$), the physics is the same for all modes with different wavenumbers, leading to scale-invariance. In this work, we explicitly break the scale invariance through the introduction of primordial features, which also introduce more arbitrary energy scales and couplings.
In scenarios studied in ~\cite{Kofman:1997yn,Chung:1999ve,Kofman:2004yc, Barnaby:2009mc, Flauger:2016idt, Kim:2021ida}, there are
non-shift-symmetric couplings of the inflaton to a heavy field. This coupling can lead to a time-dependent mass for the heavy field, facilitating its cosmological production when the effective mass of the heavy field passes through a minimum. In contrast, in our scenario the heavy field mass is a constant and its excitation is due to injection of extra energy through features in the model. Also, more importantly, our focus will be on extracting the signature of the massive field which is absent in the low energy effective theories, as those in studies of the cosmological collider physics~\cite{Chen:2009zp,Arkani-Hamed:2015bza}.

\section{Setup}\label{sec.setup}
In this section we review some relations regarding scalar fields in de Sitter space, and set up the Lagrangian that we will use for computation of correlation functions.

\paragraph{Free theory.} The quadratic Lagrangian for the fluctuations of the inflaton, $\delta\phi$, and the heavy field, $\delta\chi$, is as follows:\footnote{Note that $S = \int\d^4x\sqrt{-g}\CL = \int\d^4xa^3\CL$.}
\begin{align}\label{eq:kinetic_lagrangian}
    \CL_2 = \frac{1}{2}\dot{\delta\phi}^2 - \frac{1}{2a^2}(\partial_i \delta \phi)^2  + \frac{1}{2} \dot{\delta\chi}^2 - \frac{1}{2a^2} (\partial_i \delta \chi)^2 -\dfrac{1}{2} m_\chi^2 \delta\chi^2,
\end{align}
where an overdot $\dot{\braket{~~}}$ denotes a derivative with respect to the physical time $t$, and $a=e^{H t}$ is the scale factor during inflation. We canonically quantize the system by writing Fourier components as $\delta\phi_\kk(\eta) = u_\kk(\eta) \, b_\kk + u^*_{-\kk}(\eta) b^\dagger_{-\kk}$ and $\delta \chi_\kk(\eta) = v_\kk(\eta) \, c_\kk + v^*_{-\kk}(\eta) c^\dagger_{-\kk}$, where $b_\kk,b_\kk^\dagger$ and $c_\kk,c_\kk^\dagger$ are two sets of standard creation and annihilation operators. The mode functions  obey the equations of motion:
\begin{align}\label{eq:eoms}
    & u_\kk''  - \dfrac{2}{\eta} u_\kk' + k^2 u_\kk = 0\, ,\\
    & v_\kk''  - \dfrac{2}{\eta} v_\kk' + k^2 v_\kk + \dfrac{m_\chi^2}{H^2 \eta^2} v_\kk = 0\,,
\end{align}
where $'$ denotes a derivative with respect to conformal time $\eta$. Requiring the Bunch-Davies vacua at early times, these equations have the following solutions:
\begin{align}\label{eq:mode_function}
    u_k (\eta) =&~ \dfrac{H}{\sqrt{2 k^3}} (1 + ik\eta) e^{-ik\eta}\, ,\\
    v_k(\eta) =&~ -i e^{-\pi\mu/2 + i\pi/4} \dfrac{\sqrt{\pi}}{2}  H (-\eta)^{3/2} H^{(1)}_{i \mu}(-k \eta)\,,
\end{align}
where we have defined $\mu\equiv\sqrt{\left(\dfrac{m_\chi}{H}\right)^2-\dfrac{9}{4}}$ and $H^{(1)}_{i \mu}$ is the Hankel function of the first kind.
\paragraph{Interactions.}
Throughout this work, we parametrize the coupling between the heavy field $\chi$ and the inflaton $\phi$ using the following operator,
\begin{align}\label{eq:int_lagrangian}
    \CL_\interaction \supset \rho(1+B(t))\dot{\delta\phi}\delta\chi+\frac{\lambda}{\Lambda}(1+C(t))\left[ (\dot{\delta\phi})^2 - \frac{1}{a^2} (\partial_i \delta\phi)^2 \right]\delta\chi+\cdots.
\end{align}
Here the parameters $B(t)$ and $C(t)$ are small but time-dependent, whereas we will treat $\rho, \lambda$ and $\Lambda$ as constants, and $\lambda\sim {\cal O}(1)$.  As we will see, $B(t)$ plays a crucial role in unsuppressed production of heavy particle $\chi$, and imprinting the corresponding non-analytic, oscillatory signal in non-Gaussianity. On the other hand, $C(t)$ alone does not lead to such an unsuppressed particle production with three point function.
This is because, in the absence of $B(t)$ the quadratic vertex of the three-point in-in diagram is Boltzmann suppressed, as we will discuss in Sec.~\ref{subsec:bispectrum}. %\st{Therefore, to consider the simplest of such possibilities, we will require the presence of $B(t)$, but turn off $C(t)$ and leave its effect for future study.} 
However, one can have a scenario in which both $B(t)$ and $C(t)$ are present. We will parametrically estimate in Sec.~\ref{subsec:bispectrum} the associated three point function and show under which conditions the $C(t)$ coupling can be important, leaving a detailed numerical computation for a future study.
In Sec.~\ref{sec.applications}, we will use examples to demonstrate how the background evolution of the inflaton $\phi$ field, induced by the features in the potential or oscillations of another field coupled to it, can lead to a rapid time dependence in $B(t)$.
We consider two forms of $B(t)$, corresponding to an oscillatory feature and a sharp feature,
\begin{align}\label{eq.oscfeat}
    \text{Oscillatory feature (Sec.~\ref{sec.osc}):}& ~B_\c(t) = B_0\,(a/a_0)^{-n}\,\sin{(\omega_\c(t-t_0)+\alpha)}\theta(t-t_0),\\
    \label{eq.sfeat}
    \text{Sharp feature (Sec.~\ref{sec.sharp}):}& ~B_s(t) = B_0\theta(t-t_0).
\end{align}
Here we follow a phenomenological approach and take $B_0$ to be a constant whose magnitude we fix by respecting the current bounds on the power spectrum, as we will discuss later. We will also choose $\alpha=0$ such that the oscillatory feature turns on at $t_0$ continuously through an infinitely sharp kink, although it does not have to be the case. For persistent oscillations, the starting time $t_0$ can be sent to past infinity. The parameter $\omega_c$ determines the oscillation frequency, and as we will show it assists particle production for masses up to~$\omega_c$. The dilution parameter $n$ can obtain different values depending on how the rapid oscillations arise. For example, $n = 3/2$ corresponds to scenarios where time dependence comes from coherent oscillations of the clock field with the value $3/2$ denoting the $1/\sqrt{\text{volume}}$ dilation of its mode function.\footnote{In this scenario, a similar time dependent function $B_d(t)$ can also play a role by a direct coupling of the type $B_d(t)(\partial\phi)^2$, which is distinct from the inflaton and the massive field $\chi$ coupling. 
} The $n=0$ scenario, on the other hand, corresponds to oscillations induced by ripples in the potential, such as in the resonant models, where the amplitude of the oscillation is assumed to be constant.
More details about these examples will be discussed in Sec.~\ref{sec.applications}.

The cubic interaction $\lambda$ is time-independent, and hence it does not play any special role in on-shell production of heavy particles. However, the heavy particles can decay into inflaton quanta via this cubic coupling and hence it plays a crucial role in imprinting heavy particle signatures in the inflationary correlation functions.

We emphasize that the interaction terms in \eqref{eq:int_lagrangian} are by no means the only choice for models of classical cosmological collider physics. They are simple, effective field theory-motivated examples that we study in this paper. Other type of couplings can appear and the background oscillations may enter couplings of other terms too. 

\paragraph{Cosmological correlation functions.} Given the Lagrangian in Eq.~\eqref{eq:int_lagrangian}, we will calculate the corrections to inflaton power spectrum and look for new signatures in inflaton three-point function. In the following calculations we assume spatially flat gauge (see e.g.~\cite{Maldacena:2002vr}), in which we have non-zero inflaton fluctuations but no scalar fluctuation in the spatial part of the metric. To calculate correlation functions of the curvature perturbation, we can use the leading-order linear transformation between inflaton perturbations (in spatially flat gauge) and curvature perturbations (in uniform inflaton gauge), i.e., $\zeta \simeq - ({H}/{\dphi_0}) \delta\phi$. The inflaton $n$-point function is given by the expectation value of $\delta\phi^n$ at the end of inflation, $t_{\rm end}$. In the ``in-in'' formalism~\cite{Weinberg:2005vy} this is given by
\begin{align}\label{eq.ininmaster}
    \braket{\delta \phi^n} = \bra{0} \underbrace{\left[\bar{T}  e^{i \int_{-\infty}^{t_{\rm end}} \d t' H_\interaction(t')}\right]}_{-} \, \delta \phi^n(t_{\rm end}) \, \underbrace{\left[T e^{ -i \int_{-\infty}^{t_{\rm end}} \d t' H_\interaction(t')}\right]}_{+} \ket{0}\,,
\end{align}
where $H_\interaction(t)$ denotes the interacting part of the Hamiltonian in the interaction picture, and $T(\bar{T})$ denotes time(anti-time) ordering.
In practice, the correlation functions can be more conveniently evaluated following the diagrammatic rules summarized in \cite{Chen:2017ryl}.
\begin{figure}[tb!]
    \centering
    \includegraphics[width=15cm]{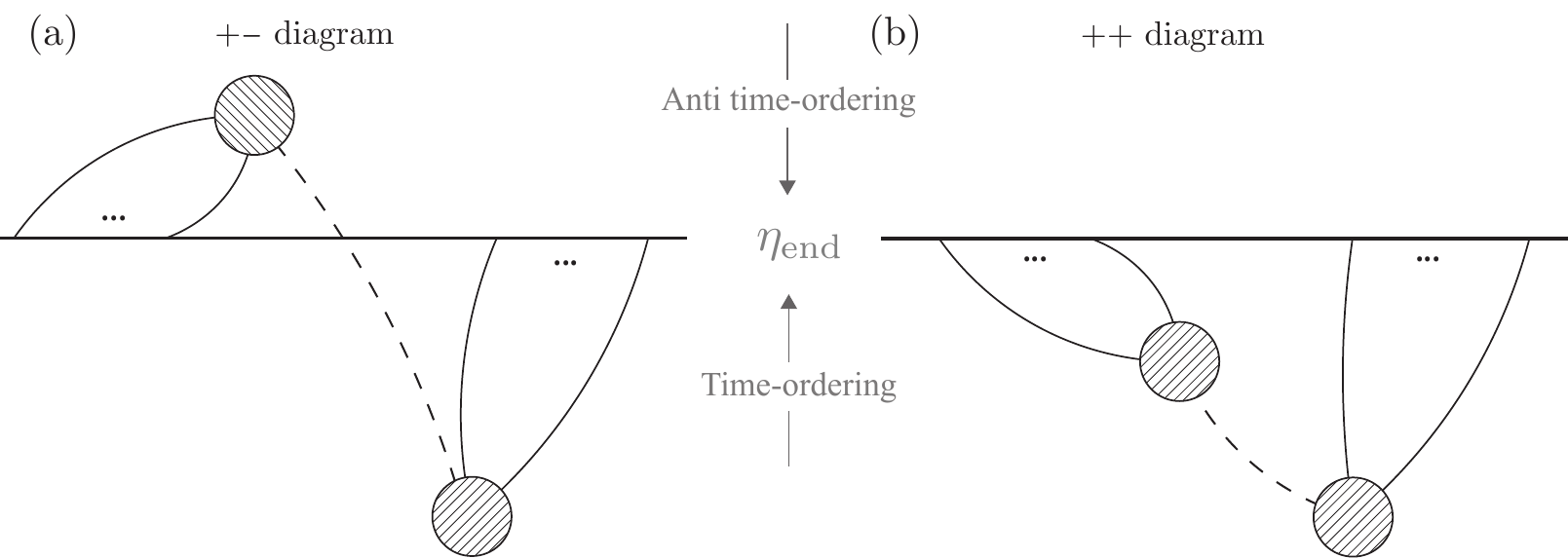}
    \caption{Schematics of the Feynman diagrams, discussed around Eq.~\eqref{eq:+-contributions_definition}: (a) $+-$ contribution, where one of the vertices is chosen from time-ordered evolution operator (shown below the end-of-inflation time slice at $\eta_{\rm end}$) and the other vertex is chosen from anti time-ordered operator (shown above the end-of-inflation time slice), (b) $++$ contribution, where both vertices are chosen from time-ordered time evolution operator. The $-+$ and $--$ diagrams are complex conjugates of the the $+-$ and $++$ diagrams respectively. The shaded blobs denote interaction vertices between the inflaton (solid) and the heavy field (dashed).}
    \label{fig:+-contributions_definition}
\end{figure}
As a notation, in the perturbative expansion of the in-in formalism time evolution operators, we denote any vertex coming from the time-ordered (anti time-ordered) part by $+$ ($-$). For example, for a diagram with two vertices, there are $2^2$ possible contributions to the correlation function depending on whether each of the vertices come from time-ordering or anti-time-ordering parts:
\begin{align}
    \braket{\delta \phi^n} = \braket{\delta \phi^n}_{++} + \braket{\delta \phi^n}_{--} + \braket{\delta \phi^n}_{+-} + \braket{\delta \phi^n}_{-+}\,,
    \label{eq:+-contributions_definition}
\end{align}
where $+$ stands for couplings in time-ordered part and $-$ stands for couplings in anti-time-ordered part. The four terms are related via complex conjugation, i.e., $\braket{\delta \phi^n}_{--} = \braket{\delta \phi^n}_{++}^*$ and $\braket{\delta \phi^n}_{-+} = \braket{\delta \phi^n}_{+-}^*$. To show these contributions diagrammatically, we draw Feynman diagrams where the time-ordered vertices are below the end-of-inflation time slice and anti time-ordered vertices are above that, following Ref.~\cite{Bodas:2020yho} (see Fig.~\ref{fig:+-contributions_definition}).

\begin{figure}[tb!]
    \centering
    \includegraphics[width=15cm]{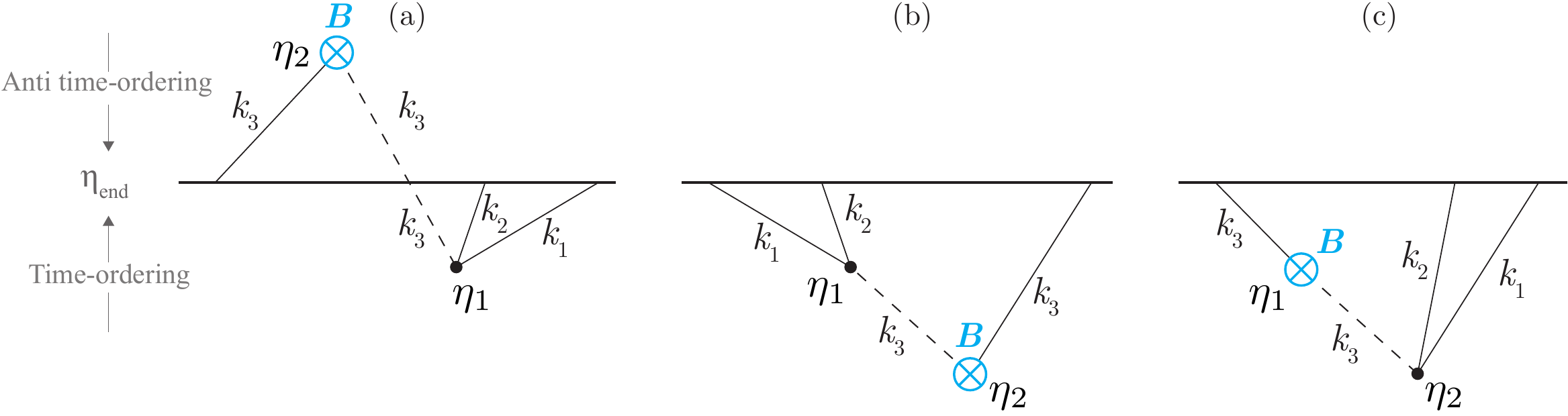}
    \caption{Feynman diagrams for bispectrum: (a) $+-$ contribution, (b) and (c) $++$ contribution. Figs.~(b) and (c) differ in whether the two point vertex happens earlier (b) or later (c) compared to the three point vertex. In particular, Fig.~(b) represents the process in which an energy injection through the coupling $B$ efficiently produces super-$H$ massive field $\chi$ quanta (dashed) at $\eta_2$ which then decays into inflaton quanta at $\eta_1$. This is the primary diagram of interest that captures the unsuppressed classical cosmological collider signal encoding the mass of $\chi$.}
    \label{fig:3ptinin}
\end{figure}

\paragraph{Bispectrum.} Specializing to the case of bispectrum and considering the Lagrangian in Eq.~\eqref{eq:int_lagrangian}, we have two independent contributions, namely the $++$ and $+-$ diagrams. They correpond to $\braket{\delta \phi^3}_{++}$ and $\braket{\delta \phi^3}_{+-}$, respectively. Graphically, these are shown in Fig.~\ref{fig:3ptinin}.
To evaluate these (and their complex conjugates) we need the inflaton three-point function, given by the expectation value of $\delta\phi^3$ at the end of inflation, $t_{\rm end}$ as in Eq.~\eqref{eq.ininmaster}, 
\begin{align}
    \braket{\delta \phi^3} = \bra{0} \underbrace{\left[\bar{T}  e^{i \int_{-\infty}^{t_{\rm end}} \d t' H_\interaction(t')}\right]}_{-} \, \delta \phi^3(t_{\rm end}) \, \underbrace{\left[T e^{ -i \int_{-\infty}^{t_{\rm end}} \d t' H_\interaction(t')}\right]}_{+} \ket{0}\,.
\end{align}

\section{Bispectrum in the presence of oscillatory features}\label{sec.osc}
In this section, we first give some qualitative estimates for both the power spectrum and the bispectrum associated with heavy particle production in the presence of oscillatory features. This will help us identify the interesting parts of the parameter space and the physical processes where signatures of heavy particles are most easily observable. This will also pave the way for an analytical computation of the non-analytic, oscillatory part of the bispectrum. Following this, we perform a numerical computation of the full bispectrum, including both the shape and scale dependence.

\subsection{Parametric estimates}\label{subsec:bispectrum} 

In this subsection, we present parametric estimates and explain essential physics of the bispectrum of our model. More detailed analytical and numerical computation will be presented in Sec.~\ref{subsec.analytics} and Sec.~\ref{subsec.numerics_osc}.

We rewrite the oscillatory feature in Eq.~\eqref{eq.oscfeat} as,
\begin{align}\label{eq:B_modeling}
    B_\c(t) =& B_0\,(a/a_0)^{-n}\,\sin{(\omega_\c(t-t_0))} \, \theta(t-t_0)\,\\
    = & \dfrac{i}{2}B_0\theta(\eta-\eta_0)\left[\left(\frac{\eta}{\eta_0}\right)^{n+i\mu_c}-\text{c.c.} \right].
\end{align}
Here $\mu_c=\omega_\c/H$, and $t_0~(\eta_0)$ denotes a fiducial physical time (conformal time) at which the background oscillation starts.
The leading Feynman diagrams for $+-$ and $++$ contributions are shown in Fig.~\ref{fig:3ptinin}.\footnote{The diagrams where $\rho B_\c(t)$ is replaced by just $\rho$ are not shown since they do not have the high-frequency energy insertion, and therefore non-analytic contributions to NG are exponentially suppressed for super-$H$ masses.}  
Using mode functions given in Eq.~\eqref{eq:mode_function}, we now estimate these diagrams. 
For simplicity, in the following estimates we only consider temporal derivatives of the inflaton fluctuations, and we include spatial derivatives in the full calculation given later in this section.

\paragraph{The $+-$ contribution.}
Diagram.~\ref{fig:3ptinin}a gives
\begin{align} \label{eq:+-contribution}
    \braket{\delta\phi^3}'_{+-} =& -  \dfrac{\rho\lambda}{\Lambda} \, u_{k_1} u_{k_2} u^*_{k_3} (\eta_{\rm end}) \, \int_{-\infty}^0 \dfrac{\d\eta_1}{(H \eta_1)^4} \, \du^*_{k_1} \du^*_{k_2} v^*_{k_3}(\eta_1) \, \int_{-\infty}^{0} \dfrac{\d\eta_2}{(H \eta_2)^4} \, \du_{k_3} v_{k_3} B_\c (\eta_2) \nonumber \\
    \propto & \underbrace{e^{\pi\mu/2}\int_0^\infty \d z_1 z_1^{3/2} \, H_{i\mu}^{(2)}(z_1) \, e^{-i p z_1}}_{\CI_3^+}   \,\underbrace{e^{-\pi\mu/2} \int_{0}^{\infty} \dfrac{\d z_2}{\sqrt{z_2}} \, H_{i\mu}^{(1)}(z_2) \, e^{i z_2} \, z_2^{n-i \mu_{\rm c}}}_{\CI_2^-} - \{\mu_c\rightarrow-\mu_c\}\,, 
\end{align}
where we have defined dimensionless parameters
\begin{align}
z_{1,2} \equiv -k_3 \eta_{1,2} ~,
\end{align}
and the momentum ratio 
\begin{align}
p \equiv (k_1 + k_2)/k_3 ~.
\end{align}
The primes on the correlators are defined through, $\braket{\delta\phi^3} \equiv \braket{\delta\phi^3}' (2\pi)^3 \delta^3(\sum \vec{k}_i)$.
Oscillations with the frequency $\mu_c$ in the mixing vertex assists particle production in the mixing vertex as long as $\mu\lesssim\mu_c$ and eliminates the exponential suppression with respect to $\mu$ in $\CI_2^-$. However, the cubic interaction vertex is suppressed for all $\mu> 1$. This is because among the two components of the massive field mode function, one of them is able to resonate with the inflaton field but its amplitude is exponentially suppressed, and the other unsuppressed component does not resonate; and both equally, in order of magnitude, contribute to a final Boltzmann-suppressed result, $\sim e^{-\pi\mu}$.
In other words, in the leading component, the energy conservation condition, which is necessary for resonance, cannot be satisfied. (See App.~A of \cite{Chen:2015lza} for details.) Therefore we do not consider the $+-$ and its complex conjugate $-+$ diagram further, although explicit form of the exponentially-suppressed result is presented in App. \ref{app:+-integrals}.

\paragraph{The $++$ contribution.} There are two possibilities for the $++$ contribution with opposite time orders of the two vertices associated with the propagator of the massive field $\chi$, which are shown in Fig.~\ref{fig:3ptinin}b and Fig.~\ref{fig:3ptinin}c, respectively. 

In Fig.~\ref{fig:3ptinin}b, to get the unsuppressed clock signal we look at the configuration where the wavelength of the massive field is much longer than those of the inflatons. In this limit, each massive field mode can resonate with the inflaton at the three-point (3pt) vertex at a specific time. This is different from the previous $+-$ case due to the opposite conjugation of the massive field mode function involved here. This boosts the integration at the 3pt vertex and generates the clock signal without suppression. (See \cite{Chen:2015lza} for more detailed explanations.) At the two-point (2pt) vertex, if without any primordial features, no resonance happens and the integration at this vertex gives rise to a Boltzmann suppression factor if $\mu>1$. However, in the scenario of this paper, an extra energy injection $\mu_c>\mu$ at the 2pt vertex excites a quantum mode of $\chi$ without suppression. The entire diagram Fig.~\ref{fig:3ptinin}b can therefore avoid the Boltzmann suppression. This is the most interesting diagram that we will evaluate below. 

On the other hand, even with the primordial feature, the clock signal in Fig.~\ref{fig:3ptinin}c remains Boltzmann-suppressed. This is because, with this timing ordering, the leading component of the $\chi$ field mode function does not allow resonance at the 3pt vertex for the reason explained in the $+-$ contribution case.

The physical interpretation of these two diagrams is also clear. Fig.~\ref{fig:3ptinin}b describes the situation where a massive field with $1<\mu<\mu_c$ is excited on-shell by the primordial feature which provides sufficient energy, and then decays to two inflatons. In Fig.~\ref{fig:3ptinin}c, the massive field has to be first created from the inflationary background which has insufficient energy to create it on-shell, and therefore gets Boltzmann suppressed.

We will also study the amplitude of the clock signal in the case where the massive particle is heavier than the energy scale of the feature, $\mu>\mu_c$. In this case, we will show that a modified Boltzmann factor appears, $\sim e^{-\pi(\mu-\mu_c)}$.

Besides the clock signal that has non-analytical dependence on momenta, these two diagrams also contribute to terms that are analytical in momenta. These are the low energy EFT terms when the massive field is integrated out, mainly contributed by configurations where the two vertices are close to each other. These contributions are power-law suppressed in $\mu$, $\sim 1/\mu^2$. This is shown explicitly in~\cite{Chen:2015lza} for $\mu_c=0$, and we will argue that here this power-law suppression starts from $\mu>\mu_c$. 

We will consider both diagrams in the numerical calculations later in Sec.~\ref{subsec.numerics_osc}. In the present subsection, for qualitative discussion, let us only consider the most interesting diagram Fig.~\ref{fig:3ptinin}b, which gives
\begin{align}\label{eq:++}
    \braket{\delta\phi^3}'_{++} &\supset %\braket{\delta\phi^3}'_{++}\bigg\rvert_{Fig.~\ref{fig:3ptinin}b} =
    \dfrac{\rho\lambda}{\Lambda} \, u_{k_1} u_{k_2} u_{k_3} (\eta_{\rm end}) \, \int_{-\infty}^0 \dfrac{\d\eta_1}{(H \eta_1)^4} \, \du^*_{k_1} \du^*_{k_2} v_{k_3}(\eta_1) \, \int_{-\infty}^{\eta_1} \dfrac{\d\eta_2}{(H \eta_2)^4} \, \du^*_{k_3} v^*_{k_3} B_\c (\eta_2) \nonumber \\
    &= \dfrac{\rho\lambda}{\Lambda} \dfrac{-\pi H^3}{32 k_1 k_2 k_3^4} \, \dfrac{i}{2}B_0 \, z_0^{-n - i \mu_{\rm c}}\,\nonumber\\
    & \times e^{-\pi\mu/2}\int_0^\infty \d z_1 z_1^{3/2} \, H_{i\mu}^{(1)}(z_1) \, e^{-i p z_1}   \,\underbrace{e^{\pi\mu/2} \int_{z_1}^{\infty} \dfrac{\d z_2}{\sqrt{z_2}} \, H_{i\mu}^{(2)}(z_2) \, e^{-i z_2} \, z_2^{n+i \mu_{\rm c}}\theta(z_0-z_2)}_{\CI_2^+(z_1)}\nonumber\\
    &- \{\mu_c\rightarrow-\mu_c\}\,, 
\end{align}
where we have defined 
\begin{align}
  z_0\equiv-k_3\eta_0 ~.
  \label{z_notations}
\end{align}
To reiterate some physics discussed above and clarify terminologies, we emphasize that there are two different types of resonances that can happen in this diagram. At the 3pt vertex, a resonance can happen between a long massive $\chi$-mode and two short inflaton modes, which generates a clock signal. At the 2pt vertex, a resonance can happen between the classical background of the primordial feature (either of oscillation or sharp feature form) and the massive $\chi$-mode, which excites unsuppressed quantum $\chi$-mode.

We now study two different regimes: $\mu_c > \mu$ and $\mu_c < \mu$. For $\mu_c > \mu$, $\CI_2^+(z_1)$ has a stationary phase point at which the resonant particle production happens. For $\mu_c < \mu$, no stationary phase exists and the production of the on-shell massive field is exponentially suppressed.
In the next subsections we discuss these two regimes separately, focusing primarily on $\mu_c>\mu$. We will also note that within this stationary phase approximation, the non-trivial nested integral in Eq.~\eqref{eq:++} simplifies and we get factorized integrals for each vertex.

\subsubsection{Stationary phase analysis for $\mu_c > \mu$}
\begin{figure}[htb!]
    \centering
    \includegraphics[width=9cm]{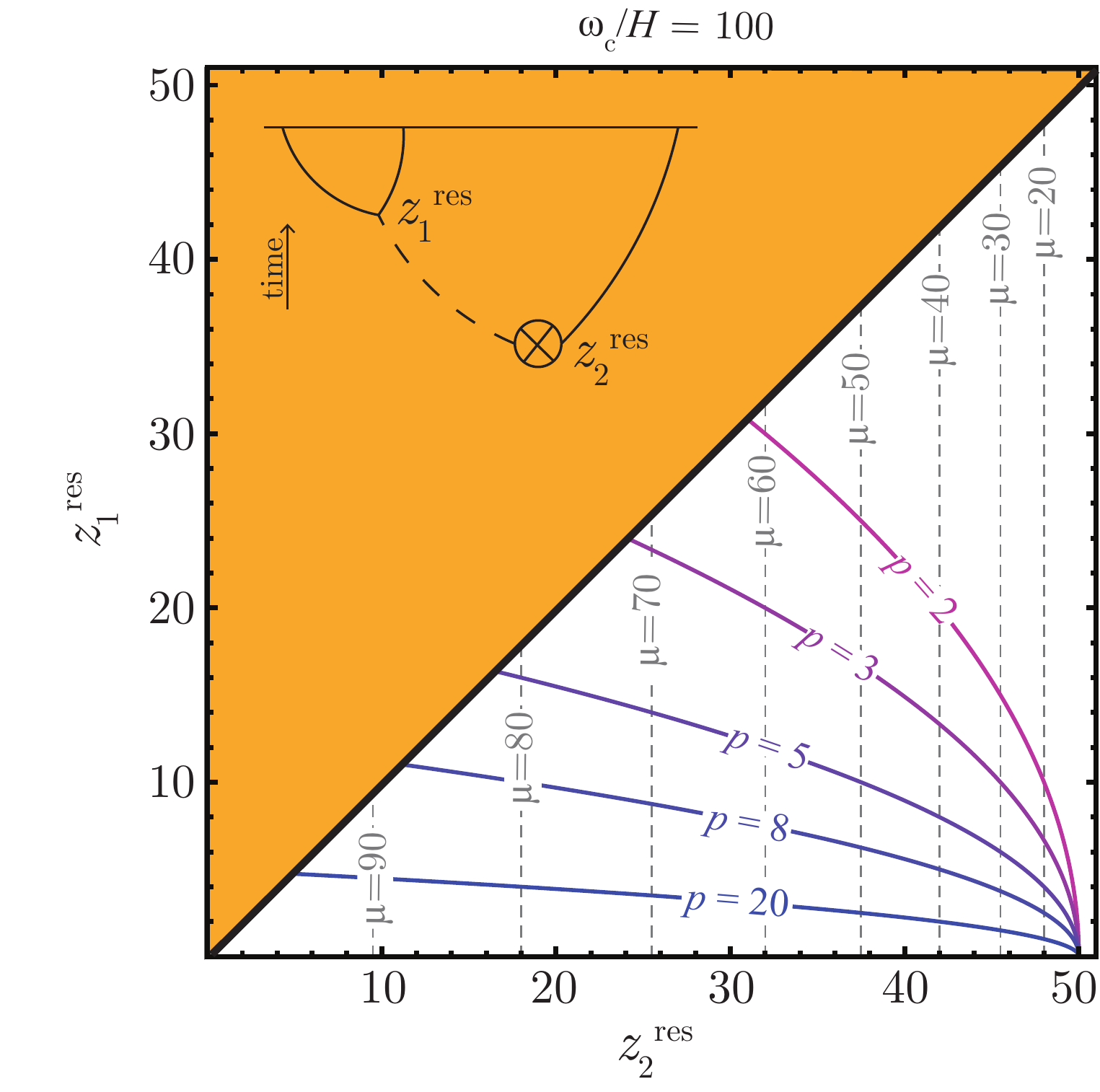}
    \caption{Resonance points in the $z_2^\res-z_1^\res$ plane, according to Eqs.~\eqref{eq:eta2res} and \eqref{eq:eta1res}. Massive field and inflaton resonate at time $z_2^{\rm res}$. At a later time $z_1^{\rm res}$, the massive field decays into a pair of inflaton quanta through resonance. In this figure, for a specific value of $\mu\simeq m_\chi/H$ and $p=(k_1+k_2)/k_3$, intersection of the corresponding vertical dashed line with the given curve for that specific value of $p$, shows the resonance points. Here we choose $\mu_c\simeq\omega_\c/H=100$. It justifies the stationary phase approximation with $z_2^{\rm res}>z_1^{\rm res}$ with only a moderate squeezing $p\simeq 5$ for most masses below $\mu_c$.}
    \label{fig:resonances}
\end{figure}

The $++$ diagram can be physically described in the following way: at the 2pt mixing vertex, due to the oscillations in $B_\c(t)$, massive field $\chi$ and inflaton $\phi$ get produced at time $\eta_2\sim\eta_2^{\rm res}$. Following this, at time $\eta_1\sim\eta_1^{\rm res}$, $\chi$ decays into a pair of inflaton quanta. The estimation of $\eta_2^{\rm res}$ and $\eta_1^{\rm res}$ can be done using simple arguments relying on stationary phase or equivalently, energy conservation. For the mixing vertex, a resonant particle production happens for $\mu_c = |k_3\eta_2^{\rm res}|+\sqrt{(k_3\eta_2^{\rm res})^2+\mu^2}$, where $\mu_c$ is the energy (in unit $H$) of the injected classical oscillation, the first term on the right hand side is the energy of the $\phi$ field, and the second is the energy of the new massive $\chi$ field. 
Adapting the notations defined in \eqref{z_notations} to $z_2\equiv-k_3\eta_2$,
this condition leads to
\begin{align}\label{eq:eta2res}
    z_2^\res \simeq \frac{\mu_c^2-\mu^2}{2\mu_c}.    
\end{align}
For the 3pt decay vertex where the background classical oscillation is absent, the resonance happens between the quantum modes of the massive field $\chi$ and the inflaton $\phi$. Using similar arguments, we get $\sqrt{(k_3\eta_1^{\rm res})^2+\mu^2}= |k_{12}\eta_1^{\rm res}|$, where $k_{12} \equiv k_1 + k_2$. This in the squeezed limit $k_3\ll k_1,k_2$ reduces to,
\begin{align}\label{eq:eta1res}
    p\,z_1^\res \equiv |k_{12}\eta_1^{\rm res}|\simeq \mu ~,
\end{align}
where we recall the definition $p\equiv k_{12}/k_3$ 
and $z_1^{\rm res}\equiv k_3 \eta_1^{\rm res}$.
We are interested in the regime $z_2^{\rm res}>z_1^{\rm res}$, because, as explained above, diagram of this time ordering is unsuppressed.
If $\mu_c\gg \mu$, $z_2^{\rm res}>z_1^{\rm res}$ can be satisfied as long as $p>2$. As $\mu$ gets close to $\mu_c$, a more squeezed configuration (i.e.~larger $p$) is required to satisfy $z_2^{\rm res}>z_1^{\rm res}$.  (An example of  
the two-dimensional plane of the resonance points are shown in Fig.~\ref{fig:resonances}.) 
Also, in this limit, because the two resonances happen independently, the nested integrals can be factorized.
(We will numerically check in Fig.~\ref{fig:shapeFULL} and Fig.~\ref{fig:scaleFULL} that this factorization approximation reproduces full numerical computation very well.)

To calculate integrals using stationary phase approximations we replace Hankel functions with proper late or early-time asymptotic forms for large $\mu$. For late-time expansion ($z \lesssim \sqrt{\mu}$) we have\footnote{Note that $H^{(2)}_{i\mu}(z) = \left(H^{(1)}_{i\mu}(z)\right)^* e^{-\pi \mu}$.}
\begin{align}\label{eq:late_time_expansion}
    H^{(1)}_{i\mu}(z) &\simeq e^{-i \pi/4} \sqrt{\dfrac{2}{\pi \mu}} e^{\pi\mu/2} e^{i\mu(1-\log \mu)} \left( \dfrac{z}{2} \right)^{i \mu},\nonumber\\
    H^{(2)}_{i\mu}(z) &\simeq e^{+i \pi/4} \sqrt{\dfrac{2}{\pi \mu}} e^{-\pi\mu/2} e^{-i\mu(1-\log \mu)} \left( \dfrac{z}{2} \right)^{-i \mu}.
\end{align}
And for early-time expansion ($z \gtrsim \mu^2$) we have
\begin{align}\label{eq:early_time_expansion}
    H^{(1)}_{i\mu}(z) &\simeq e^{-i \pi/4} \sqrt{\dfrac{2}{\pi z}}  e^{\pi \mu /2}  \, e^{i z},\nonumber\\
    H^{(2)}_{i\mu}(z) &\simeq e^{+i \pi/4} \sqrt{\dfrac{2}{\pi z}}  e^{-\pi \mu /2}e^{-i z}.
\end{align}
Given the asymptotic forms, we can analytically calculate $\CI_2(z_1)$ in two limiting cases: early-time resonance and late-time resonance, which happens for the case $\mu_c\gg \mu$ and $\mu_c \approx \mu$, respectively.

If the mixing vertex resonance happens at an early time when the above early-time expansion approximation \eqref{eq:early_time_expansion} is valid, i.e., $z_2^\res \gtrsim \mu^2$, the stationary phase approximation yields
\begin{align}\label{eq:I2_early}
    \CI_2^{\rm early} \simeq&~ \sqrt{\dfrac{2}{\pi }} \, e^{i \pi/4} \, \int_{\mu^2}^{\infty} \d z_2 \, z_2^{n - 1 + i \mu_c} \, e^{- 2 i z_2}\nonumber\\
    \simeq&~  e^{- i \mu_c} \dfrac{2}{\sqrt{\mu_c}}\left(\dfrac{\mu_c}{2}\right)^{n+i\mu_c}\,, 
\end{align}
where the stationary phase is at $z_2^\res = \mu_c/2$ which is consistent with the energy conservation expectations Eq.~\eqref{eq:eta2res}, for $\mu_c \gg \mu$. In the first line we used Eq.~\eqref{eq:early_time_expansion} and in the second line we used the stationary phase approximation.\footnote{The stationary phase approximation is given by $\int \d z ~g(z) e^{i f(z)} \simeq g(z_\res) e^{i f(z_\res) \pm i \pi/4} \sqrt{2\pi /|f''(z_\res)|}$ for stationary phase point $f'(z_\res) = 0$. In this subsection we are interested in integrals of the following form:
\begin{align}
    \int \d z z^{x+iy} e^{-iz} \simeq \sqrt{2\pi y}\,y^x e^{-i\pi/4 +iy(\log y -1)},
\end{align}
where the result is obtained using stationary phase approximation. One can also calculate the full result
\begin{align}
    \int_0^\infty \d z z^{x+iy} e^{-iz} = -i e^{-i\pi(x+iy)/2}\Gamma(1+x+iy)\,.
\end{align}
This result reproduces the stationary phase approximation in the large $y$ limit using \eqref{eq:gamma_asymptotic}, where the second-order correction is $\mathcal{O}\left((1+6x+6x^2)/(12y)\right)$ smaller than the first order. The conditions for the stationary phase approximation are $1/y \ll 1$ and $x^2/y \ll1$. In this approximation, we expand $f(z)$ to the second order and the integration picks up its main contribution in the region $\Delta z \sim \sqrt{y}$ around the resonant point $z= y$. The condition $1/y \ll 1$ ensures that the higher order terms in the expansion of $f(z)$ are negligible. The condition $x^2/y \ll 1$ ensures that the change of the prefactor $g(z)$ from $g(z=y)$ over the region $\Delta z \sim \sqrt{y}$ is negligible, because $\Delta \ln{g(z)}\sim (x/y) \Delta z$. Both conditions are satisfied in our examples.
}

If the mixing vertex resonance happens at a late time when the above late-time expansion approximation \eqref{eq:late_time_expansion} is valid, i.e., $z_2^\res \lesssim \sqrt{\mu}$, the stationary phase approximation gives
\begin{align}\label{eq:I2_late}
    \CI_2^{\rm late} \simeq&~ \sqrt{\dfrac{2}{\pi \mu}} \, e^{i \pi/4} \, 2^{i\mu}\, e^{-i \mu(1-\log\mu)} \, \int_0^{\sqrt{\mu}} \d z_2 \, z_2^{n -1/2 + i\mu_c- i \mu} \, e^{-i z_2} \nonumber\\
    \simeq &~ \dfrac{2}{\sqrt{\mu}} e^{-i \mu_c} (2\mu)^{i \mu} (\mu_c -\mu)^{n+i(\mu_c - \mu)},
\end{align}
in which the stationary phase happens at $z_2^\res \simeq \mu_c - \mu$, the $\mu_c \simeq \mu$ limit of the energy conservation expectation Eq.~\eqref{eq:eta2res}. In the first line we used Eq.~\eqref{eq:late_time_expansion} and in the second line we used the stationary phase approximation.

For the integration at the cubic vertex, the stationary point is given by Eq.~\eqref{eq:eta1res}, i.e., $z_1^\res=\mu/p$. This satisfies the late-time expansion condition $z_1^{\rm res} \lesssim \sqrt{\mu}$ in the squeezed limit $p\gtrsim \sqrt{\mu}$, where we can use the approximation \eqref{eq:late_time_expansion}. The result is
\begin{align}\label{eq:I3_stationary}
    \simeq&~ \sqrt{\dfrac{2}{\pi \mu}} \, e^{-i \pi/4} \, 2^{-i\mu}\, e^{i \mu(1-\log\mu)} \, \int_0^{\sqrt{\mu}} \d z_1 \, z_1^{3/2 + i\mu} \, e^{-i p z_1}\nonumber\\
    \simeq&~  -i 2^{1-i\mu} \mu^{3/2} p^{-5/2-i\mu}\,.
\end{align}
%Where again we used Eq.~\eqref{eq:early_time_expansion} in the first line and stationary phase approximation in the second line.
Therefore, as long as $\mu_c/2,\mu_c-\mu>\mu/p$ so that $z_2^{\rm res}>z_1^{\rm res}$ is satisfied, none of the resonant contributions Eqs.~\eqref{eq:I2_late}-\eqref{eq:I3_stationary} has any exponentially suppression factors and, consequently, the $\chi$-mediated bispectrum in our case, unlike the pure quantum case, is not suppressed by any Boltzmann factor. While we present the full squeezed-limit calculation in Sec.~\ref{subsec.analytics} beyond the stationary phase results given above, we can already extract the momentum dependence of the bispectrum. This gives the generic signature of the classical cosmological collider signal of the heavy particle $\chi$ as follows: 
\begin{align}
    \braket{\delta\phi^3}'_{++} \xrightarrow{k_r\ll k_3\ll k_1} \dfrac{1}{k_1 k_2 k_3^4} \, \left(\dfrac{k_3}{k_r}\right)^{-n - i \mu_{\rm c}} p^{-5/2-i\mu},
\end{align}
where $k_r\sim\mu_{\rm c}k_0$ with $k_0=-1/\eta_0$ being the scale corresponding to the sharp feature. We will give a more precise definition of $k_r$ in the following discussion. The scale-dependent component $(k_3/k_r)^{-n-i\mu_c}$ comes from the $z_0^{-n-i\mu_c}$ factor in Eq.~\eqref{eq:++}.

Momentum dependence of bispectrum is usually characterized by dimensionless shape function
\begin{align}
    \mathcal{S}(k_1,k_2,k_3) \propto k_1^2k_2^2k_3^2\braket{\delta\phi^3}'\,.
\end{align}
Thus, in the squeezed limit we have
\begin{align}\label{eq:fsimple}
    \mathcal{S}(k_1,k_2,k_3) \xrightarrow{k_r \ll k_3\ll k_1} \left(\dfrac{k_3}{k_r}\right)^{-n - i \mu_{\rm c}} \left(\dfrac{k_1}{k_3}\right)^{-1/2-i\mu}
    + {\rm c.c.}
    \,.
\end{align}
The first factor shows the scale dependence of the result coming from the temporal oscillations of the classical source. This is a unique feature of the classical cosmological collider signal, not present in the conventional quantum analog where classical oscillations do not play a role. The extra decay factor, for non-zero $n$, signifies the decay of the classical source. For example, $n=3/2$ is the $1/\sqrt{\text{volume}}$ dilution of the classical oscillations of the sourcing massive field.
The second factor shows a familiar oscillatory shape-dependent clock signal, which is a signature of the quantum fluctuation of the massive field $\chi$.

The momentum dependence of this first factor can be intuitively understood as follows. 
In evaluating the correlation function, there are two oscillation factors, namely the mode function and the background oscillation, which resonate with each other at some point.
With the inflationary background and fixed-frequency background oscillation, the mode function with momentum $k$ picks up a $k$-independent phase due to the resonance, $\exp(-i k \eta_{\rm res})=\exp(i\mu_c)$ or $\exp(i\mu/2)$. Therefore, the final momentum-dependent phase is determined by that of the background oscillation.  For simplicity, focusing on the case $\mu\ll \mu_c$, Eq.~\eqref{eq:eta2res} implies that $|\eta_2^{\rm res}|\simeq \mu_c/(2 k_3)$ is the time when the oscillating source produces the $\phi$ and $\chi$ quanta. This implies that, between $\eta_0$ (the onset of the sharp feature that induces the oscillation) and $\eta_2^{\rm res}$, the only non-trivial dynamics is the oscillation (and dilution for $n\neq 0$) of the source, which goes as $(\eta_2^{\rm res}/\eta_0)^{n+i\mu_c}\sim (\mu_c/ (k_3|\eta_0|))^{n+i\mu_c}$. Thus, using $k_r\sim \mu_ck_0$ and $k_0=-1/\eta_0$, we get the above momentum dependence.
After the $\chi$-field is resonantly produced, its quantum fluctuations serves an oscillatory source for the resonant production of the $\phi$-fields in the 3pt vertex.
Similarly, the second factor in \eqref{eq:fsimple} encapsulates the temporal oscillations of the produced $\chi$-field between $\eta_2^{\rm res}$ and $\eta_1^{\rm res}$.

We define and calculate full squeezed-limit form of the shape function $\CS$ analytically with the appropriate proportionality factors in Sec.~\ref{subsec.analytics}. We include complete numerical results for more general momentum configuration in Sec.~\ref{subsec.numerics_osc}.

As an aside, let us estimate the parametric dependence of the three point function in the scenario where both $B(t)$ and $C(t)$ are present. We focus on the case where the two point function takes place earlier than the three point function.
The mixing vertex can be estimated as before since it is mediated by $B(t)$, and parametric dependence is as in Eq.~\eqref{eq:I2_early}. However, the cubic vertex can have a resonance dictated by energy injection due to $C(t)$. To estimate its effect, we can first write $C(t) = C_0 e^{i \mu_c t} + {\rm c.c.}$. Then we note that the oscillating frequency would change $z_1^{i\mu} \rightarrow z_1^{i\mu \pm i\mu_c}$ in the integrand of Eq.~\eqref{eq:I3_stationary}. With the $+$ sign in the exponent, one can have a resonance and the result would be given by,
\begin{align}
    C_0 \frac{(\mu+\mu_c)^2}{\sqrt{\mu}} p^{-5/2-i\mu - i\mu_c}.
\end{align}
Therefore, combining both the two point and the three point vertex, we can obtain the parametric dependence of the answer on $\mu,\mu_c$ (dropping some unimportant prefactors and phases):
\begin{align}
    \langle \delta \phi \rangle'_{++} \propto C_0  \mu_c^{n+i\mu_c} \frac{(\mu+\mu_c)^2}{\sqrt{\mu \mu_c}} p^{-5/2-i\mu - i\mu_c}.
\end{align}
This still has the relevant clock signal. To compare to the scenario where $C(t)$ is absent,
\begin{align}
    \frac{\langle \delta \phi \rangle'_{++}\big\rvert_{C_0\neq 0}}{\langle \delta \phi \rangle'_{++}\big\rvert_{C_0 = 0}} \propto C_0\frac{(\mu + \mu_c)^2}{\mu^2}.
\end{align}
Therefore, depending on the magnitude of $C_0$, this contribution can additionally enhance the signals or be subdominant. To capture the basic mechanism in a simple way, however, we still focus on $C_0 = 0$ in the rest of the work, leaving a detailed analysis for nonzero $C(t)$ for future work.

\subsubsection{IR expansion analysis for $\mu_c < \mu$}\label{subsec.IRexpansion}
\label{subsec:irexpansion}
For $\mu_c < \mu$ there is no stationary phase for the inner integral, as can be seen from the energy conservation argument in Eq.~\eqref{eq:eta2res}. The leading contribution comes from the late-time integration and is power-law suppressed, 
\begin{align}\label{eq:final_scaling}
   ++ \text{diagram}\sim \dfrac{1}{\mu^2} \simeq \dfrac{H^2}{m_\chi^2}\,.
\end{align}
However, this term does not have the oscillatory signal encoding mass of the $\chi$ field.
This is expected since the background oscillation is the source of energy injection in this scenario and, if its frequency is smaller than the particle mass, the particle can not be produced on-shell. In that case, it only contributes via an off-shell contact term $\propto 1/m_\chi^2$. We describe the details of this computation in App.~\ref{app:IRexpansion}.

The oscillatory signal encoding the mass of the $\chi$ field, on the other hand, is expected to be exponentially suppressed if $\mu_c<\mu$. The suppression again comes from the 2pt vertex, namely, the inner integral of \eqref{eq:++}. The massive field mode function $v_{k_3}^*$ contains two components. One of them oscillates as $\sim e^{i\mu H t_2}$. Together with the background oscillation from $B_c$ and the massless mode function $u'^*_{k_3}$, the inner integrand behaves as $\sim e^{i(\mu\pm \mu_c)Ht_2} e^{ik_3\eta_2}$. Such an integral contains no resonance and, as demonstrated in a similar integral in App.~A of \cite{Chen:2015lza}, is of order $e^{-\pi(\mu-\mu_c)}$. The other component of $v_{k_3}^*$ oscillates as $\sim e^{-\pi\mu} e^{-i\mu H t_2}$. With the background oscillation and massless mode function, the inner integrand behaves as $\sim e^{-\pi\mu} e^{-i(\mu\pm \mu_c)Ht_2} e^{ik_3\eta_2}$. The oscillation part now resonates, hence this integral is of order $\sim e^{-\pi\mu}$. Combining both components of $v_{k_3}^*$, we see the appearance of a modified Boltzmann suppression factor $e^{-\pi(\mu-\mu_c)}$, which for $\mu\gg\mu_c$ approaches the familiar Boltzmann suppression factor $e^{-\pi\mu}$. 
We give more detailed derivations of this modified Boltzmann factor in Sec.~\ref{subsec.analytics} and App.~\ref{app:IRexpansion}.

\subsection{Corrections to power spectrum}\label{subsec:powerspectrum}
Having discussed the contributions to the bispectrum that constitute the primary observable in this work, now we discuss the contributions to the power spectrum as well. Our goal is to ensure that the parameter space, where the bispectrum is observably large, is not ruled out based on the existing searches for the signatures in power spectrum. On the other hand, the signatures in the power spectrum will likely provide the first hints of the existence of the phenomena studied in this work, because it is typically easier to experimentally search for these feature signals in power spectrum than bispectrum.

As before, the power spectrum in the in-in formalism can be calculated using Eq.~\eqref{eq.ininmaster},
\begin{align}
    \braket{\delta \phi^2} = \bra{0} \left[\bar{T}  e^{i \int_{-\infty}^{t_{\rm end}} \d t' H_\interaction(t')}\right] \, \delta \phi^2(t_{\rm end}) \, \left[T e^{ -i \int_{-\infty}^{t_{\rm end}} \d t' H_\interaction(t')}\right]\ket{0}\,.
\end{align}
There are three relevant classes of diagrams which contribute to the power spectrum. We discuss each of them below. 

\begin{figure}[tb!]
    \centering
    \includegraphics[width=15cm]{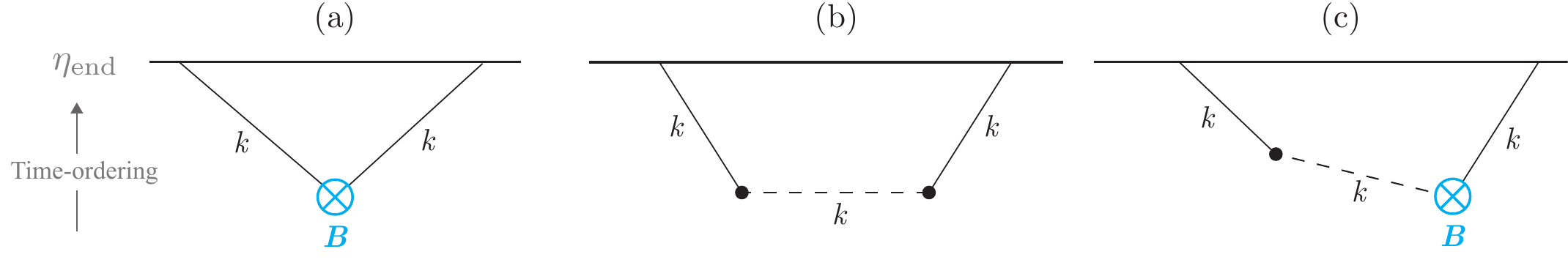}
    \caption{Feynman diagrams contributing to power spectrum: (a) direct feature contribution, (b) heavy field contribution ($++$ diagram), (c) ``feature-heavy field'' contribution ($++$ diagram).}
    \label{fig:powerspectrum}
\end{figure}

\paragraph{Direct feature contribution.} Signatures of the classical oscillations in the power spectrum has been studied in previous works~\cite{Chen:2011zf,Chen:2014cwa}. The direct feature contribution, shown in diagram~\ref{fig:powerspectrum}a, originates from a mixing terms such as $B_d(t) (\partial\phi)^2$ and this does not directly involve the heavy field of interest $\chi$. This diagram gives the correction
\begin{align}
   \dfrac{\Delta P_\zeta}{P_\zeta}=-2i\int_{-\infty}^0\dfrac{\d\eta}{(H\eta)^4}\left(\du_{k}^{2}-\dfrac{k^2}{a^2}u_{k}^2\right)B_d(\eta)+ \text{c.c.}\,.
\end{align}
Assuming $B_d \simeq B_c$ (as we will motivate in Sec.~\ref{sec.applications}), this yields
\begin{align}\label{eq:power_correction_direct}
    \dfrac{\Delta P_\zeta}{P_\zeta}=\sqrt{2\pi}\dfrac{B_0}{2}\mu_c^{1/2} \left(\dfrac{2k}{k_r}\right)^{-n-i\mu_c+i\alpha_1}+\text{c.c.}\,,
\end{align}
where $k_r\sim\mu_{\rm c}k_0/2=-\mu_{\rm c}/(2\eta_0)$ as before and $\alpha_1$ is a $k$-independent phase. The momentum dependence involving $k_r$ is the same as in Eq.~\eqref{eq:fsimple} and captures source dynamics between the time of the sharp feature $\eta_0$ and the time of production, $\mu_c/(2k)$, of inflaton with momentum $k$. See Ref.~\cite{Braglia:2021sun,Braglia:2021rej} for a recent analysis of constraining such clock signals using the {\it Planck} power spectrum residuals, as well as forecasts for future CMB experiments. 

\paragraph{Heavy field contribution.} The heavy field mediated correction without any oscillatory feature contribution (i.e. $\mu_c=0$) has also been studied in previous works \cite{Chen:2012ge}. The $+-$ contribution is shown to be exponentially suppressed and the dominant contribution is from the $++$ contribution shown in diagram~\ref{fig:powerspectrum}b. This diagram gives the correction:
\begin{align}
    \dfrac{\Delta P_\zeta}{P_\zeta}=\rho^2\int\dfrac{\d\eta_1}{(H\eta_1)^4}\du_k^*v_k(\eta_1)\int\dfrac{\d\eta_2}{(H\eta_2)^4}\du^*_kv_k^*(\eta_2)+\text{c.c.}\,.
\end{align}
Complete analytic solution to this integral can be found in Ref.~\cite{Chen:2012ge}. The leading contribution comes from late time ($|k\eta|<\sqrt{\mu}$) part of the integral, which gives
\begin{align}\label{eq:power_heavy}
    \dfrac{\Delta P_\zeta}{P_\zeta}\simeq\dfrac{\rho^2}{H^2}\dfrac{1}{\mu^2}\,.
\end{align}
This result can be understood as coming from integrating out the heavy field with mass $\mu$ which generates the standard $\propto 1/\mu^2$ correction as in flat space processes. 

\paragraph{Feature-heavy field contribution.} This contribution similarly consists of $+-$ and $++$ parts (and their conjugates) as two vertices can be chosen from time-ordered or anti time-ordered operators. The $+-$ contribution reduces to two separate integrals. One of the integrals contains no background oscillations and is exponentially suppressed, just like the case of bispectrum. Explicit result of this $+-$ contribution is given in Eq.~\eqref{eq:+-power}. Let us therefore focus on the $++$ contribution, shown in diagram~\ref{fig:powerspectrum}c. This contribution is
\begin{align}
&\dfrac{\Delta P_\zeta}{P_\zeta}\bigg\rvert_{++}=\rho^2\int\dfrac{\d\eta_1}{(H\eta_1)^4}\du^*_kv_k(\eta_1)\int\dfrac{\d\eta_2}{(H\eta_2)^4}\du^*_k v^*_kB_\c(\eta_2)\\
&= -\dfrac{\pi}{4}\dfrac{\rho^2}{H^2}\dfrac{i}{2}B_0 z_0^{-n - i \mu_{\rm c}} \, e^{-\pi\mu/2}\int_0^\infty \dfrac{\d z_1}{\sqrt{z_1}}H_{i\mu}^{(1)}(z_1) \, e^{-i z_1}   \,\underbrace{e^{\pi\mu/2} \int_{z_1}^{\infty} \dfrac{\d z_2}{\sqrt{z_2}} \, H_{i\mu}^{(2)}(z_2) \, e^{-i z_2} \, z_2^{n+i \mu_{\rm c}}\theta(z_0-z_2)}_{\CI_2^+(z_1)}\nonumber\\
&- \{\mu_c\rightarrow-\mu_c\}\,.
\end{align}
The integral $\CI_2^+(z_1)$ was previously calculated and it has a resonant contribution: early and late resonance stationary phase results, given in Eqs.~\eqref{eq:I2_early} and \eqref{eq:I2_late}. Note that the $z_1$ vertex happens after the $z_2$ vertex. Thus, with the resonance point contribution to the inner integral at $z_2^\res$, the outer integral can be simplified as follows:
\begin{align}\label{eq:outerintegral_clockHeavyField_powerspectrum}
    e^{-\pi\mu/2}\int_0^{z_2^{\res}} \dfrac{\d z_1}{\sqrt{z_1}}H_{i\mu}^{(1)}(z_1) \, e^{-i z_1}\,,
\end{align}
where the integration is confined to $0$ and $z_2^\res$. Integral of Eq.~\eqref{eq:outerintegral_clockHeavyField_powerspectrum} can be calculated analytically as
\begin{align}
    2e^{-\pi\mu/2}\sqrt{z_2^{\res}}\left[(1+\rm{coth}(\pi\mu))\CA_+-\rm{csch}(\pi\mu)\CA_-\right]\,,
    \label{eq:outerIntegral_clockheavyfield_powerspectrum}
\end{align}
where
\begin{align}
    \CA_\pm = \left(\dfrac{z_2^{\res}}{2}\right)^{\pm i\mu} \dfrac{1}{(1\pm2i\mu)\Gamma(1\pm i\mu)}~_2F_2(1/2\pm i\mu,1/2\pm i\mu;3/2\pm i\mu,1\pm2i\mu;-2iz_2^{\res})\,.
\end{align}
It can be numerically checked that Eq.~\eqref{eq:outerIntegral_clockheavyfield_powerspectrum} is power-law suppressed as $\sim 1/\mu^{3/2}$. By combining this result and the early resonant contribution Eq.~\eqref{eq:I2_early}, we get the following contribution to power spectrum:
\begin{align}\label{eq:power_correction_mixed}
    \dfrac{\Delta P_\zeta}{P_\zeta}\sim\dfrac{\rho^2}{H^2\sqrt{\mu^3\mu_c}}B_0\mu_c^{1/2}\left(\dfrac{2k}{k_r}\right)^{-n-i\mu_c}+\text{c.c.}\sim\dfrac{\rho^2}{H^2\sqrt{\mu^3\mu_c}}\times\text{Eq.~\eqref{eq:power_correction_direct}}\,.
\end{align}
So this contribution is similar to the direct clock contribution, but it is more suppressed since we will require $\rho/H <\mu, \mu_c $ in the following. We note that for the above estimate we used $B_d(t)\simeq B_\c(t)$ for simplicity and we will give two examples of this in Sec.~\ref{sec.applications}. However, for other scenarios $B_d(t)$ can very well be parametrically different from $B_\c(t)$ and correspondingly the two corrections to the power spectrum would be different as well. In the parameter space where $\CI_2$ has late-time stationary point as in Eq.~\eqref{eq:I2_late}, the correction is also suppressed. 

While the above is a parametric estimate, we have also confirmed numerically for $\mu=5$, $\mu_c=10$, that the proportionality factor in Eq.~\eqref{eq:power_correction_mixed} is $\mathcal{O}(1)$. Hence we continue to use the above parametric estimate. Given the smallness of this feature-heavy field contribution, we only consider constraints on direct contribution, shown in diagram~\ref{fig:powerspectrum}a, in the following discussions.

\subsection{Analytical computation in the squeezed limit}\label{subsec.analytics}
In this subsection we present approximate analytic results (with all terms included) in the squeezed limit for the interesting part of the parameter space $\mu< \mu_c$, where the oscillatory signatures of a heavy field are present. In the limit of large squeezing $p\gg 1$, according to Eqs.~\eqref{eq:eta2res} and \eqref{eq:eta1res} we have
\begin{align}
    z_2^\res \gg z_1^\res\,.
\end{align}
Since the stationary phases for inner and outer integrands are then at sufficiently separated temporal points, we are justified to treat the nested integral in Eq.~\eqref{eq:++} as two \textit{factorized} integrals over the full range $z_{1,2}=0$ to $\infty$. To illustrate exactly how much squeezing is necessary for given values of $\mu$, we showed in Fig.~\ref{fig:resonances} the locations of the stationary points for $\mu_c=100$. It shows that already with moderate squeezing $p\simeq 5$, non-Gaussianity due to massive particles up to $\mu\lesssim 80$ can be calculated using this factorization approximation. We will compare this approximation with a full numerical calculation in Fig.~\ref{fig:shapeFULL} and Fig.~\ref{fig:scaleFULL}.

\paragraph{The mixing vertex.}
This integral has the structure,
\begin{align}
(-i)e^{\pi\mu/2}\frac{-\sqrt{\pi}}{4}\dfrac{i}{2}B_0\frac{\rho}{k_3^{3/2}}\left(\frac{1}{|k_3\eta_0|}\right)^{n+i\mu_c}\int_0^{\infty}\d z z^{n+i\mu_c-1/2}e^{-iz} H_{i\mu}^{(2)}(z)-\{\mu_c\rightarrow -\mu_c\},   
\end{align}
which can be evaluated to be,
\begin{align}
\dfrac{i}{2}B_0\frac{\rho}{4}\frac{(+i)}{k_3^{3/2}}\frac{(-i/2)^{n-1/2+i\mu_c}}{(|k_3\eta_0|)^{n+i\mu_c}}\frac{\Gamma(n+1/2+i\mu_c+i\mu)\Gamma(n+1/2+i\mu_c-i\mu)}{\Gamma(n+1+i\mu_c)}-\{\mu_c\rightarrow -\mu_c\}.  
\end{align}
To get an analytic expression, here we have not considered the step function $\theta(z_0-z_2)$ appearing in Eq.~\eqref{eq:++}. For sufficiently large $z_0$ this is justified since for large argument the integrand falls of as $e^{-2z}$ already. The scale $\eta_0$ enters through the phase and envelop of the background oscillation $B_c(t)$.

\paragraph{The cubic vertex.}
This integral also has a similar structure except for the extra appearance of spatial derivatives in the vertex,
\begin{align}
(-i)\dfrac{\lambda}{\Lambda}\frac{1}{4k_1^3k_2^3k_3^{1/2}}\frac{\sqrt{\pi}}{2}e^{-\pi\mu/2}\int^{\infty}_0\frac{\d z}{\sqrt{z}}H_{i\mu}^{(1)}(z)\mathcal{D}e^{-ip z}.    
\end{align}
Here the operator $\mathcal{D}$ takes care of the spatial derivatives,
\begin{align}\label{eq:D}
\mathcal{D}=k_1^2k_2^2\partial_{k_{12}}^2+(-\vec{k}_1\cdot \vec{k}_2)(1-k_{12}\partial_{k_{12}}+k_1 k_2 \partial_{k_{12}}^2),
\end{align}
where the partial derivative $\partial_{k_{12}}$ acts on the $k_{12}\equiv k_1+k_2$. This form of $\mathcal{D}$ can be derived by writing the inflaton contractions in the compact form,
\begin{align}
\langle \delta\phi(\eta_{\rm end},\vec{k}_1)\delta\phi(\eta_{\rm end},\vec{k}_2)\partial_\mu\delta\phi(\eta,-\vec{k}_1) \partial^\mu\delta\phi(\eta,-\vec{k}_2)\rangle =\frac{\eta^2}{4k_1^3k_2^3}\mathcal{D}e^{i(k_1+k_2)\eta},
\end{align}
in which at the end of inflation $\eta_{\rm end}\rightarrow 0$. In the squeezed limit, i.e., $p\gg 1$, the above integral for the cubic vertex can be evaluated to be (see Ref.~\cite{Kumar:2017ecc}),
\begin{align}
\dfrac{1}{32\sqrt{2}}\dfrac{\lambda}{\Lambda}\frac{p^2k_3^{3/2}}{k_1^3 k_2^3}e^{-3i\pi/4}  \frac{\Gamma(-2i\mu)\Gamma(1/2+i\mu)}{\Gamma(1/2-i\mu)}\left(-\frac{p}{2}\right)^{-\frac{1}{2}-i\mu}\left(\frac{3}{2}+i\mu\right)  \left(\frac{5}{2}+i\mu\right) +\{ \mu\rightarrow -\mu \}.
\end{align}

\paragraph{The $++$ contribution.} Now we can put the above results together to get the full $++$ contribution corresponding to Fig.~\ref{fig:3ptinin}b in the squeezed limit,
\begin{align} \label{eq:dphi3}
\braket{\delta\phi^3}'_{++}
\approx
\frac{H^3}{k_1^3 k_3^3}\left(\dfrac{k_3}{k_r}\right)^{-n - i \mu_{\rm c}} \left(\dfrac{k_1}{k_3}\right)^{-3/2-i\mu}\mathcal{F}(\mu,\mu_c,n),
\end{align}
with
\begin{align}
\mathcal{F}(\mu,\mu_c,n) = &\frac{1}{16}  \frac{\rho\lambda}{\Lambda}\dfrac{B_0}{2}e^{\pi\mu}\left(\dfrac{-i\mu_c}{\mu_c^2-\mu^2}\right)^{n+i\mu_c} \left(\frac{3}{2}+i\mu\right)\left(\frac{5}{2}+i\mu\right)\times \nonumber\\ &\frac{\Gamma(n+\frac{1}{2}+i\mu_c+i\mu)\Gamma(n+\frac{1}{2}+i\mu_c-i\mu)}{\Gamma(n+1+i\mu_c)}\frac{\Gamma(-2i\mu)\Gamma(\frac{1}{2}+i\mu)}{\Gamma(\frac{1}{2}-i\mu)}+\cdots,
\end{align}
where the $\cdots$ represents subleading corrections and $k_r = k_0(\mu_c^2-\mu^2)/(2\mu_c)$. 
For the above result to be valid, we require $\mu<\mu_c$ following from $|\eta_2^{\rm res}|\geq |\eta_3|$.
We also note that, if $n>0$, the result \eqref{eq:dphi3} diverges as $k_3\to 0$ because we have not used the step function $\theta(z_0-z_2)$ appearing in Eq.~\eqref{eq:++}. Once this cutoff is imposed,
the resonance for the $k_3$ mode, given in Eq.~\eqref{eq:eta2res}, must happen after the time corresponding to the sharp feature. This implies, 
\begin{align}
    |\eta_2^{\rm res}|\leq |\eta_0| \equiv 1/k_0,
\end{align}
leading to $k_3\geq k_r$, which becomes the validity region of the result. Besides imposing the validity region for this clock signal, a sharp feature such as this cutoff also contributes to the final signal. We will show the net effect numerically in Sec.~\ref{subsec.numerics_osc}. We will discuss the effect of sharp features in Sec.~\ref{sec.sharp}.

\paragraph{Asymptotic Form.} For $\mu_c\gg\mu\gg1$, we find that
\begin{align}
    \mathcal{F}(\mu,\mu_c,n) \propto \sqrt{\dfrac{\mu^3}{\mu_c}}\,,
\end{align}
where, we used \eqref{eq:gamma_asymptotic} to expand Gamma functions. For large $p$, the above factorization approximation works for $\mu>\mu_c$ as well, because large $p$ delays the resonant time for the 3pt vertex. Since in that case the injected frequency is smaller than the mass of the heavy particle $H\mu$, we expect NG would be exponentially suppressed. This can be checked explicitly by taking large $\mu,\mu_c$ limit of $\mathcal{F}$ with $\mu_c<\mu$ and use the asymptotic expansion  $|\Gamma(x+iy)|\sim \sqrt{2\pi}|y|^{x-1/2}e^{-\pi|y|/2}$,
\begin{equation}\label{eq.large_mu_ana}
\begin{aligned}
\mathcal{F}(\mu,\mu_c,n) \sim e^{-\pi(\mu-\mu_c)}.
\end{aligned}
\end{equation}
In the above we have kept only the exponential dependence of $\mathcal{F}$ on $\mu$ and $\mu_c$.

\paragraph{The three point function.} Finally, we can combine the above results to get the dimensionless three point function, (using $\zeta \simeq - \dfrac{H}{\dphi_0} \delta\phi$)
\begin{align}\label{eq:shapefunc_def}
    \mathcal{S}(k_1,k_2,k_3)= \dfrac{(k_1k_2k_3)^2}{A^2}\langle\zeta(\vec{k}_1)\zeta(\vec{k}_2)\zeta(\vec{k}_3)\rangle'\,,
\end{align}
where $\langle\zeta(\vec{k})\zeta(-\vec{k})\rangle'=P_\zeta=A/k^3$ and $A=H^4/(2\dot{\phi}_0^2)$. For $k_1=k_2$ and the factorized analytic approximation, this reads as
\begin{align}\label{eq:fullF_fnl}
\mathcal{S}(k_1,k_2,k_3)\bigg\rvert_{\text{analytic}}
&
\xrightarrow{k_r \ll k_3\ll k_1}
-\frac{4\dot{\phi}_0}{H^2}\left(\dfrac{k_3}{k_r}\right)^{-n - i \mu_{\rm c}} \left(\dfrac{k_3}{k_1}\right)^{1/2+i\mu}\mathcal{F}(\mu,\mu_c,n)+\text{c.c.}\nonumber \\ & \equiv f_{\rm NL} \left(\dfrac{k_3}{k_r}\right)^{-n - i \mu_{\rm c}} \left(\dfrac{k_3}{k_1}\right)^{1/2+i\mu}+\text{c.c.}.
\end{align}

The parameter $f_{\rm NL}$ quantifies the strength of NG. This summarizes our squeezed-limit analytic computation of Fig.~\ref{fig:3ptinin}b. To study more general momentum configuration and include Fig.~\ref{fig:3ptinin}c as well, we now move on to a numerical computation. We will see for the full shape function $\mathcal{S}(k_1,k_2,k_3)$ matches $\mathcal{S}(k_1,k_2,k_3)\bigg\rvert_{\text{analytic}}$ quite well for $p\gtrsim 5$.

\subsection{Numerical computation: shape and scale dependence}
\label{subsec.numerics_osc}

Away from the squeezed limit, the factorization described above does not apply and hence we need to evaluate the bispectrum numerically. Numerical calculation can also reveal some minor details that we did not focus on in the analytical approximation, such as the effect of the sharp feature (i.e.~the step function in this example) at the onset of the background oscillation.

We first write the $++$ contribution from Fig.~\ref{fig:3ptinin}b in a form similar to Eq.~\eqref{eq:++}. Since in Eq.~\eqref{eq:++} we only kept the time derivative contributions for simplicity, it amounted to keeping only the first term in Eq.~\eqref{eq:D}. Now taking into account the full operator $\mathcal{D}$, the $++$ contribution reads as,

\begin{align}\label{eq:++full}
    &\braket{\delta\phi^3}'_{++}\bigg\rvert_{\rm{Fig.}~\ref{fig:3ptinin}b} 
    =\dfrac{\rho\lambda}{\Lambda} \dfrac{-\pi H^3}{32 k_1 k_2 k_3^4} \, \dfrac{i}{2}B_0 \, z_0^{-n - i \mu_{\rm c}} \,\nonumber\\
    & \times \int_0^\infty \d z_1 \left(T_1(z_1)+T_2(z_1)+T_3(z_1)\right) H_{i\mu}^{(1)}(z_1) e^{-i p z_1}\int_{z_1}^{\infty} \dfrac{\d z_2}{\sqrt{z_2}} \, H_{i\mu}^{(2)}(z_2) e^{-i z_2} z_2^{n+i \mu_{\rm c}}\theta(z_0-z_2)\nonumber\\
    & \hspace{1.5cm}- \{\mu_c\rightarrow-\mu_c\}\,.
\end{align}
Here the different terms in Eq.~\eqref{eq:D} contribute as,
\begin{align}
T_1(z) & = z^{3/2}\left(1-\frac{\vec{k}_1\cdot\vec{k}_2}{k_1k_2}\right)=z^{3/2}\left(2-\dfrac{2}{p^2}\right),\nonumber\\
T_2(z) & = z^{-1/2}\left(\frac{\vec{k}_1\cdot\vec{k}_2 k_3^2}{k_1^2k_2^2}\right)=-z^{-1/2}\dfrac{4}{p^2}\left(1-\dfrac{2}{p^2}\right),\nonumber\\
T_3(z) & = z^{1/2}\left(\frac{\vec{k}_1\cdot\vec{k}_2 k_3^2}{k_1^2k_2^2}\right)\left(i\frac{k_{12}}{k_3}\right)=-z^{1/2}\dfrac{4i}{p}\left(1-\dfrac{2}{p^2}\right),
\end{align}
where we have assumed the special case of $k_1=k_2$ and $p=(k_1+k_2)/k_3$ in the second equalities.

In addition, we now also keep the contribution from Fig.~\ref{fig:3ptinin}c in which the 3pt vertex happens earlier. That reads as,
\begin{align}\label{eq:++full_2nd}
    &\braket{\delta\phi^3}'_{++}\bigg\rvert_{\rm{Fig.}~\ref{fig:3ptinin}c} 
    =\dfrac{\rho\lambda}{\Lambda} \dfrac{-\pi H^3}{32 k_1 k_2 k_3^4} \, \dfrac{i}{2}B_0 \, z_0^{-n - i \mu_{\rm c}} \,\nonumber\\
    & \times \int_0^\infty \dfrac{\d z_1}{\sqrt{z_1}} \, H_{i\mu}^{(2)}(z_1) e^{-i z_1} z_1^{n+i \mu_{\rm c}}\theta(z_0-z_1) \int_{z_1}^{\infty} \d z_2 \left(T_1(z_2)+T_2(z_2)+T_3(z_2)\right) H_{i\mu}^{(1)}(z_2) e^{-i p z_2} \nonumber\\
    & \hspace{1.5cm}- \{\mu_c\rightarrow-\mu_c\}\,.
\end{align}
Therefore, the full contribution from $++$ diagram is given by
\begin{align}
\braket{\delta\phi^3}'_{++} = \braket{\delta\phi^3}'_{++}\bigg\rvert_{\rm{Fig.}~\ref{fig:3ptinin}b} + \braket{\delta\phi^3}'_{++}\bigg\rvert_{\rm{Fig.}~\ref{fig:3ptinin}c} ~.
\end{align}
In the following, we present results of the first term (Fig.~\ref{fig:3ptinin}b) while we provide numerical result and discussion of the second term (Fig.~\ref{fig:3ptinin}c) in App.~\ref{app.comparison}. We verify that Fig.~\ref{fig:3ptinin}c does not contain the $\mu$-dependent clock signal at leading order as argued in the discussion above \eqref{eq:++}.

\paragraph{Numerical convergence.} Numerical evaluation of these cosmological correlation functions may suffer from UV and/or IR spurious divergences. In our integrals, the schematic IR behavior of Eq.~\eqref{eq:++full} is as follows
\begin{align}
    \int\d z_1 z^{a+i\mu}\int\d z_2 z_2^{n-1/2-i\mu+i\mu_c}\,,
\end{align}
where $a=3/2, -1/2,$ and $1/2$ for each of the $T_1, T_2,$ and $T_3$ contributions. Thus, IR contribution is convergent. The schematic UV behavior is
\begin{align}
    \int_{0}^\infty\d z_1 e^{-i(p-1)z_1}\int_{z_1}^\infty\d z_2 e^{-2iz_2}\theta(z_0-z_2)\,.
\end{align}
Both  the $z_1$ and $z_2$ integrals are highly oscillatory at UV. To handle such UV oscillation one can Wick rotate integration parameters as $z_{1,2}\rightarrow -iy_{1,2}$ \cite{Chen:2009zp}. However, in the present case the $\theta$ function itself cuts off the UV contribution in the inner integral and we compute this inner integral analytically. Subsequently, to handle the oscillations of the outer integral, we do a Wick rotation. More details on these computations are given in App.~\ref{app.full_bispec}.
Using the final expression for the shape function in Eq.~\eqref{eq:shapefunc_numeric_3integrals}, we now study the shape and scale dependence.

\begin{figure}[t!]
    \centering
    \includegraphics[width=15cm]{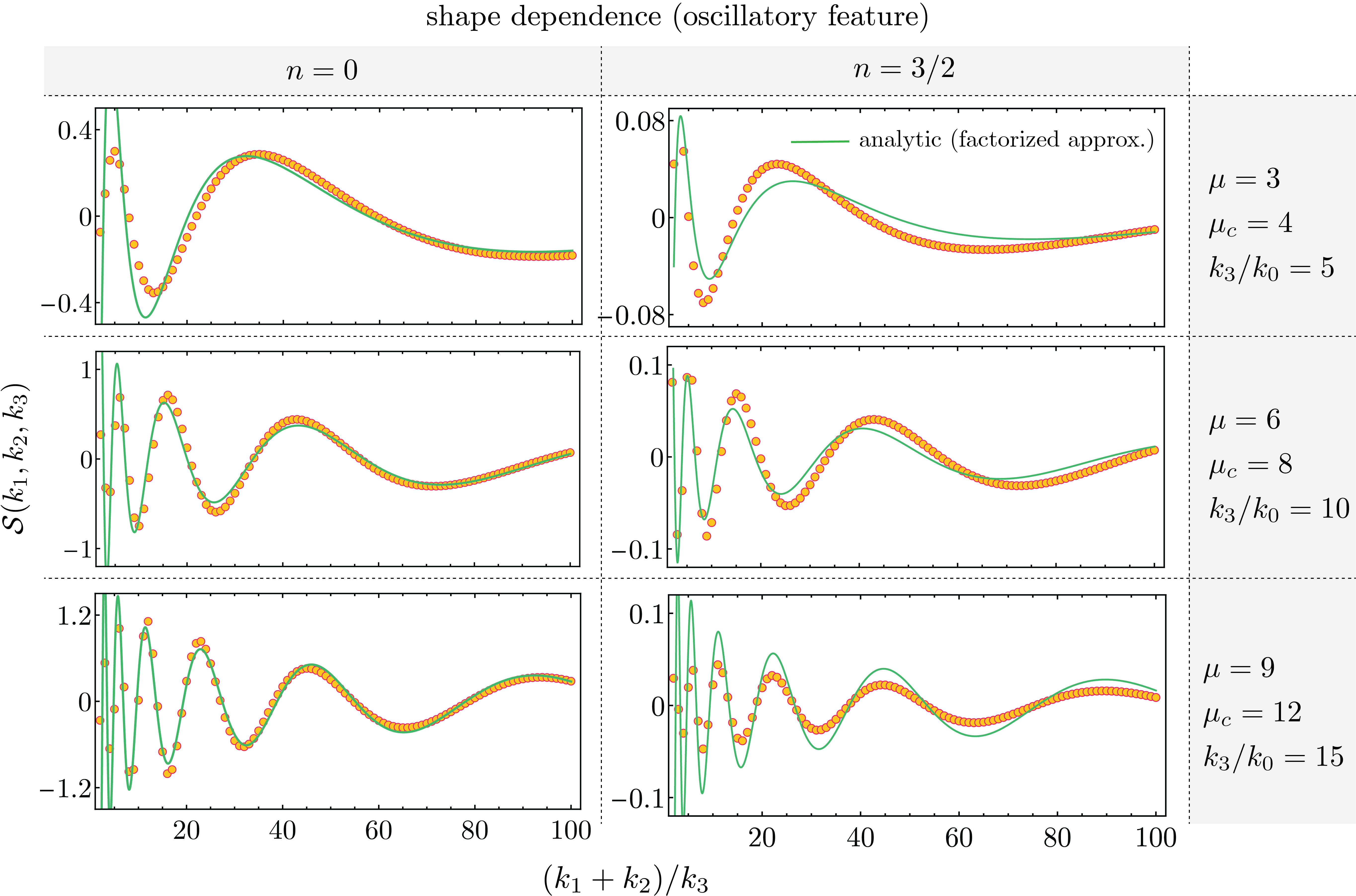}
    \caption{Oscillatory-feature-induced shape function as a function of squeezing parameter $p=(k_1+k_2)/k_3$, for the $n=0$ case ({\it Left}) and the $n=3/2$ case ({\it Right}).
    The orange points are full numeric results. The green curves are the analytic result using factorization approximation \eqref{eq:fullF_fnl}, including the values of $f_{\rm NL}$. 
    The differences between the numeric results and analytical templates are mainly due to the effects from the sharp turn-on of the oscillatory feature by the step function $\theta(t-t_0)$ in Eq.~\eqref{eq.oscfeat}, which are not fully taken into account in the analytical approximation. These effects both affect the excitation of massive field (shown here) and generate a component of scale-dependent sharp-feature signal (shown in Fig.\,\ref{fig:scaleFULL}).
    The difference is more apparent for $n=3/2$ compared to $n=0$ at larger $k_3/k_0$, because the oscillatory-feature-induced signal decays towards larger $k_3/k_0$ values for $n=3/2$, while the $\theta$-function contribution decays more slowly due to the sharpness of this feature (see Fig.\,\ref{fig:scaleFULL}). 
    We have chosen $\lambda=1$, $\rho/\Lambda=0.03$ and $B_0=5\times 10^{-3}$ respecting the bounds from the power spectrum searches and EFT consistency (see Sec.~\ref{sec.benchmark}).
    }
    \label{fig:shapeFULL}
\end{figure}

\paragraph{Shape dependence.} We evaluate the shape dependence fixing several values of $k_3/k_0$ and plotting $\mathcal{S}(k_1,k_2,k_3)$ as a function of $(k_1+k_2)/k_3$ in Fig.~\ref{fig:shapeFULL} for $n=0$ and $n=3/2$. We also show the analytical result from Eq.~\eqref{eq:fullF_fnl} with a dependence $\mathcal{S}\propto \sin(\mu\log((k_1+k_2)/k_3) + {\rm phase})$. 
The feature in Eq.~\eqref{eq.oscfeat} contributes in two ways: first, through the oscillation frequency $\omega_c$ itself, and second, through the turn-on of the feature at $t=t_0$ via the $\theta$ function. For $n=0$, the oscillatory feature does not decay with time and therefore it dominates over the particle production from the $\theta$ function itself. On the contrary, for $n=3/2$ the oscillatory feature decays faster and the contribution from the $\theta$ function becomes comparable.\footnote{The amplitude of the non-Gaussian signal is proportional to the amplitude of the background oscillations at the time of resonance, therefore, for $n=3/2$ the resonant signal is smaller than that for $n=0$. However, particle production due to the sharp feature happens at $t=t_0$ so for both $n=3/2$ and $0$ the signal amplitude is the same.} Since in our factorized analytic approximation given in Eq.~\eqref{eq:fullF_fnl} we did not consider this $\theta$ function contribution, we see the match between analytic and numerical results are better for $n=0$. We also note that this match is good already for a squeezing parameter for $(k_1+k_2)/k_3\gtrsim 5$. Finally, for both these cases, we see the tell-tale oscillations corresponding to the mass of the massive field $\mu H$.

\begin{figure}[t!]
    \centering
    \includegraphics[width=15cm]{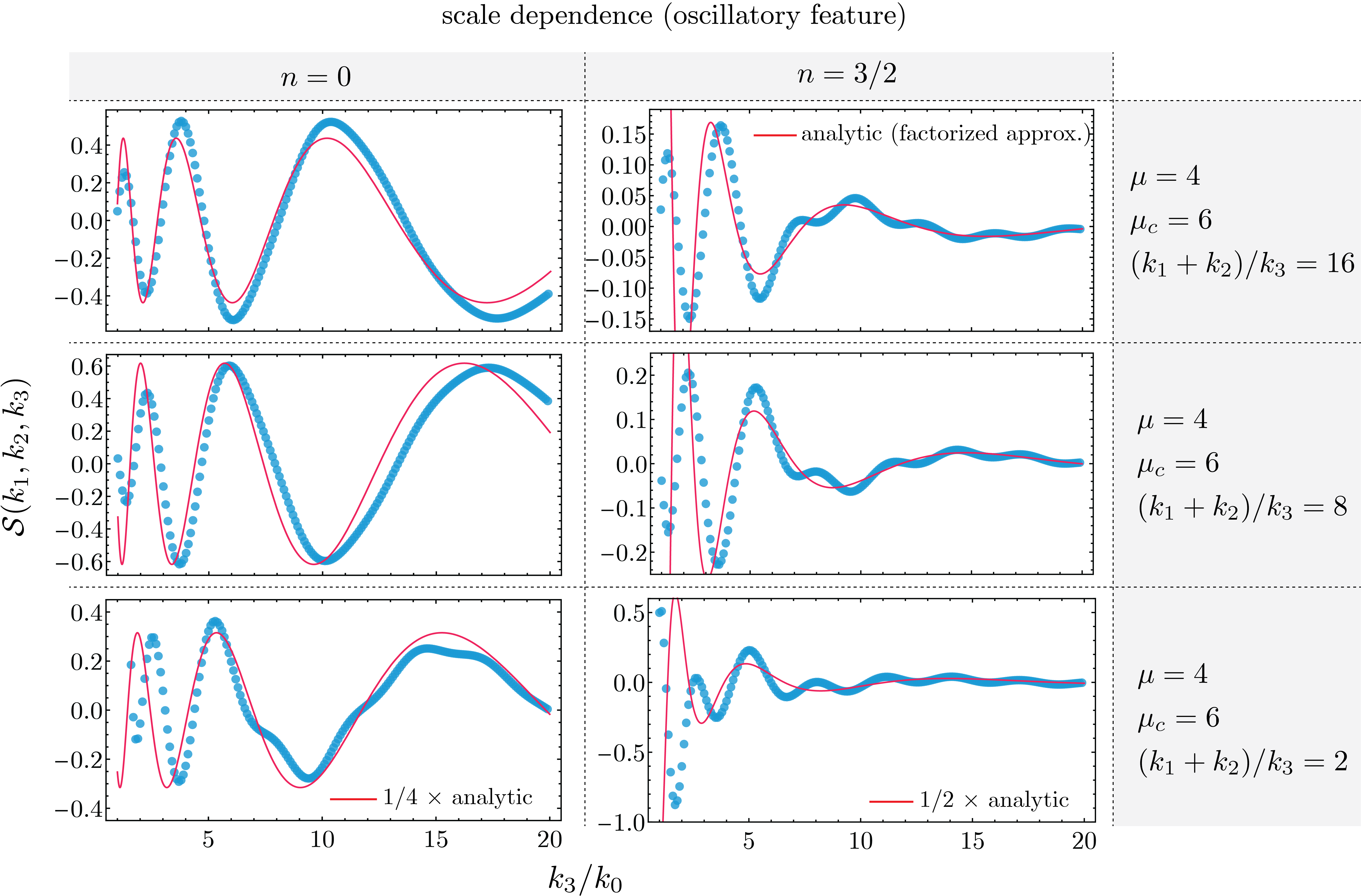}
    \caption{Scale dependence in oscillatory-feature-induced bispectrum. We observe the expected $\sin(\mu_c\log(k_3/k_0) + {\rm phase})$ scale-dependence coming from background oscillatory feature with frequency $\mu_c=\omega_c/H$. However, the scale dependence induced by the abrupt onset of the background oscillation, which is $\propto\sin(2k_3/k_0)$, is also evident in the full result. Especially, for $n=3/2$, where the oscillation-induced signal decays with time, $\theta$-function-induced signal becomes dominant at larger $k_3/k_0$. Note that in a realistic scenario where the onset of clock oscillation is smooth, we expect the contribution from the step to be smaller. In the bottom row, we have scaled the analytic results (as shown) to match the overall shape. The difference in overall normalization is expected since this is evaluated in the equilateral limit, away from our squeezed-limit analytic result. We have chosen $\lambda=1$, $\rho/\Lambda=0.03$ and $B_0=5\times 10^{-3}$ respecting the bounds from the power spectrum searches and EFT consistency (see Sec.~\ref{sec.benchmark}).}
    \label{fig:scaleFULL}
\end{figure}

\paragraph{Scale dependence.} Along with oscillations of the massive field, in our scenario we also expect a non-trivial scale dependence coming from the primordial feature. To show this, we plot $\mathcal{S}(k_1,k_2,k_3)$ as a function of $k_3/k_0$ for some fixed choices of $(k_1+k_2)/k_3$ in Fig.~\ref{fig:scaleFULL}. The scale-dependent oscillatory features behave as $\mathcal{S}\propto (k_3/k_0)^{-n} \sin(\mu_c\log(k_0/k_3) + {\rm phase})$ as in Eq.~\eqref{eq:fullF_fnl}. Note that, unlike the shape-dependent features, the frequency here, $\mu_c$, is determined by the background oscillation.
An effect of the step function, that turns on the background oscillation, can also be clearly seen here and is not captured in the analytical approximation. Namely, the sharp feature generates a component with sinusoidal scale-dependence $\sim \sin(2k_3/k_0)$.
Because this kink feature in the background $B_c(t)$ has been artificially made infinitely sharp, the effect decays slowly towards large $k_3/k_0$. This can be mostly easily seen for $n=3/2$ at large values of $k_3/k_0$, since by the time of mode exit of $k_3$, the resonant contribution of the oscillatory background has decayed.

\section{Bispectrum in the presence of sharp features}
\label{sec.sharp}
\begin{figure}[t!]
    \centering
    \includegraphics[width=14cm]{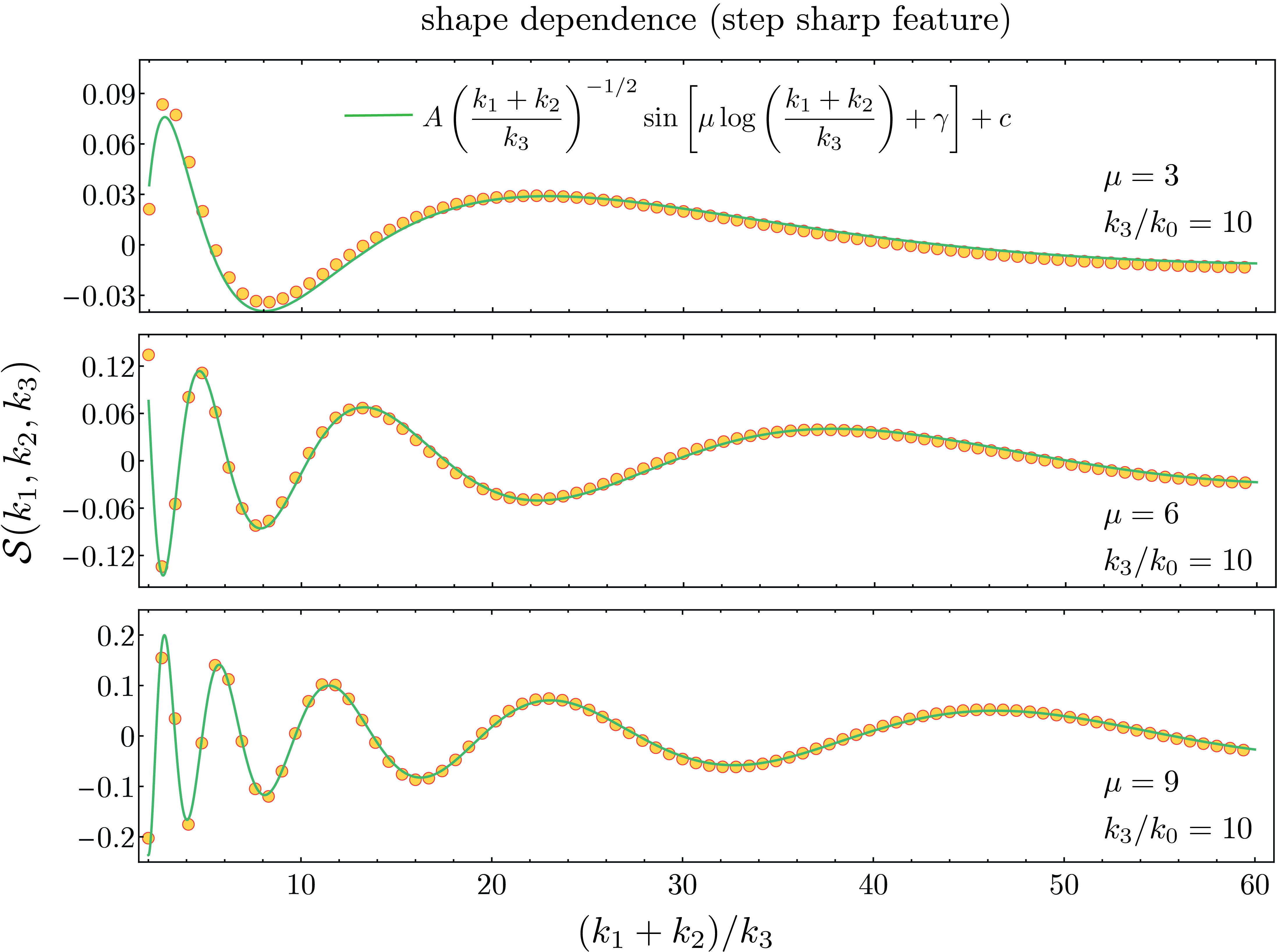}
    \caption{Sharp-feature-induced shape function as a function of squeezing parameter $p=(k_1+k_2)/k_3$. We have chosen $\lambda=1$, $\rho/\Lambda=0.03$ and $B_0=5\times 10^{-3}$ respecting the bounds from the power spectrum searches and EFT consistency (see Sec.~\ref{sec.benchmark}). We observe a good agreement between the oscillation frequency $\mu$ in the full numeric result (orange points) and the analytical expectation (green curve). The parameters $A,\gamma,c$ are chosen to match the numerical result.
    % {\it Top:} $A=0.125, \gamma=-1.68,\rm{~and~}c=0.008$. {\it Middle:} $A=0.243, \gamma=-1.46,\rm{~and~}c=0.002$. {\it Bottom:} $A=0.335, \gamma=-1.60,\rm{~and~}c=0.001$.
    }
    \label{fig:shapeSTEP}
\end{figure}

In this section we focus on the other type of primordial feature described in Eq.~\eqref{eq.sfeat},
\begin{align}
B_s(t) = B_0\theta(t-t_0).
\end{align}
This form of a step function can arise in Eq.~\eqref{eq:int_lagrangian} if we have a coupling of the type $s \dot{\delta\phi}\delta\chi$ with a time-dependent VEV, $\langle s\rangle$ for the field $s$. For example, if $s$ undergoes a sharp transition in the landscape, we can have a step function coupling as above. While this type of step function does not have a specific oscillation frequency $\mu_c$ associated with it, we note that the Fourier transform of $\theta(t)$ is given by $\dfrac{1}{2}\left(\delta(\omega)-\dfrac{i}{\pi\omega}\right)$. Therefore, such a step function has a decreasing, but non-zero support to excite heavier and heavier masses through energy $\omega$, and we expect non-exponentially suppressed production even for $\mu\gg 1$. Beyond production of massive $\chi$ quanta, a sharp feature will also produce inflaton quanta. 
These modes produced at the time of sharp feature introduce a scale-dependent sinusoidal signal in cosmological correlation functions~$\sim\cos(K\eta_0+{\rm phase})$, where $\eta_0$ is the conformal time of the feature and $K$ is the sum of wave numbers that are connected to the vertex acted by the sharp feature. We confirm these expectations by computing the shape function $\mathcal{S}(k_1,k_2,k_3)$ numerically following the same procedure as above, except with this different choice $B_s(t)$. 
\paragraph{Shape dependence.}
In Fig.~\ref{fig:shapeSTEP} we show the shape dependence as a function of $(k_1+k_2)/k_3$ for several choices of $k_3/k_0$. We see unsuppressed, characteristic oscillations for several choices of heavy field mass $\mu$. While we do not present an analytic computation for full shape function for this step feature scenario, we do expect based on Eq.~\eqref{eq:fullF_fnl} that the shape function would approximately be given by $\sin(\mu\log((k_1+k_2)/k_3) + {\rm phase})$ as before. We overlay that analytic form to show the match.

\begin{figure}[t!]
    \centering
    \includegraphics[width=14cm]{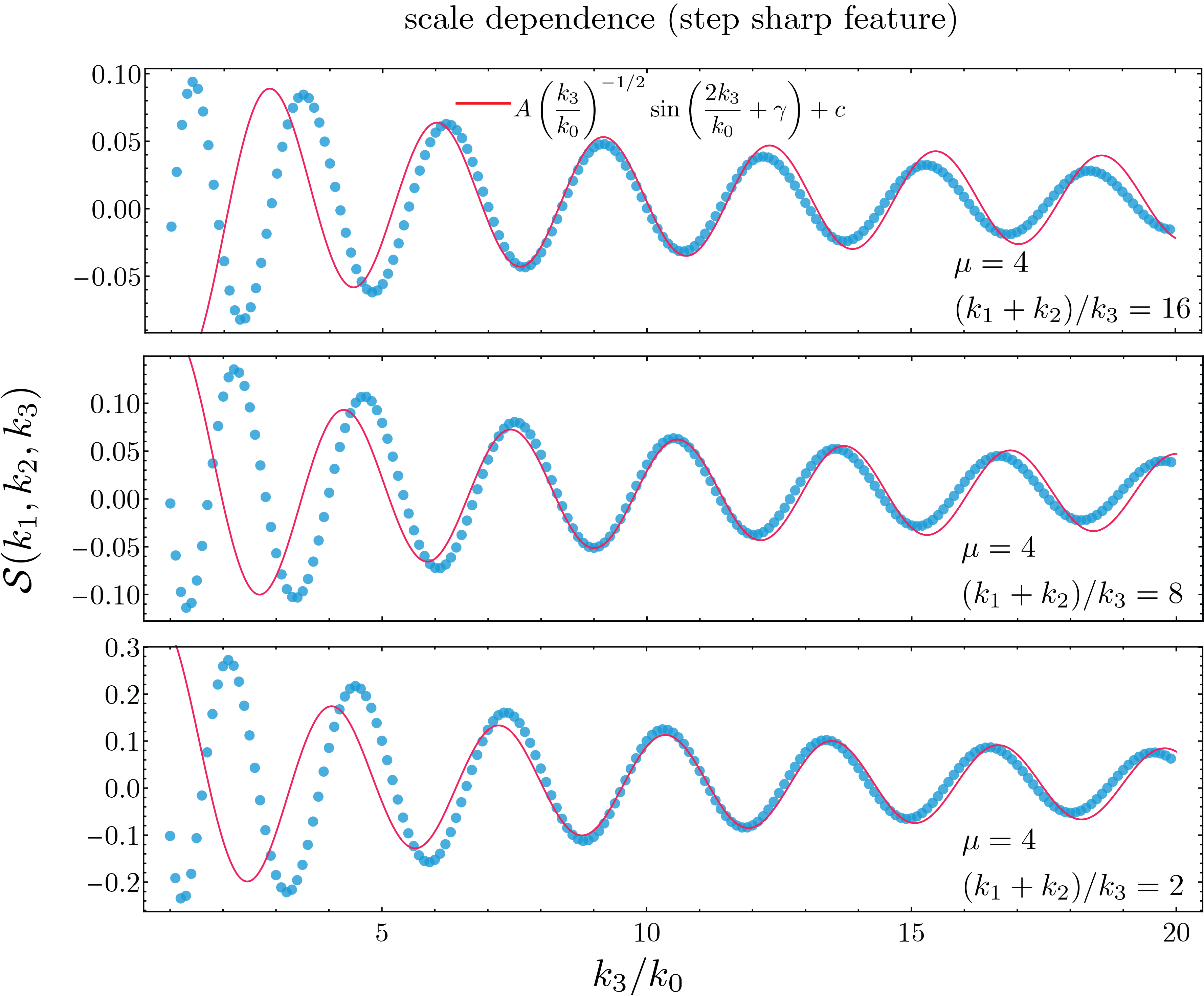}
    \caption{Scale dependence in sharp-feature-induced bispectrum. We have chosen $\lambda=1$, $\rho/\Lambda=0.03$ and $B_0=5\times 10^{-3}$ respecting the bounds from the power spectrum searches and EFT consistency (see Sec.~\ref{sec.benchmark}). The oscillation frequency in the numerical results (blue) are well reproduced by the analytical expectation (red). We choose amplitude, phase and offset of the analytic function to match the numerical result.
    % {\it Top:} $A=0.139, \gamma=2.04,\rm{~and~}c=0.007$. {\it Middle:} $A=-0.177, \gamma=2.39,\rm{~and~}c=0.007$. {\it Bottom:} $A=-0.330, \gamma=2.83,\rm{~and~}c=0.010$.
    }
    \label{fig:scaleSTEP}
\end{figure}

\paragraph{Scale dependence.} The scale dependence corresponding to such a sharp feature is given by $\sim\sum_{i=1,2,3}\sin(2k_i\eta_0 +{\rm phase}) \sim \sum_{i=1,2,3} \sin(2k_i/k_0)$. 
From our numerical computation in Fig.~\ref{fig:scaleSTEP} we see this explicitly. 
Note that the summation over different $k_i$ is outside the sinusoidal function, because the sharp feature only acts at the 2pt vertex in our model. This is different from the examples in Ref.~\cite{Chen:2008wn} where the summation is inside the sinusoidal function.
We find that the sharp feature signal is unsuppressed for $k_3\gtrsim k_0\equiv-1/\eta_0$ --- the horizon scale when sharp feature takes place. This is manifestation of the fact that a sharp feature is able to excite modes at or below horizon scale. 
A more realistic sharp feature will not be infinitely sharp, and the above signal will decay more quickly towards larger scales.

\section{Example implementations involving oscillatory features}
\label{sec.applications}

In the discussion so far we have focused on two examples of primordial features: oscillatory and sharp features. To describe how these features assist in producing heavy particles, we have focused on a simplified Lagrangian in Eq.~\eqref{eq:int_lagrangian}. Subsequently we investigated the shape and scale dependence, both analytically and numerically, of the bispectrum to understand how the information of mass of the excited heavy particle is encoded in this correlation function. In the following, applying the more general results derived in the previous sections, we will discuss two examples of the scenario with oscillatory feature. The first example is an oscillatory feature that is directly introduced in the inflaton potential. The second example is closely related to the sharp feature case, but we focus on the subsequent effect of a classically oscillating clock field on heavy field $\chi$ of interest. We also describe these couplings within an EFT framework to determine both theoretical and observational constraints on the parameters in Eq.~\eqref{eq:int_lagrangian}.

\subsection{EFT parametrization and constraints}
To capture the leading effects, we will write down an EFT containing the inflaton $\phi$, the clock field $\sigma$ and the heavy scalar of interest, $\chi$.
A simple coupling between $\phi$ and $\chi$ (with mass $m_\chi$), respecting the shift symmetry of the inflaton, can be chosen as
\begin{align}
    \CL_{\rm int} = f(\chi) g^{\mu\nu} \partial_\mu \phi \partial_\nu \phi\,.
    \label{eq:resonant2}
\end{align}
Such interactions may be realized in terms of a turning trajectory \cite{Chen:2009zp} with the $\chi$-field being the massive field perpendicular to the turning inflaton trajectory. Considering the case where the $\chi$-field is classically resting at the bottom of its potential and neglecting its classical oscillation due to backreaction, Eq.~\eqref{eq:resonant2} leads to the interactions between the perturbations which we can parametrize similarly as in Eq.~\eqref{eq:int_lagrangian}. For concreteness, we take $f(\chi)\propto\chi$ and that gives rise to a dimension-5 operator,
\begin{align}
\frac{c_1}{\Lambda_\chi}(\partial\phi)^2\chi.    
\end{align}
Here $c_1$ is an $\mathcal{O}(1)$ dimensionless coefficient and $\Lambda_\chi$ denotes an energy scale above which the EFT description breaks down. Motivated by this parametrization, we can also describe the interactions between $\phi$ and $\sigma$ (with mass $m_\sigma$), 
\begin{align}
\frac{c_2}{\Lambda_\sigma}(\partial\phi)^2\sigma.   
\end{align}
Since $\sigma$ oscillates about its minimum, the cutoff $\Lambda_\sigma$ should be bigger than both the amplitude and frequency of such oscillations for a controlled EFT description.

Combining such interactions, the leading terms in the EFT up to dimension-6 are given by (respecting $\phi$-shift symmetry),
\begin{align}\label{eq:EFT}
\frac{c_1}{\Lambda_\chi}(\partial\phi)^2\chi + \frac{c_2}{\Lambda_\sigma}(\partial\phi)^2\sigma + %\frac{c_3}{\Lambda_\chi\Lambda_\sigma}(\partial\phi)^2\chi\sigma +
\frac{c_3}{\Lambda_\chi^2}(\partial\phi)^2\chi^2 + \frac{c_4}{\Lambda_\sigma^2}(\partial\phi)^2\sigma^2+\cdots.
\end{align}
For simplicity, we have assumed that there is no direct coupling between $\sigma$ and $\chi$ at the tree level.
We now describe the constraints on the terms in \eqref{eq:EFT} assuming all $c_i\sim\mathcal{O}(1)$:
\begin{itemize}
\item The terms involving $c_1$ and $c_2$ give rise to tree-level corrections to the scalar power spectrum, whose order of magnitude is quoted in \eqref{eq:power_heavy}. Requiring those to be small implies,
\begin{align}
    \Lambda_\chi &\gtrsim \dot{\phi}_0/m_\chi,\label{eq:con1}\\
    \Lambda_\sigma &\gtrsim \dot{\phi}_0/m_\sigma\label{eq:con2}.
\end{align}
\item The terms involving $c_3$ and $c_4$ give rise to mass corrections of the $\chi$ and $\sigma$ fields when $\phi$ is set to $\phi_0$. Requiring these to be smaller than their tree-level values, we get the same constraints as the above.
\item In the above we have written terms involving $(\partial\phi)^2$. To ensure a subdominant correction from terms with $(\partial\phi)^{n>2}$, we require \cite{Creminelli:2003iq} 
\begin{align}\label{eq.EFTcutoff1}
\Lambda_\chi &> \sqrt{\dot{\phi}_0},\\
\Lambda_\sigma &> \sqrt{\dot{\phi}_0}.
\end{align}
\item Finally, we can also have loops of $\chi$ and/or $\sigma$ giving rise to inflaton self-interactions $\propto (\partial\phi)^4$. However, such corrections are small once the above constraints are obeyed.
\end{itemize}
Therefore, in the following we will impose the constraints in Eqs.~\eqref{eq:con1} and \eqref{eq:con2} to determine the viable parameter space.

\subsection{Inflation models with oscillatory features in the potential}
One way a small time-dependent oscillatory component in the background evolution can arise is simply via periodic features in the inflaton potential or its internal space. As an example, such features in the inflaton potential can be generically written as
\begin{align}
    \CL_{\rm resonant~model} = V_{\rm sr}(\phi) \left[ 1 + c_{\rm osc} \sin\left( \frac{\phi}{\phi_r} + \beta \right) \right],
    \label{eq:resonant1}
\end{align}
where $V_{\rm sr}(\phi)$ denotes any slow-roll potential for the inflaton field $\phi$, and the small correction in the squared bracket is the oscillatory feature ($c_{\rm osc}$, $\phi_r$ and $\beta$ are constants).
These models are called the resonant models because the background periodic oscillation, induced by the features, resonates with the quantum fluctuations of the massless inflaton field mode by mode, generating resonant features in the power spectrum and non-Gaussianities  \cite{Chen:2008wn,Flauger:2009ab,Flauger:2010ja,Chen:2010bka}. The high-energy-physics motivated model-building realization of this class of models can be found in the axion-monodromy inflation \cite{Flauger:2009ab} as a large-field example and brane inflation \cite{Bean:2008na} as a small-field example.

In this work, we consider another effect of this background oscillation in addition to its effect on the inflaton field. Namely, this oscillation introduces a new energy scale that can be used to quantum-mechanically excite spectator massive fields, which leave their signatures at the cosmological collider physics.

Given Eq.~\eqref{eq:resonant1}, a subdominant, but high-frequency, inflaton oscillation $\phi_1(t)$ would get induced on top of the slow-roll dynamics determined by $\phi_0(t)$,
\begin{align} \label{phi_decomp}
\phi_{\rm background} = \phi_0(t) + \phi_1(t).
\end{align}
A precise form of $\phi_1(t)$ can be determined once Eq.~\eqref{eq:resonant1} is specified. Here, we generically write following the EFT expansion in Eq.~\eqref{eq:EFT},
\begin{align}\label{eq:clock2}
\frac{c_1}{\Lambda_\chi}(\partial\phi)^2\delta\chi \supset 
-2 \frac{c_1 (\dot{\phi}_0 + \dot{\phi}_1) }{\Lambda_\chi} \dot{\delta\phi}\delta\chi -\frac{c_1}{\Lambda_\chi}\left[ (\dot{\delta\phi})^2 - \frac{1}{a^2} (\partial_i \delta\phi)^2 \right]\delta\chi+\cdots.
\end{align}
To match with Eq.~\eqref{eq:int_lagrangian} we parametrize
\begin{align}
-2 \frac{c_1 \dot{\phi}_0}{\Lambda_\chi} \equiv \rho,~~
-2 \frac{c_1\dot{\phi}_1}{\Lambda_\chi} \equiv \rho B(t)=\rho B_0 \, \sin{(\omega_\c (t-t_0))}=\frac{i}{2}\rho B_0 \left(\frac{\eta}{\eta_0}\right)^{i\mu_c}+\text{c.c.}, 
\end{align}
and take $\lambda=-c_1$ and $\Lambda = \Lambda_\chi$. Thus $B(t)=\dot{\phi}_1/\dot{\phi}_0$.

We now discuss how $B(t)$ is related to the fundamental parameters $c_{\rm osc}$ and $\phi_r$. Using the $\phi_1$ equation of motion, we can estimate
\begin{align}
\ddot{\phi}_1 \sim c_{\rm osc} V_{\rm sr} \cos(\omega_\c t)/\phi_r.
\end{align}
Here we have denoted $\phi/\phi_r \approx \dot{\phi}_0 t/\phi_r \equiv \omega_\c t$ and ignored the subdominant friction  term since we are interested in $\omega_\c \gg H$. We have also dropped unimportant phase factors. Therefore, $\phi_1$ oscillates with a frequency $\omega_\c$ due to the source term and is given by
\begin{align}
\phi_1 \sim \frac{c_{\rm osc} V_{\rm sr}}{\phi_r \omega_\c^2} \cos(\omega_\c t).   
\end{align}
Using this we can estimate $B(t)$
\begin{align}
B(t) = \frac{\dot{\phi}_1}{\dot{\phi}_0} \sim \frac{c_{\rm osc} V_{\rm sr}}{\phi_r \omega_\c \dot{\phi}_0} \sin(\omega_\c t) \sim \frac{c_{\rm osc} V_{\rm sr}}{\phi_r^2 \omega_\c^2} \sin(\omega_\c t).
%\sim \frac{c_{\rm osc} V_{\rm sr}}{\dot{\phi}_0^2} \sin(\omega t) \sim \frac{c_{\rm osc}}{\epsilon} \sin(\omega t). 
\end{align}
% In the last relation we have used $\dot{\phi}_0^2 = 2\epsilon H^2 M_{\rm pl}^2$ and $V_{\rm sr} = 3 H^2 M_{\rm pl}^2$ with $\epsilon$ being the slow roll parameter $\epsilon = |\dot{H}|/H^2$.

To estimate the correction to power spectrum due to the resonant feature, we can expand $\phi= \phi_0+\delta\phi$. This implies, 
\begin{align}
\mathcal{L}_{\rm resonant~model} \supset c_{\rm osc} V_{\rm sr} \sin(\phi_0/\phi_r) \frac{(\delta\phi)^2}{\phi_r^2}.  
\end{align}
Since inflaton fluctuations are produced with energy $\omega_\c$, we can use $\dot{\delta\phi} \sim \omega_\c\delta\phi$ to get,
\begin{align}
\mathcal{L}_{\rm resonant~model} \supset \frac{c_{\rm osc}V_{\rm sr}}{\omega_\c^2\phi_r^2} \sin(\phi_0/\phi_r) \dot{\delta\phi}^2 \sim B(t) \dot{\delta\phi}^2.  
\end{align}
Thus we see the same oscillatory source $B(t)$ that contributes to the NG, also controls the correction to power spectrum in this model. Consequently, we can choose $B(t)$ based on the observational constraint on power spectrum and determine how large NG can be.

\subsection{Inflation models with classically oscillating fields}

From a UV-completion point of view, the trajectory of the inflaton is determined by the low energy valleys of the inflationary potential landscape. This landscape can be formed by many fields and can have a very large dimension, so it is natural to expect that these valleys are not always straight and that there are sharp features which, from time to time, disturb the inflaton away from its attractor solution. If the features are sharp enough, massive fields that are orthogonal to the valley may be excited and can start to oscillate classically.
Although the mechanism underlying the sharp feature can be very model-dependent, once a massive field starts to oscillate, its evolution is quite model-independent. If the mass of such a field is a constant $m_\sigma>3H/2$, its oscillation is approximately a harmonic one with a decaying amplitude due to the spatial expansion, namely,
\begin{align}\label{eq:SharpBackground}
\sigma_\c(t) = \sigma_\s (a/a_0)^{-3/2} \sin(m_\sigma (t-t_\sigma)) ~\theta(t-t_0) \,,
\end{align}
where $(a/a_0)^{-3/2}$ is due to the Hubble dilution of the field amplitude, $t_0$ is the time of the sharp feature and $t_\sigma$ specifies the phase of the oscillation.
Through the direct coupling between $\sigma$ and $\phi$, this oscillation induces a classical oscillation in $\phi$. Parametrizing $\phi = \phi_0+\phi_1$ as before, using $\phi_1$ equation of motion we have,
\begin{align}
\ddot{\phi}_1 + \dot{\phi}_0 \frac{\dot{\sigma}_c}{\Lambda} \approx 0.   
\end{align}
This determines,
\begin{align}
B(t) = \frac{\dot{\phi}_1(t)}{\dot{\phi}_0(t)} \approx -\frac{\sigma_c}{\Lambda}.   
\end{align}

Next, we estimate the correction to the power spectrum. In this example, such an effect comes from $(\partial\phi)^2\sigma/\Lambda$. Therefore this correction is determined by $\sigma_c/\Lambda$, or equivalently, the same oscilatory source $B(t)$ that controls NG. Consequently, we can use observational constraints on $B(t)$ through power spectrum measurments, to determine how large NG can be.

% \begin{align}\label{eq:SharpBackground}
% \phi_1(t) = \phi_s (a/a_0)^{-3/2} \sin(m_\sigma (t-t_\phi)) ~\theta(t-t_0) \,.
% \end{align}
% The rest is similar to what is discussed after \eqref{phi_decomp} for the resonant model. 

Because of our assumption in \eqref{eq:EFT} that $\sigma$ and $\chi$ are approximately decoupled, the classical oscillation induced in $\chi$ is of higher order in $1/\Lambda$ than those in $\sigma$ and $\phi$ and therefore neglected here.\footnote{This classical oscillation can also lead to observational signatures, especially in power spectrum. Introducing a direct coupling between $\sigma$ and $\chi$ will further enhance such signatures. We leave these more complicated cases for future study.}

\subsection{Benchmarks for classical cosmological collider physics}\label{sec.benchmark}
Now we discuss the viable parameter space given the EFT constraints and the existing power spectrum measurements. As given in Eq.~\eqref{eq:fullF_fnl}, the NG contribution is proportional to $\rho B_0\lambda/\Lambda$. We focus on each of these factors now.

We set $\lambda=c_1=1$ for simplicity.
To obey the constraint~\eqref{eq:con1}, we require a sufficiently small magnitude of the mixing parameter $\rho$,
\begin{align}
\rho < m_\chi.    
\end{align}
This also ensures a small heavy-field contribution to the power spectrum as discussed in Eq.~\eqref{eq:power_heavy}. Combining the above mentioned constraints from EFT considerations and power spectrum, we get
\begin{align}
\dfrac{\rho}{\Lambda}<\text{min}\left\{\dfrac{m_\chi^2}{\dot{\phi}_0}, 1\right\}.   \end{align}
Here we note that as $m_\chi > \sqrt{\dot{\phi}_0}$, a stronger bound $\rho<\Lambda$, originating from Eq.~\eqref{eq.EFTcutoff1}, comes into play. Overall, this constraint weakens as we go to larger masses. However, given our focus on $m_\chi \gg H$ and to be conservative, we consider a benchmark $\rho/\Lambda=0.03$ which is obeyed for all $m_\chi\gtrsim 10H$. 

Next we discuss the constraints on $B_0$. We saw in the previous two subsections that $B_0$ appears as a correction to power spectrum via a term schematically of the type
$B_0 \sin(\omega_c t) (\partial\phi)^2$, whose effect was 
described in Eq.~\eqref{eq:power_correction_direct}. %that the parameter $B_0$ gives rise to a direct correction to the power spectrum as well.
Observational constraints on such corrections from CMB were discussed in Ref.~\cite{Planck:2018jri, Braglia:2021rej, Achucarro:2022qrl} which gives $B_0\lesssim 5\times10^{-3}$ for $n=0$ and $\mu_c\sim 20$. For $n=3/2$, the constraint is a factor of few weaker. For higher values of $\mu_c$ that we are interested in, the constraint on $B_0$ is expected to weaken as well. However, to be conservative we still enforce the same restriction $B_0\lesssim 5\times10^{-3}$ for all $\mu_c$ and for both $n=0$ and $n=3/2$. These choices of $B_0$ and $\rho$ also ensure a small correction to the power spectrum due to feature-heavy field contribution in Eq.~\eqref{eq:power_correction_mixed}. To summarize, the above discussions justify our choice of $B_0=5\times10^{-3}$ and $\rho/\Lambda=0.03$ in Fig.~\ref{fig:shapeFULL} through Fig.~\ref{fig:scaleSTEP}.

\begin{figure}[tb!]
    \centering
    \includegraphics[width=10cm]{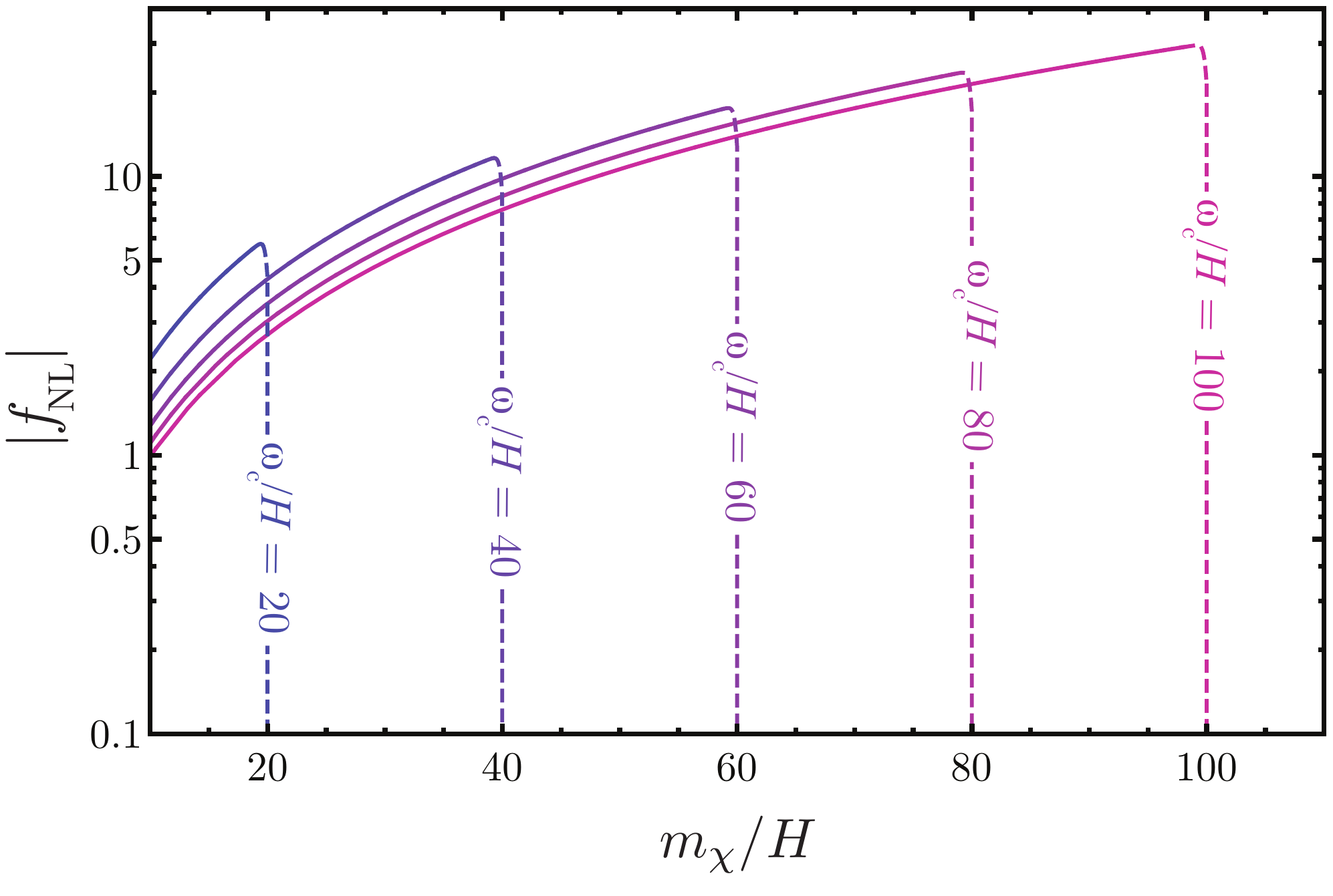}
    \caption{Strength of non-Gaussianity (see \eqref{eq:fullF_fnl}), $f_{\rm NL}\equiv(-4\dot{\phi}_0/H^2)\mathcal{F}(\mu,\mu_c,n)$ as a function of heavy scalar mass~$m_\chi$. The solid lines indicate the results for parameter space $\mu<\mu_c$ where the factorization approximation is appropriate. For $\mu\geq\mu_c$, however, this approximation is not fully accurate, and hence, we have shown the corresponding estimate via dashed lines. This is sufficient for our purpose since the non-analytic signature decays exponentially for $\mu>\mu_c$.
    We have chosen $\lambda=1$, $\rho/\Lambda=0.03$ and $B_0=5\times10^{-3}$ respecting the bounds from the power spectrum searches and EFT consistency.}
    \label{fig:3pt}
\end{figure}

%$\rho\lambda/\Lambda$ to lie between $0.01-0.1$ and $\tilde{B}=10^{-3}$ and 
With these choices we show the resulting strength of NG in Fig.~\ref{fig:3pt}. This shows that large NG can arise for masses $\sim 100 H$, or in principle, even larger. Due to the introduction of the classical source, this overcomes the upper limit of $\dot{\phi}_0^{1/2}\approx 60 H$ present in earlier studies involving current-coupling.

\section{Summary and discussions}\label{sec.summary}
From a UV perspective, the inflationary landscape can very well consist of many fields with the inflaton being one light degree of freedom that rolls along some particular direction in the landscape. It is natural to imagine that such a landscape is not flat everywhere and  non-shift symmetric features can easily arise. Such primordial features can lead to various interesting effects in cosmological correlation functions and have been studied extensively in the literature.

In this work, we have explored a novel connection between the study of primordial features and the cosmological collider physics program. Namely, these primordial features naturally establish a cosmological collider with super-Hubble energy. In particular, we have studied how such primordial features can lead to quantum mechanical production of heavy particles. Since the energy scale corresponding to such features can be much larger than the Hubble scale during inflation, primordial features can efficiently excite very heavy degrees of freedom, extending the reach of the existing cosmological collider program by a few orders of magnitude. Such features are uniquely relevant in producing heavy, \textit{real} scalar particles which otherwise are not excited by standard shift-symmetric inflaton couplings.

\begin{figure}[t]
    \centering
    \includegraphics[width=15cm]{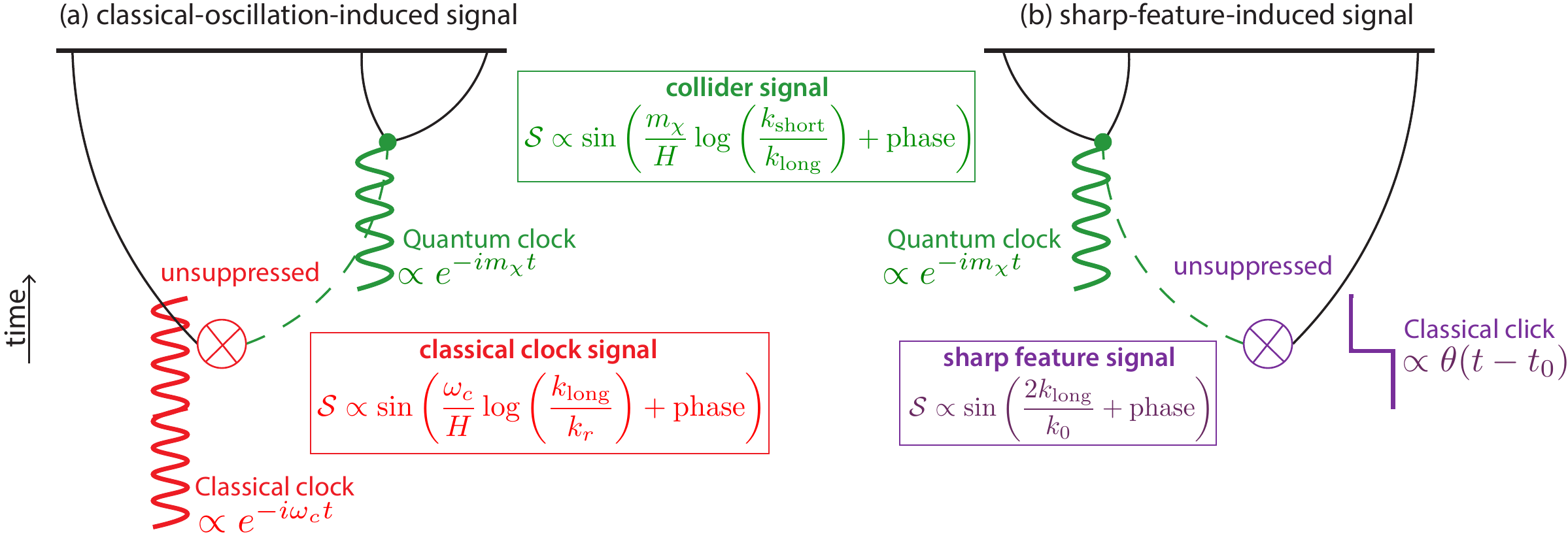}
    \caption{Physical mechanisms underlying (a) classical-oscillation-induced signal and (b) sharp-feature-induced signal (see text for details).  }
    \label{fig:clock_sharpfeature_expl}
\end{figure}

Concretely, we have focused on two classes of mechanisms. In the first scenario, summarized in Fig.~\ref{fig:clock_sharpfeature_expl}a, we focus on oscillatory primordial features. Such features can, for example, come from coherent oscillations of another massive field or subdominant oscillations of the inflaton itself in resonance models. These classical oscillations can induce a \textit{scale}-dependent NG of the type $\mathcal{S}\propto \sin\left(\frac{\omega_c}{H}\log\left(\frac{k_{\rm long}}{k_r}\right)\right)$ that characterizes the dynamics of the feature itself before the production of $\chi$ particles at the vertex marked with the red cross. Once the $\chi$ particles are produced, they oscillate quantum-mechanically with their characteristic frequency $e^{-i m_\chi t}$ until their decay at the vertex marked with the green dot. This leads to \textit{shape}-dependent NG of the type $\mathcal{S}\propto \sin\left(\frac{m_\chi}{H}\log\left(\frac{k_{\rm short}}{k_{\rm long}}\right)\right)$ which constitutes the cosmological collider signal. As long as $\omega_c > m_\chi$, the oscillatory feature is energetic enough to lead to an unsuppressed NG.

In the second scenario, summarized in Fig.~\ref{fig:clock_sharpfeature_expl}b, the primordial feature that is responsible for $\chi$ particle production is a non-oscillatory, sharp feature. While there can be various ways in which a sharp feature can arise, we consider a simple parametrization in which the sharp feature is a step function in time. Since such a step function has a Fourier support over a wide frequency range, including super-$H$ scales, unsuppressed production of particles with $m_\chi\gg H$ can occur. While the shape-dependent NG is same as in the first scenario, the scale dependence is different due to the different nature of the feature. Unlike the case of continuous oscillations, a step function acts for a short period of time around $t_0$. Therefore, the scale-dependent NG is of the sharp feature type $\mathcal{S}\propto \sin\left(\frac{2k_{\rm long}}{k_0}\right)$.

We find that both scenarios can give rise to strong enough NG signatures that may be observable in the future and can correlate with signals in the power spectrum.

There remain various interesting directions that can be pursued in the future. For simplicity, we have considered production of scalar fields. It would be interesting to generalize these mechanisms to include particles with non-zero spins. 
We have considered a simple parametrization for the clock field oscillation and parametrized the couplings in an EFT framework. It would be interesting to consider more complicated possibilities where the background features enter the couplings and investigate how they modify the predictions. It would be also useful to consider a complete model with a sharp feature that describes the entire dynamics of the classical field-inflaton system and evaluate the resulting non-Gaussian contributions.
Furthermore, we have seen that in the presence of primordial features, the non-Gaussianity signatures have both a \textit{shape} and \textit{scale} dependence that are correlated. New kinds of templates are needed to analyze these simultaneously present signatures. In particular, it would be very interesting to revisit the existing large-scale structure and 21-cm cosmology forecasts to see what values of heavy masses can be probed in the future.

\acknowledgments
We thank Arushi Bodas, Borna Salehian, Raman Sundrum and Alireza Talebian for useful conversations. RE is supported in part by the University of Maryland Quantum Technology Center. This work used computing resources at the University of Maryland Quantum Technology Center. SK is supported in part by the NSF grant PHY-1915314 and the U.S. DOE Contract DE-AC02-05CH11231. SK thanks Aspen Center of Physics, supported by NSF grant PHY-1607611, for its hospitality while this work was in progress. 
\appendix

\section{IR expansion analysis for $\mu>\mu_c$}\label{app:IRexpansion}
Here we give an explicit computation for the parametric dependence of the bispectrum for $\mu>\mu_c$ as mentioned in Sec.~\ref{subsec.IRexpansion}. We utilize the IR expansion method developed in \cite{Chen:2012ge} to evaluate Eq.~\eqref{eq:++}. 

Let us define UV region as $z>\sqrt{\mu}$ and IR region as $z<\sqrt{\mu}$. We show that the UV contribution, considering UV asymptotic forms of the mode functions, is exponentially suppressed. Also, extending the IR asymptotic forms to the UV region gives an exponentially suppressed contribution. However, integrating IR asymptotic forms over the \textit{full} region yields a power-law suppressed result. These justify that the main contribution is from the late-time integration and the IR expansion method captures that leading contribution. In the remainder of this section we perform these calculations. While the details are somewhat technical, the final result of this subsection is given in Eq.~\eqref{eq:final_scaling} which is easy to understand from an effective field theory point of view as we explain there.

We first do a Wick rotation in Eq.~\eqref{eq:++} to better handle the oscillatory components of the integral. Using the variables $y_1=+i z_1$ and $y_2=+i z_2$, the Wick-rotated integral read as,
\begin{align}
\int_0^\infty \d z_1 z_1^{3/2} \, H_{i\mu}^{(1)}(z_1) \, e^{-i p z_1} \int_{z_1}^{\infty} \dfrac{\d z_2}{\sqrt{z_2}} \, H_{i\mu}^{(2)}(z_2) \, e^{-i z_2} \, z_2^{n+i \mu_{\rm c}}\nonumber \\
= (-i)^{n+i\mu_c+3}\int_0^\infty dy_1 y_1^{3/2}H_{i\mu}^{(1)}(-i y_1)e^{-py_1}\int_{y_1}^{\infty}dy_2 H_{i\mu}^{(2)}(-i y_2) e^{-y_2} y_2^{n+i\mu_c-\frac{1}{2}}.
\end{align}

\paragraph{UV contribution using UV asymptotics.} The scaling of the UV contribution using the early-time asymptotic expansion in Eq.~\eqref{eq:early_time_expansion} is proportional to
\begin{align}
(-i)^{n+i\mu_c+2} \, \int_{\sqrt{\mu}}^\infty \d y_1 y_1 \, e^{- p y_1 + y_1} \, \int_{y_1}^{\infty} {\d y_2} \, y_2^{n - 1 +i \mu_{\rm c}}\, e^{- 2 y_2}\,.    
\end{align}
Note that we have extended the range of validity of the early-time expansion from $z\sim \mu^2$ down to $\sqrt{\mu}$. Below $z\sim \mu^2$ the UV asymptotic from is not fully accurate, but for the purpose of finding the parametric dependence this is sufficient. Because of the exponential decay in the inner integral, the main contribution is from $y_2\simeq y_1$, so we have
\begin{align}\label{eq:UV_using_UV}
    e^{\pi \mu_c/2} \, \int_{\sqrt{\mu}}^\infty \d y_1 y_1^{n+i \mu_c} \, e^{-(1+ p) y_1}\, \propto e^{\pi \mu_c/2} \, e^{-(1+p)\sqrt{\mu}}\,.
\end{align}
This shows that the UV contribution is exponentially suppressed for large $\mu$ values.

\paragraph{UV contribution using IR asymptotics.} Next we integrate the late-time asymptotic form Eq.~\eqref{eq:late_time_expansion} over the UV region and the result is proportional to,
\begin{align}
\dfrac{1}{\mu} (-i)^{n + 3 + i \mu_c}\, \int_{\sqrt{\mu}}^\infty \d y_1 y_1^{3/2 + i \mu} \, e^{-p y_1} \, \int_{y_1}^{\infty} {\d y_2} \, y_2^{n - 1/2 +i \mu_{\rm c}-i \mu}\, e^{-y_2}.
\end{align}
We again consider the main contribution of the inner integral from $y_2 \simeq y_1$, so we have 
\begin{align}
    \dfrac{1}{\mu} \, (-i)^{n + 3 + i \mu_c} \, \int_{\sqrt{\mu}}^\infty \d y_1 y_1^{3/2 + i \mu} \, e^{- p y_1} \, y_1^{n - 1/2 +i \mu_{\rm c}-i \mu}\, e^{- y_1} \propto e^{\pi \mu_c/2} \, e^{-(1+p)\sqrt{\mu}}\,.
\end{align}
Thus, we find that this integral is also exponentially suppressed with respect to $\mu$. 

\paragraph{Full contribution using IR asymptotics.} Next step is to integrate the IR asymptotic form over full range. If the result is power-law suppressed instead of exponentially, we can trust that this result is the leading result. We will see that this is indeed the case. So we calculate
\begin{align}
    \dfrac{1}{\mu} \, z_0^{-n - i \mu_{\rm c}} \, (-i)^{n + 3 + i \mu_c} \, \int_0^\infty \d y_1 y_1^{3/2 + i \mu} \, e^{- p y_1} \, \int_{y_1}^{\infty} {\d y_2} \, y_2^{n - 1/2 +i \mu_{\rm c}-i \mu}\, e^{- y_2}\,.
    \label{eq:IRcontribution_integral}
\end{align}
The inner integral can be written in terms of incomplete gamma functions:
\begin{align}\label{eq:IR_full_range_integral}
    \dfrac{1}{\mu} \, z_0^{-n - i \mu_{\rm c}} \, (-i)^{n + 3 + i \mu_c} \, \int_0^\infty \d y_1 y_1^{3/2 + i \mu} \, e^{- p y_1} \, \Gamma(n+1/2 + i \mu_c - i \mu, y_1).
\end{align}
To perform the last integral, we expand the outer integrand in a power series and integrate each component, and then we resum the series. Series expansion of the outer integrand is
\begin{align}
    y_1^{3/2+i \mu} e^{- p y_1} = \sum_{w\geq0} y_1^{3/2+w+i \mu} \dfrac{(-p)^w}{w!}\,.
\end{align}
We rewrite the integral in Eq.~\eqref{eq:IR_full_range_integral} as
\begin{align}
    \sum_{w} \, \int_0^\infty \d y_1 y_1^{3/2+w+i \mu} \dfrac{(-p)^w}{w!} \, \Gamma(n+1/2 + i \mu_c - i \mu, y_1).
\end{align}
Now the integral can be done analytically
\begin{align}
    \dfrac{1}{\mu} \, z_0^{-n - i \mu_{\rm c}} \, (-i)^{n + 3 + i \mu_c} \, \sum_{w} \, \dfrac{(-p)^w}{w!} \, \dfrac{\Gamma(n + 3 +w + i \mu_c)}{5/2 + w + i \mu}.
\end{align}
Next we sum over $w$ and the result can be written in terms of a Hypergeometric function
\begin{align}
    \dfrac{1}{\mu} \, z_0^{-n - i \mu_{\rm c}} \, e^{-i\pi(n+3)/2}\, e^{\pi\mu_c/2} \dfrac{\Gamma(n+3+i\mu_c)}{5/2 + i \mu} ~_2F_1(5/2+i\mu, n+3 + i \mu_c; 7/2+i \mu; - p).
\end{align}
In the squeezed limit we can expand the Hypergeometric function, leading to
\begin{align}
    \sim&~ \dfrac{\sqrt{2\pi}}{\mu}\,z_0^{-n-i\mu_{\rm c}}\,e^{-i\pi/4-i\mu_c}~\dfrac{\mu_c^{n + 5/2 + i \mu_c}}{-n-1/2+i\mu-i \mu_c} ~p^{-(n+3+i\mu_c)}\nonumber\\%\label{eq:largemu_first}
    +&~\dfrac{\sqrt{2\pi}}{\mu}\,z_0^{-n-i\mu_{\rm c}}\,e^{-i\pi/4-i\mu_c}\dfrac{\mu_c^{n+5/2+i\mu_c}}{5/2+i\mu_c} \dfrac{\Gamma(7/2+i\mu) \Gamma(n+1/2+i\mu_c-i\mu)}{\Gamma(n+3+i\mu_c)} ~p^{-(5/2+i\mu)}.\label{eq:largemu_second}
\end{align}
In the $\mu\gg\mu_c$ these two terms have scaling with $\mu$ as follows\footnote{Using Eq.~\eqref{eq:gamma_asymptotic}, we find that
\begin{align}
    \text{Eq.~\eqref{eq:largemu_second}}\sim&-i\dfrac{\sqrt{2\pi}}{\mu^2}\,z_0^{-n-i\mu_{\rm c}}\,e^{-i\pi/4-i\mu_c}~\mu_c^{n+5/2+i\mu_c} ~p^{-(n+3+i\mu_c)}\nonumber\\
    &-i2\pi\,e^{-\pi(\mu-\mu_c)}\,z_0^{-n-i\mu_{\rm c}}\,e^{-in\pi-i\mu_c}\mu^{n+2} ~p^{-(5/2+i\mu)}\,.\nonumber
\end{align}}:
\begin{align}
    \text{First line of Eq.~\eqref{eq:largemu_second}} &\propto \mu^{-2}\\
    \text{Second line of Eq.~\eqref{eq:largemu_second}} &\propto e^{-\pi(\mu-\mu_c)}\mu^{n+2}\,p^{-(5/2+i\mu)}.\label{eq:largemu_suppressed_collider}
\end{align}
Thus we see that the result is power-law suppressed as mentioned. This parametric dependence simply encodes the fact that we are integrating out a heavy field with mass $m_\chi$ to get a contact contribution to the bispectrum. Therefore, it does not have the unsuppressed  non-local, non-analytic oscillatory bispectrum signature. While this leading dependence is robust, the subleading term Eq.~\eqref{eq:largemu_suppressed_collider} $\propto p^{-i\mu}$ is exponentially suppressed and should be compared to the exponentially suppressed UV contribution.  In \eqref{eq:UV_using_UV}, we found UV contribution $\propto e^{-p\sqrt{\mu}}$ but the choice of cut-off $\sqrt{\mu}$ there is somewhat arbitrary. Even so, for the squeezed limit $p>\sqrt{\mu}$ and $\mu\gtrsim\mu_c$ we expect the second line of \eqref{eq:largemu_second} to be the dominant collider signal encoding $\mu$. This parametric dependence also matches with Eq.~\eqref{eq.large_mu_ana}, the expectation from the squeezed limit analytic computation.

\section{$+-$ diagrams}\label{app:+-integrals}
\begin{figure}[ht!]
    \centering
    \includegraphics[width=15cm]{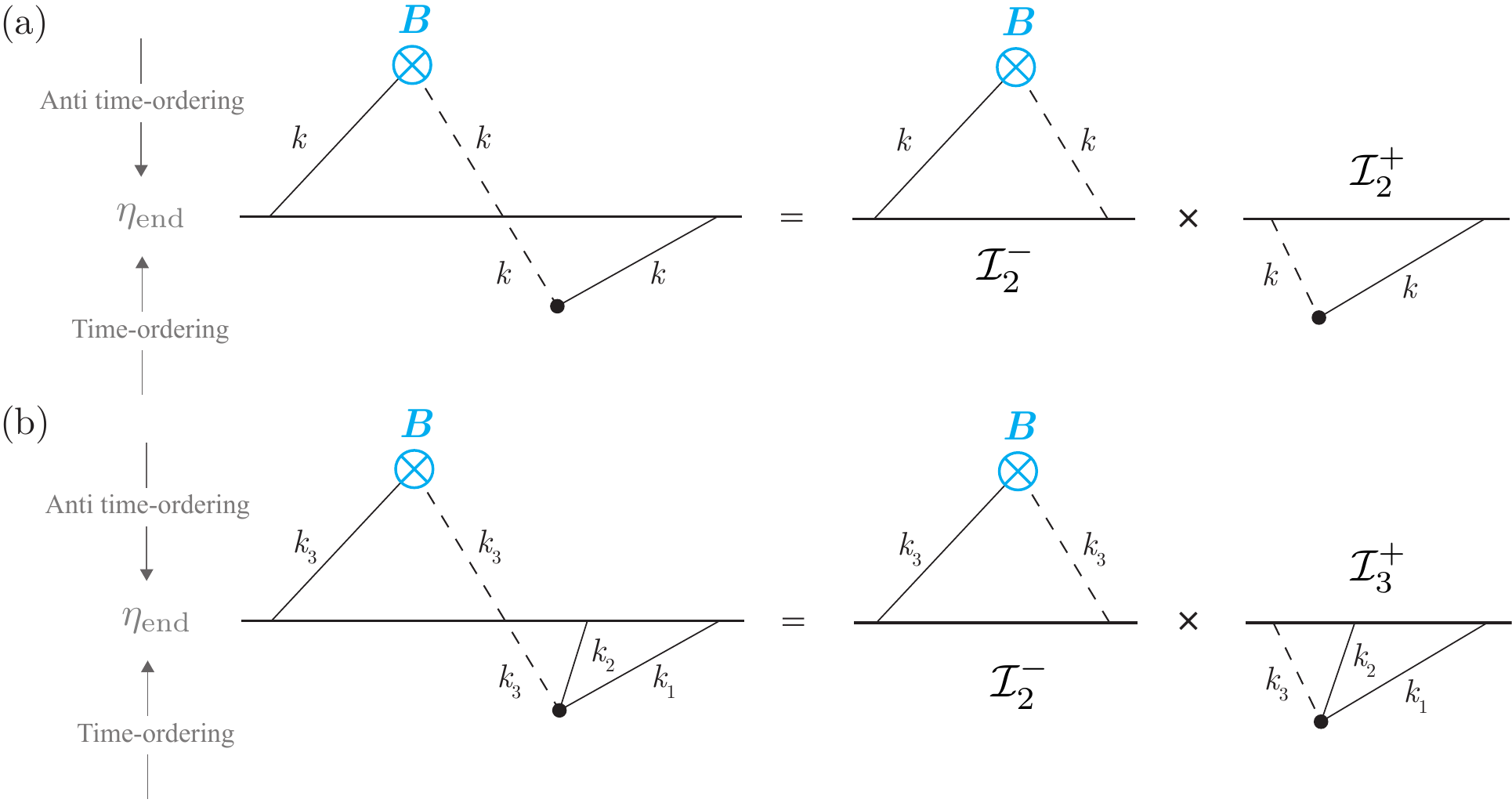}
    \caption{Feynman diagrams for $+-$ contributions to (a) power spectrum and (b) bispectrum under classical background oscillations. These diagrams can be factorized into separate single vertex diagrams, denoted by $\CI_{2}^-$, $\CI_2^+$, and $\CI_3^+$, and can be individually computed.}
    \label{fig:+-}
\end{figure}

In this appendix we show why the $+-$ diagrams are always exponentially suppressed in our scenario. The $+-$ diagrams can be written as multiplication of separate, independent integrals as shown Fig.~\ref{fig:+-}. We have
\begin{align}
    \text{diagram~\ref{fig:+-}a} &\propto \CI_{2}^- \times \CI_{2}^+\,,\label{eq:+-power}\\
    \text{diagram~\ref{fig:+-}b} &\propto \CI_{2}^- \times \CI_{3}^+\,.\label{eq:+-bispectrum}
\end{align}
These integrals can be solved analytically as follows: 
\begin{align}
    \CI_{2}^-=&~e^{-\pi\mu/2} \int_{0}^{\infty} \dfrac{\d z_2}{\sqrt{z_2}} \, H_{i\mu}^{(1)}(z_2) \, e^{i z_2} \, z_2^{n-i \mu_{\rm c}} + \{\mu_c\rightarrow-\mu_c\}\nonumber\\
    &=\dfrac{(i/2)^{n-1/2-i\mu_c}}{\sqrt{\pi}}\dfrac{\Gamma(n+1/2+i\mu-i\mu_c)\Gamma(n+1/2-i\mu-i\mu_c)}{\Gamma(n+1-i\mu_c)}- \{\mu_c\rightarrow-\mu_c\}\,,\\
    \CI_{2}^+=&e^{\pi\mu/2}\int_0^\infty\dfrac{\d z_1}{\sqrt{z_1}}H^{(2)}_{i\mu}(z_1)e^{-iz_1}=\sqrt{\dfrac{2i}{\pi}}\,\Gamma(1/2+i\mu)\Gamma(1/2-i\mu)\,,\\
    \CI_3^+=&~e^{\pi\mu/2}\int_0^\infty \d z_1 z_1^{3/2} \, H_{i\mu}^{(2)}(z_1) \, e^{-i p z_1}= e^{\pi \mu/2}(i p)^{-5/2} \left[\dfrac{i\Gamma(i\mu)}{\pi} \CX_{-i \mu} - \dfrac{\coth(\pi \mu) -1}{\Gamma(1+i\mu)} \CX_{+i \mu} \right]\,,
\end{align}
where
\begin{align}
     \CX_{\pm i \mu} =~ (2ip)^{\mp i\mu} \Gamma(5/2\pm i\mu) \, _2F_1[\dfrac{1}{2}(5/2\pm i \mu),\dfrac{1}{2}(7/2\pm i \mu);1\pm i \mu;\dfrac{1}{p^2}]\,.
\end{align}
While $\CI_{2}^-$ is not exponentially suppressed, we have $\CI_2^+$ and $\CI_3^+ \propto e^{-\pi\mu}$. In summary, the $+-$ diagram is overall exponentially suppressed for both power spectrum and bispectrum.

\section{Full ++ contribution to the bispectrum}\label{app.full_bispec}
In this Appendix, we present details of the numerical results shown in Figures \ref{fig:shapeFULL} to \ref{fig:scaleSTEP}. We discuss the method for general choice of $\mu_c$ and $n$ as shown in the background model \eqref{eq:B_modeling}. To compute results of the sharp feature scenario (Fig. \ref{fig:shapeSTEP} and Fig. \ref{fig:scaleSTEP}), we set $\mu_c=n=0$ in the following formalism.\footnote{More precisely, instead of $\int_0^\infty \d z_1\cdots\int_{z_1}^{\infty}\d z_2\cdots(i/2)\left(z_2/z_0\right)^{n+i \mu_{\rm c}}\theta(z_0-z_2)-\{\mu_c\rightarrow-\mu_c\}$ we calculate $\int_0^\infty \d z_1\cdots\int_{z_1}^{\infty}\d z_2\cdots\theta(z_0-z_2)$.}

\paragraph{Setup.} As we showed in App.~\ref{app:+-integrals}, the $+-$ contribution (diagram \ref{fig:3ptinin}a) is exponentially suppressed for heavy massive fields $\mu\gg1$. Therefore, we disregard it in the following calculation and focus on the $++$ contributions: diagrams \ref{fig:3ptinin}b and \ref{fig:3ptinin}c, i.e., 
\begin{align}
    \braket{\delta\phi^3}'_{++}\simeq\rm{diagram\,\ref{fig:3ptinin}b}+ \rm{diagram\,\ref{fig:3ptinin}c}\,.
\end{align}
We define two dimensionless integrals $\mathcal{I}_{\rm diag.\ref{fig:3ptinin}b}$ and $\mathcal{I}_{\rm diag.\ref{fig:3ptinin}c}$ corresponding to each of the diagrams,
\begin{align}
    &\rm{diagram\,\ref{fig:3ptinin}b} 
    \equiv\dfrac{\rho\lambda}{\Lambda}\dfrac{-\pi H^3}{32 k_1 k_2 k_3^4}B_0~\mathcal{I}_{\rm diag.\ref{fig:3ptinin}b}\,,\\
    & \rm{diagram\,\ref{fig:3ptinin}c} 
    \equiv\dfrac{\rho\lambda}{\Lambda}\dfrac{-\pi H^3}{32 k_1 k_2 k_3^4}B_0~\mathcal{I}_{\rm diag.\ref{fig:3ptinin}c}\,.
\end{align}
where we have
\begin{align}
    \mathcal{I}_{\rm diag.\ref{fig:3ptinin}b}=\int_0^\infty \d z_1 \left[T_1(z_1)+T_2(z_1)+T_3(z_1)\right] H_{i\mu}^{(1)}(z_1) e^{-i p z_1}&\nonumber\\
    \times\int_{z_1}^{\infty}\dfrac{\d z_2}{\sqrt{z_2}} \, H_{i\mu}^{(2)}(z_2) e^{-i z_2} \dfrac{i}{2}\left(\dfrac{z_2}{z_0}\right)^{n+i \mu_{\rm c}}\theta(z_0-z_2)&-\{\mu_c\rightarrow-\mu_c\}\,.\label{eq:first++Integral}\\
    \mathcal{I}_{\rm diag.\ref{fig:3ptinin}c}=\int_0^\infty\dfrac{\d z_1}{\sqrt{z_1}} \, H_{i\mu}^{(1)}(z_1) e^{-i z_1} \dfrac{i}{2}\left(\dfrac{z_1}{z_0}\right)^{n+i \mu_{\rm c}}\theta(z_0-z_1)&\nonumber\\
    \times\int_{z_1}^{\infty}\d z_2\left[T_1(z_2)+T_2(z_2)+T_3(z_2)\right] H_{i\mu}^{(2)}(z_2) e^{-i p z_2}&-\{\mu_c\rightarrow-\mu_c\}\,.\label{eq:second++Integral}
\end{align}
Here the different terms resulting from inflaton derivative structure in \eqref{eq:D} contribute as,
\begin{align}
T_1(z) & = z^{3/2}\left(1-\frac{\vec{k}_1\cdot\vec{k}_2}{k_1k_2}\right)=z^{3/2}\left(2-\dfrac{2}{p^2}\right),\nonumber\\
T_2(z) & = z^{-1/2}\left(\frac{\vec{k}_1\cdot\vec{k}_2 k_3^2}{k_1^2k_2^2}\right)=-z^{-1/2}\dfrac{4}{p^2}\left(1-\dfrac{2}{p^2}\right),\nonumber\\
T_3(z) & = z^{1/2}\left(\frac{\vec{k}_1\cdot\vec{k}_2 k_3^2}{k_1^2k_2^2}\right)\left(i\frac{k_{1}+k_2}{k_3}\right)=-z^{1/2}\dfrac{4i}{p}\left(1-\dfrac{2}{p^2}\right),
\end{align}
where we have assumed special case of $k_1=k_2$ and $p=(k_1+k_2)/k_3$ in the second equalities.

We recast \eqref{eq:second++Integral} to a form similar to \eqref{eq:first++Integral}, so we can compute them using the same method. For this purpose, we switch the inner and outer integrand. This can be expressed as
\begin{align}
   \mathcal{I}_{\rm diag.\ref{fig:3ptinin}c}=&\int_0^\infty \d z_1 \left[T_1(z_1)+T_2(z_1)+T_3(z_1)\right] H_{i\mu}^{(2)}(z_1) e^{-i p z_1}\nonumber\\
    &\times\int_{0}^{z_1}\dfrac{\d z_2}{\sqrt{z_2}} \, H_{i\mu}^{(1)}(z_2) e^{-i z_2} \dfrac{i}{2}\left(\dfrac{z_2}{z_0}\right)^{n+i \mu_{\rm c}}\theta(z_0-z_2)-\{\mu_c\rightarrow-\mu_c\}\,.\label{eq:second--Integral}
\end{align}
To implement the $\theta(z_0-z_2)$ in \eqref{eq:second--Integral} and for further simplification in the next steps, we further split the integration range to $z_1\in[z_0,\infty]$ and $z_1\in[0,z_0]$ which schematically reads as,
\begin{align}
    &\int_0^\infty\d z_1 \cdots \int_0^{z_1}\d z_2 \cdots \theta(z_0-z_2)=\nonumber\\
    &\int_{z_0}^\infty\d z_1 \cdots \int_0^{z_1}\d z_2 \cdots \theta(z_0-z_2)+\int_0^{z_0}\d z_1 \cdots \int_0^{z_1}\d z_2 \cdots \theta(z_0-z_2)=\nonumber\\
    &\underbrace{\int_{z_0}^\infty\d z_1 \cdots \int_0^{z_0}\d z_2 \cdots}_{\mathcal{I}^{\rm 1st}_{\rm diag.\ref{fig:3ptinin}c}}+\underbrace{\int_0^{z_0}\d z_1 \cdots \int_0^{z_1}\d z_2\cdots}_{\mathcal{I}^{\rm 2nd}_{\rm diag.\ref{fig:3ptinin}c}}.
\end{align}
Following this structure we can express the two contributions as follows:
\begin{align}
   \mathcal{I}^{\rm 1st}_{\rm diag.\ref{fig:3ptinin}c}=\int_{z_0}^\infty \d z_1 \left[T_1(z_1)+T_2(z_1)+T_3(z_1)\right] H_{i\mu}^{(2)}(z_1) e^{-i p z_1}&\nonumber\\
    \times\int_{0}^{z_0}\dfrac{\d z_2}{\sqrt{z_2}} \, H_{i\mu}^{(1)}(z_2) e^{-i z_2} \dfrac{i}{2}\left(\dfrac{z_2}{z_0}\right)^{n+i \mu_{\rm c}}&-\{\mu_c\rightarrow-\mu_c\}\,,\label{eq:second--Integral_1st}\\
    \mathcal{I}^{\rm 2nd}_{\rm diag.\ref{fig:3ptinin}c}=\int_0^{z_0} \d z_1 \left[T_1(z_1)+T_2(z_1)+T_3(z_1)\right] H_{i\mu}^{(2)}(z_1) e^{-i p z_1}&\nonumber\\
    \times\int_{0}^{z_1}\dfrac{\d z_2}{\sqrt{z_2}} \, H_{i\mu}^{(1)}(z_2) e^{-i z_2} \dfrac{i}{2}\left(\dfrac{z_2}{z_0}\right)^{n+i \mu_{\rm c}}&-\{\mu_c\rightarrow-\mu_c\}\,.\label{eq:second--Integral_2nd}
\end{align}

\paragraph{The three integrals.} To summarize, the non-Gaussianity signal can be expressed in terms of \eqref{eq:first++Integral}, \eqref{eq:second--Integral_1st}, and \eqref{eq:second--Integral_2nd} as
\begin{align}
    \braket{\delta\phi^3}'\simeq\dfrac{\rho\lambda}{\Lambda}\dfrac{-\pi H^3}{32 k_1 k_2 k_3^4}B_0\left[\mathcal{I}_{\rm diag.\ref{fig:3ptinin}b}+\mathcal{I}^{\rm 1st}_{\rm diag.\ref{fig:3ptinin}c}+\mathcal{I}^{\rm 2nd}_{\rm diag.\ref{fig:3ptinin}c}\right]+{\rm c.c.}\,.
\end{align}
Therefore, using \eqref{eq:shapefunc_def}, the shape function is
\begin{align}\label{eq:shapefunc_numeric_3integrals}
    \mathcal{S}(k_1,k_2,k_3)\simeq\dfrac{\rho\lambda}{\Lambda}\dfrac{\pi}{32}\dfrac{4\dot{\phi}}{H^2}B_0\dfrac{k_1k_2}{k_3^2}\left[\mathcal{I}_{\rm diag.\ref{fig:3ptinin}b}+\mathcal{I}^{\rm 1st}_{\rm diag.\ref{fig:3ptinin}c}+\mathcal{I}^{\rm 2nd}_{\rm diag.\ref{fig:3ptinin}c}\right]+{\rm c.c.}\,.
\end{align}

\paragraph{Analytic result for the inner integrals.} Using the analytic result of the inner integrals, we simplify the nested structure of \eqref{eq:first++Integral}, \eqref{eq:second--Integral_1st}, and \eqref{eq:second--Integral_2nd} to three one-layer integrals. Indefinite form of the inner integrals are
\begin{align}
    \mathcal{F}_{\rm\ref{eq:first++Integral}}^{\rm inner}(z)=\int\dfrac{\d z_2}{\sqrt{z_2}} \, H_{i\mu}^{(2)}(z_2) e^{-i z_2} \dfrac{i}{2}\left(\dfrac{z_2}{z_0}\right)^{n+i \mu_{\rm c}}=&\dfrac{i}{2}\left[(1-\coth\pi\mu)~\mathcal{F}_{i\mu}+\text{csch}\pi\mu~\mathcal{F}_{-i\mu}\right],\\
    \mathcal{F}_{\rm\ref{eq:second--Integral_1st}/\ref{eq:second--Integral_2nd}}^{\rm inner}(z)=\int\dfrac{\d z_2}{\sqrt{z_2}} \, H_{i\mu}^{(1)}(z_2) e^{-i z_2} \dfrac{i}{2}\left(\dfrac{z_2}{z_0}\right)^{n+i \mu_{\rm c}}=&\dfrac{i}{2}\left[(1+\coth\pi\mu)~\mathcal{F}_{i\mu}-\text{csch}\pi\mu~\mathcal{F}_{-i\mu}\right],
\end{align}
where
\begin{align}
    \mathcal{F}_{i\mu}\equiv\left(\dfrac{z}{2}\right)^{i\mu}\left(\dfrac{z}{z_0}\right)^{n+\frac{1}{2}+i\mu_c}\dfrac{_2F_2\left(i \mu +\frac{1}{2},n+i \mu +i \mu_c+\frac{1}{2};2 i \mu +1,n+i \mu +i \mu_c+\frac{3}{2};-2 i
   z\right)}{\left(i\mu +i \mu_c+n+\frac{1}{2}\right)\Gamma(1+i\mu)}\,.
\end{align}
Substituting this results back in \eqref{eq:first++Integral}, \eqref{eq:second--Integral_1st}, and \eqref{eq:second--Integral_2nd} yields
\begin{align}
    \mathcal{I}_{\rm diag.\ref{fig:3ptinin}b}=&\int_0^\infty \d z_1 \left[T_1(z_1)+T_2(z_1)+T_3(z_1)\right] H_{i\mu}^{(1)}(z_1) e^{-i p z_1} \left[\mathcal{F}_{\rm\ref{eq:first++Integral}}^{\rm inner}(z_0)-\mathcal{F}_{\rm\ref{eq:first++Integral}}^{\rm inner}(z_1)\right]\nonumber\\
    &-\{\mu_c\rightarrow-\mu_c\}\,,\label{eq:first++Integral_1layer}\\
    \mathcal{I}^{\rm 1st}_{\rm diag.\ref{fig:3ptinin}c}=&\int_{z_0}^\infty \d z_1 \left[T_1(z_1)+T_2(z_1)+T_3(z_1)\right] H_{i\mu}^{(2)}(z_1) e^{-i p z_1} \mathcal{F}_{\rm\ref{eq:second--Integral_1st}/\ref{eq:second--Integral_2nd}}^{\rm inner}(z_0)\nonumber\\
    &-\{\mu_c\rightarrow-\mu_c\}\,,\label{eq:second--Integral_1st_1layer}\\
    \mathcal{I}^{\rm 2nd}_{\rm diag.\ref{fig:3ptinin}c}=&\int_0^{z_0} \d z_1 \left[T_1(z_1)+T_2(z_1)+T_3(z_1)\right] H_{i\mu}^{(2)}(z_1) e^{-i p z_1} \mathcal{F}_{\rm\ref{eq:second--Integral_1st}/\ref{eq:second--Integral_2nd}}^{\rm inner}(z_1)\nonumber\\
    &-\{\mu_c\rightarrow-\mu_c\}\,.\label{eq:second--Integral_2nd_1layer}
\end{align}

\paragraph{Wick rotation.} To help numerical convergence, by eliminating UV oscillations, we perform following transformations in \eqref{eq:first++Integral_1layer} and \eqref{eq:second--Integral_1st_1layer}:
\begin{itemize}
    \item \eqref{eq:first++Integral_1layer}: Wick rotation of the integration parameters as $z_{1}\rightarrow -iy_{1}$.
    \item \eqref{eq:second--Integral_1st_1layer}: First, constant shift as $z_{1}\rightarrow \tilde{z}_{1}+z_{0}$. Then, Wick rotation of the integration parameters as $\tilde{z}_{1}\rightarrow -iy_{1}$.
\end{itemize}
Note that \eqref{eq:second--Integral_2nd_1layer} is convergent due to the integration cutoff at $z_0$. So we numerically compute the following integrals:

\begin{align}
    \mathcal{I}_{\rm diag.\ref{fig:3ptinin}b}=&-i\int_0^\infty \d y_1 \left[T_1(-iy_1)+T_2(-iy_1)+T_3(-iy_1)\right] H_{i\mu}^{(1)}(-iy_1) e^{-py_1}\nonumber\\
    &\times \left[\mathcal{F}_{\rm\ref{eq:first++Integral}}^{\rm inner}(z_0)-\mathcal{F}_{\rm\ref{eq:first++Integral}}^{\rm inner}(-iy_1)\right]-\{\mu_c\rightarrow-\mu_c\}\,,\label{eq:first++Integral_1layer_transformed}\\
    \mathcal{I}^{\rm 1st}_{\rm diag.\ref{fig:3ptinin}c}=&-i\int_{0}^\infty \d y_1 \left[T_1(-iy_1+z_0)+T_2(-iy_1+z_0)+T_3(-iy_1+z_0)\right] H_{i\mu}^{(2)}(-iy_1+z_0)\nonumber\\
    &\times e^{-py_1-ipz_0}\mathcal{F}_{\rm\ref{eq:second--Integral_1st}/\ref{eq:second--Integral_2nd}}^{\rm inner}(z_0)-\{\mu_c\rightarrow-\mu_c\}\,,\label{eq:second--Integral_1st_1layer_transformed}\\
    \mathcal{I}^{\rm 2nd}_{\rm diag.\ref{fig:3ptinin}c}=&~\eqref{eq:second--Integral_2nd_1layer}\,.\label{eq:second--Integral_2nd_1layer_transformed}
\end{align}
Finally, we place the results in \eqref{eq:shapefunc_numeric_3integrals} to compute the bispectrum.

% \paragraph{Wick rotation.} To help numerical convergence we Wick rotate integration parameters as $z_{1}\rightarrow -iy_{1}$ \cite{Chen:2009zp}. So we numerically compute the following integrals:

% \begin{align}
%     \mathcal{I}_{\rm diag.\ref{fig:3ptinin}b}=&-i\int_0^\infty \d y_1 \left[T_1(-iy_1)+T_2(-iy_1)+T_3(-iy_1)\right] H_{i\mu}^{(1)}(-iy_1) e^{-py_1}\nonumber\\
%     &\times \left[\mathcal{F}_{\rm\ref{eq:first++Integral}}^{\rm inner}(z_0)-\mathcal{F}_{\rm\ref{eq:first++Integral}}^{\rm inner}(-iy_1)\right]-\{\mu_c\rightarrow-\mu_c\}\,,\label{eq:first++Integral_1layer}\\
%     \mathcal{I}^{\rm 1st}_{\rm diag.\ref{fig:3ptinin}c}=&-i\int_{z_0}^\infty \d y_1 \left[T_1(-iy_1)+T_2(-iy_1)+T_3(-iy_1)\right] H_{i\mu}^{(2)}(-iy_1) e^{-py_1} \mathcal{F}_{\rm\ref{eq:second--Integral_1st}/\ref{eq:second--Integral_2nd}}^{\rm inner}(z_0)\nonumber\\
%     &-\{\mu_c\rightarrow-\mu_c\}\,,\label{eq:second--Integral_1st_1layer}\\
%     \mathcal{I}^{\rm 2nd}_{\rm diag.\ref{fig:3ptinin}c}=&-i\int_0^{z_0} \d y_1 \left[T_1(-iy_1)+T_2(-iy_1)+T_3(-iy_1)\right] H_{i\mu}^{(2)}(-iy_1) e^{-py_1} \mathcal{F}_{\rm\ref{eq:second--Integral_1st}/\ref{eq:second--Integral_2nd}}^{\rm inner}(-iy_1)\nonumber\\
%     &-\{\mu_c\rightarrow-\mu_c\}\,.\label{eq:second--Integral_2nd_1layer}
% \end{align}
% Finally, we place the results in \eqref{eq:shapefunc_numeric_3integrals} to compute the bispectrum.

\section{Comparison of dominant and subdominant contributions}
\label{app.comparison}
\begin{figure}[htb!]
    \centering
    \includegraphics[width=15cm]{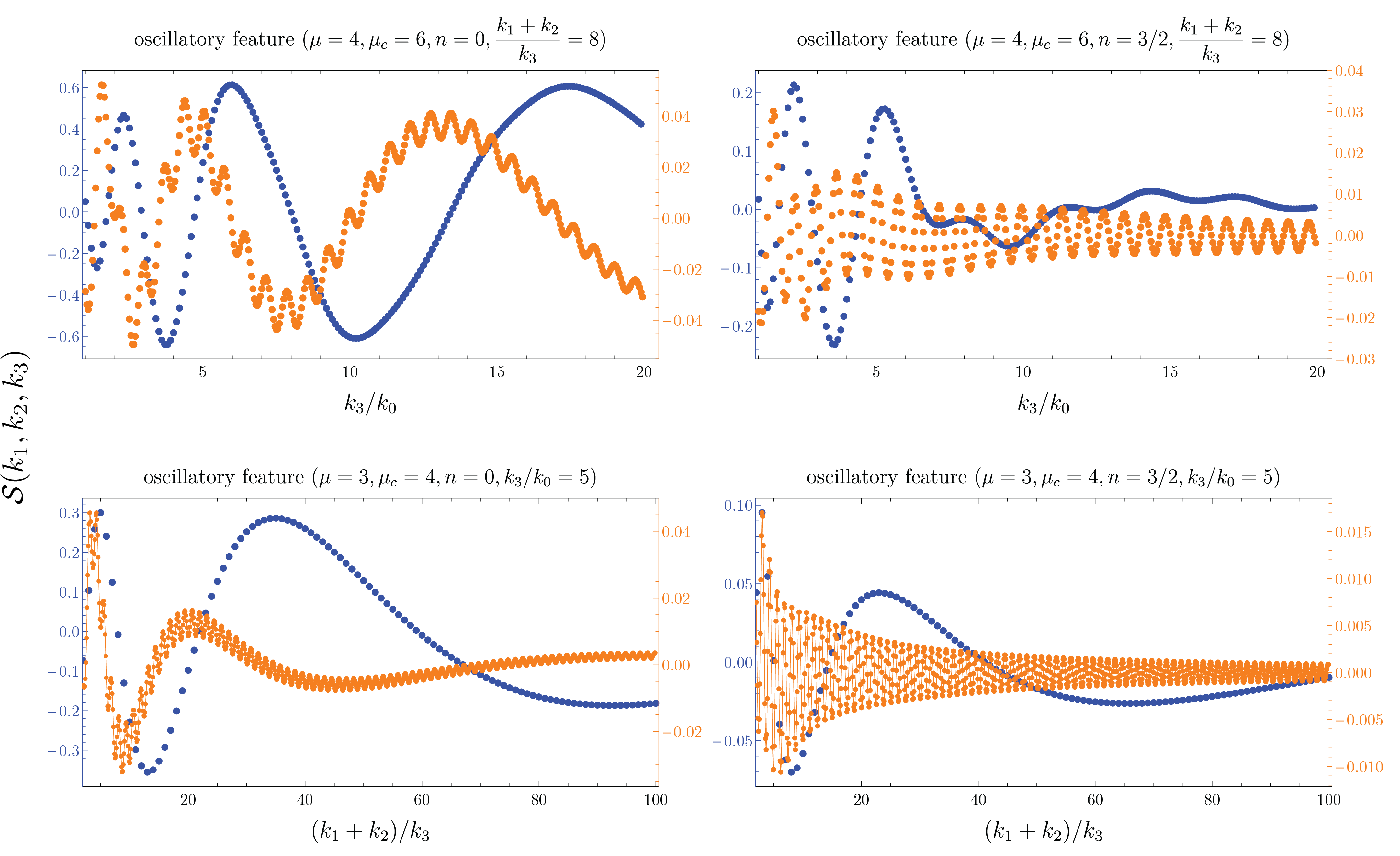}
    \caption{Contribution to the shape function $\mathcal{S}(k_1,k_2,k_3)$ for the oscillatory feature scenario from Fig.\,\ref{fig:3ptinin}b (blue) and Fig.\,\ref{fig:3ptinin}c (orange). The vertical scale corresponding to the blue (orange) curve is shown on the left (right) axis for each panel.
    The left and the right panels correspond to $n=0$ and $n=3/2$, respectively.
    The fist row shows the dependence of $\mathcal{S}$ on $k_3/k_0$ with $(k_1+k_2)/k_3$ fixed;
    the second row shows the dependence of $\mathcal{S}$ on $(k_1+k_2)/k_3$ with $k_3/k_0$ fixed.
    Note that the orange curves are power-law suppressed in $\mu$ and do not contain the clock signal with mass $\mu H$. Their scale-dependent oscillation patterns are determined by the background frequency $\mu_c$ as $\mathcal{S}\propto\sin[\mu_c\log(K/k_0)]$ for fixed $(k_1+k_2)/k_3$ (upper) and $k_3/k_0$ (lower). Here $K\equiv k_1+k_2+k_3$. In addition, we observe sinusoidal running in orange curves due to the sharp feature, given by $\mathcal{S}\propto\sin(K/k_0)$ for fixed $(k_1+k_2)/k_3$ (upper) and $k_3/k_0$ (lower). 
    }
    \label{fig:comparisonFULL}
\end{figure}

In this appendix, we compare the contributions to $\mathcal{S}(k_1,k_2,k_3)$ from Fig.~\ref{fig:3ptinin}b and \ref{fig:3ptinin}c. We first focus on Fig.~\ref{fig:comparisonFULL} where we plot $\mathcal{S}(k_1,k_2,k_3)$ for the oscillatory scenario corresponding to Eq.~\eqref{eq.oscfeat}. The left and the right panels correspond to $n=0$ and $n=3/2$, respectively. The results shown in blue correspond to Fig.~\ref{fig:3ptinin}b and their properties were qualitatively explained in the earlier sections. Therefore we now focus on results shown in orange corresponding to Fig.~\ref{fig:3ptinin}c. As evident from the right-$y$ axis of each panel, the contributions from Fig.~\ref{fig:3ptinin}c are subdominant compared to Fig.~\ref{fig:3ptinin}b.

Fig.~\ref{fig:3ptinin}c captures the case where the 3pt vertex happens before the 2pt vertex. 
Since there is no resonance at this 3pt vertex, an on-shell production of $\chi$ is exponentially suppressed.
The leading contribution is dominated by the process where $\chi$ is produced off-shell and we can replace its propagator by a contact interaction proportional to $1/\mu^2$. This implies Fig.~\ref{fig:3ptinin}c essentially reduces to a diagram where all the inflaton fluctuations are produced from the same vertex. This corresponds to an effective interaction of the type,
\begin{align}
\propto B(t)\dot{\delta\phi} \left[(\dot{\delta\phi})^2 - \frac{1}{a^2}(\partial_i\delta\phi)^2\right].   
\end{align}
This is very similar to the type of models considered in~\cite{Chen:2010bka}, where the interaction terms contain both higher derivatives and scale-dependent couplings which lead to the presence of both non-trivial shape-dependence and scale-dependence in bispectra. The nontrivial $\mathcal{S}(k_1,k_2,k_3)$ then comes from two contributions: first, the oscillation of the source from time $t_0$ (or $\eta_0$) till the time of inflaton production $\eta_{\rm res}$, and second, the standard shape function corresponding to 3pt inflaton interaction. 
The first contribution is similar to what was calculated in the resonant models \cite{Chen:2008wn,Flauger:2009ab,Flauger:2010ja,Chen:2010bka},
\begin{align}
\left(\frac{\eta_0}{\eta_{\rm res}}\right)^{-n-i\mu_c} \sim  \left(\frac{\eta_0}{\mu_c/(k_1+k_2+k_3)}\right)^{-n-i\mu_c} \sim \left(\frac{k_1+k_2+k_3}{k_0}\right)^{-n-i\mu_c}.
\end{align}
Here the time of production is determined by the energy conservation/resonance relation $|(k_1+k_2+k_3)\eta_{\rm res}| = \mu_c$ and $k_0 = 1/|\eta_0|$. The second contribution is determined by the inflaton derivative interactions, similar to what was calculated in~\cite{Creminelli:2003iq}.
Denoting this component as $\langle\zeta^3\rangle_{\rm contact}$, we get the full shape function as,
\begin{align}
\mathcal{S} \propto (k_1 k_2 k_3)^2 \langle\zeta^3\rangle_{\rm contact} \left(\frac{k_1+k_2+k_3}{k_0}\right)^{-n} \sin\left(\mu_c\log\left(\frac{k_1+k_2+k_3}{k_0}\right)\right).
\end{align}
In general, $\langle\zeta^3\rangle_{\rm contact}$ is a detailed function of $k_1,k_2,k_3$~\cite{Creminelli:2003iq}. However, in the squeezed limit, $\langle\zeta^3\rangle_{\rm contact}\propto 1/(k_1^5 k_3)$, where we have used $k_1\approx k_2\gg k_3$. Therefore, the full shape function also simplifies,
\begin{align}
\mathcal{S}_{\rm sq. limit} \propto \frac{k_3}{k_1} \left(\frac{k_1+k_2+k_3}{k_0}\right)^{-n} \sin\left(\mu_c\log\left(\frac{k_1+k_2+k_3}{k_0}\right)\right).    
\end{align}
Therefore, the function $\mathcal{S}(k_1,k_2,k_3)$ oscillates as $\sin\left(\mu_c\log\left(\frac{k_1+k_2+k_3}{k_0}\right)\right)$ with an envelope given by $((k_1+k_2+k_3)/k_0)^{-n}\times (k_3/k_1)$ in the squeezed limit. This functional form explains all the orange curves in Fig.~\ref{fig:comparisonFULL}. On top of this resonant signal, NG induced by oscillatory feature also exhibits sinusoidal running, $\mathcal{S}\propto\sin(K/k_0)$, due to the abrupt onset of the feature at $t=t_0$ (see \cite{Chen:2014cwa} for more explanation), which is manifest in Fig.~\ref{fig:comparisonFULL} orange curves.

\begin{figure}[htb!]
    \centering
    \includegraphics[width=15cm]{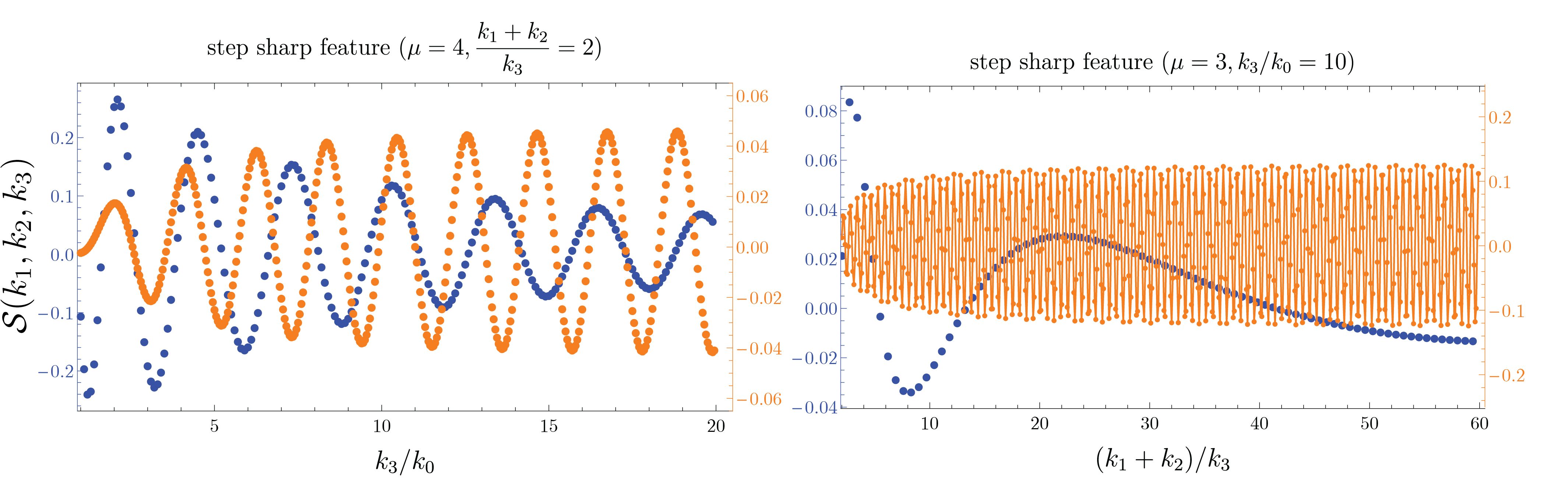}
    \caption{Contribution to the shape function $\mathcal{S}(k_1,k_2,k_3)$ for the sharp feature scenario from Fig.\,\ref{fig:3ptinin}b (blue) and Fig.\,\ref{fig:3ptinin}c (orange). The vertical scale corresponding to the blue (orange) curve is shown on the left (right) axis for each panel. The sinusoidal running in orange curves is determined by $\mathcal{S}\propto\sin((k_1+k_2+k_3)/k_0)$ which reduces to the observed frequencies for fixed $(k_1+k_2)/k_3$ (left) and $k_3/k_0$ (right). We note that, for the right panel, the Fig.\,\ref{fig:3ptinin}c (orange) contribution is larger than Fig.\,\ref{fig:3ptinin}b (blue) contribution due to the infinite sharpness of the feature in the setup.
    }
    \label{fig:comparisonSTEP}
\end{figure}

\begin{figure}[t!]
    \centering
    \includegraphics[width=15cm]{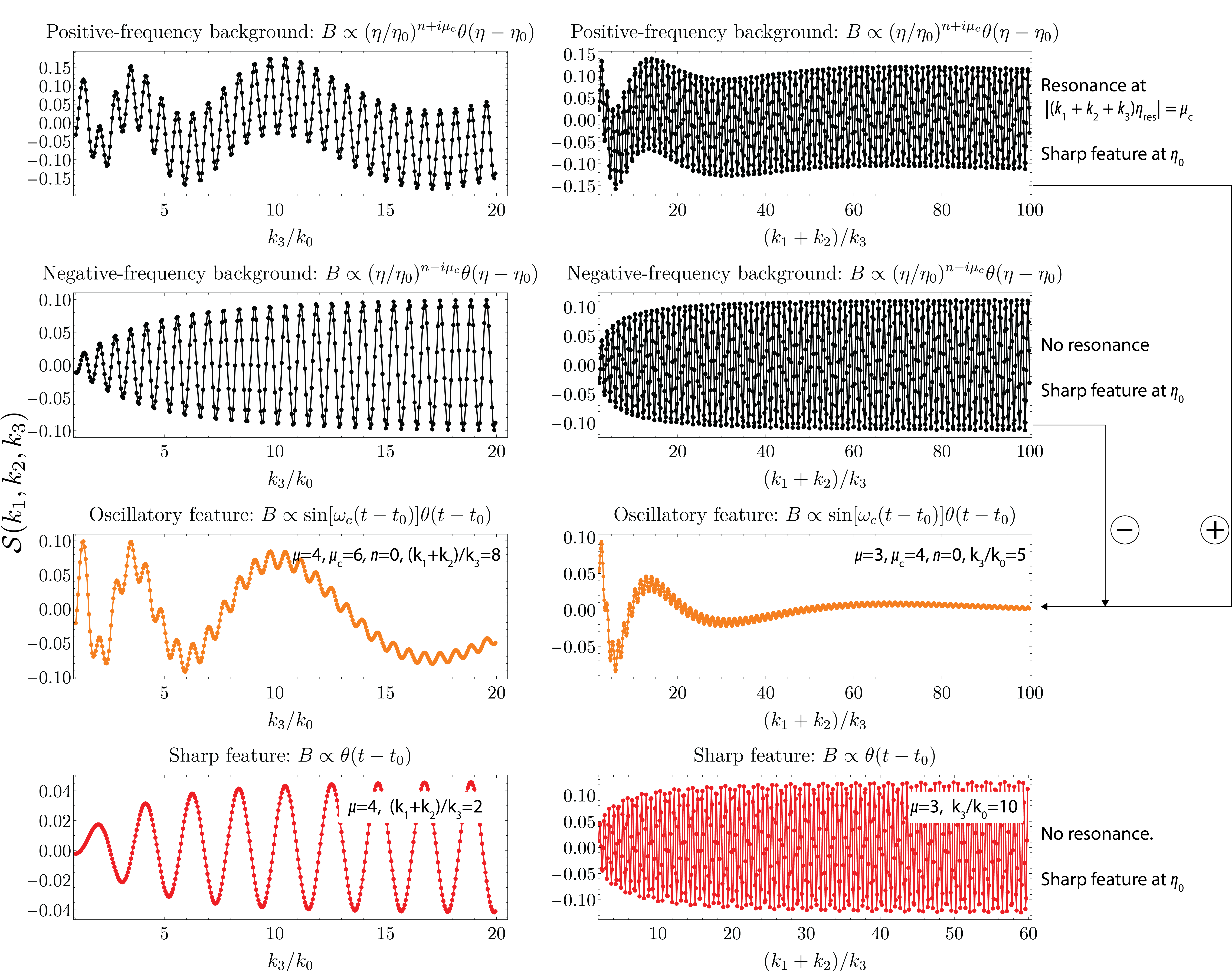}
    \caption{Amplitudes of the feature-induced NG in the contact interaction limit. {\it First row:} NG induced by the positive frequency component of the oscillatory feature. This component has sinusoidal running due to the step function as well as resonant signal due to the oscillatory feature. {\it Second row:} NG induced by negative frequency component of the oscillatory feature which only contains sinusoidal running as this component has no resonance point. {\it Third row:} Subtraction of the second and first rows, corresponding to an oscillatory feature $\propto\sin(t-t_0)\theta(t-t_0)$. The amplitude of sinusoidal running is reduced compared to each of the positive and negative frequency components as the sharp feature is less sharp, i.e., the background feature $B(t)$ is continuous but with an abrupt onset. {\it Fourth row:} Step sharp feature exhibits similar behavior as negative frequency component of the oscillatory feature since in both of these cases only sharp feature contributes to NG.     
    }
    \label{fig:comparison_pmFreq}
\end{figure}

Now we move on to Fig.~\ref{fig:comparisonSTEP} where we show results for the sharp feature case given in Eq.~\eqref{eq.sfeat}. As before, Fig.~\ref{fig:3ptinin}c again reduces to a contact interaction which we can roughly parametrize as a sharp feature acting on the contact vertex,
\begin{align}
\propto B_0 \theta(t-t_0)\dot{\delta\phi} \left[(\dot{\delta\phi})^2 - \frac{1}{a^2}(\partial_i\delta\phi)^2\right].   
\end{align}
Given this structure, we can immediately notice that the oscillation of $\mathcal{S}$ will be given by $\propto \sin\left((k_1+k_2+k_3)/k_0\right)$, which explains both oscillation patterns of the orange curves in Fig.~\ref{fig:comparisonSTEP}.
%Therefore, the oscillation frequency is given by $((k_1+k_2)/k_3) + 1$ or $k_3/k_0$ as a function of $k_3/k_0$ or $(k_1+k_2)/k_3)$, respectively. 
We notice from the right panel that the Fig.~\ref{fig:3ptinin}c contribution can actually be comparable or bigger than the  Fig.~\ref{fig:3ptinin}b contribution in some kinematic configurations. This is likely a consequence of an infinitely sharp feature and in realistic models this contribution is expected to become smaller. Fig.~\ref{fig:comparison_pmFreq} contains some further comparison between the oscillatory feature and sharp feature case.

\section{Useful formulae}\label{app:formula}
\paragraph{Asymptotic expansions:}
\begin{itemize}
    \item Gamma function at $y\gg1$:
        \begin{align}\label{eq:gamma_asymptotic}
            \Gamma(x\pm iy) \simeq \sqrt{2\pi}e^{\mp i (\pi/4 -\pi x/2 + y)} y^{x-1/2\pm i y} e^{- \pi y/2}\,,
        \end{align}
        where $x$ and $y$ are positive real numbers.
    \item Hypergeometric function at $p \gg 1$:
            \begin{align}\label{eq:hypergeometric_large_argument}
                _2F_1(a,b;c;p) \simeq \dfrac{\Gamma(c) \Gamma(b - a)}{\Gamma(b) \Gamma(c-a)} (-p)^{-a} +   \dfrac{\Gamma(c) \Gamma(a - b)}{\Gamma(a) \Gamma(c-b)} (-p)^{-b} \,.
            \end{align}
            % \begin{align}\label{eq:hypergeometric22_large_argument}
            %     _2F_2(a,b;c,d;p) \simeq \dfrac{\Gamma(c) \Gamma(d)}{\Gamma(a) \Gamma(b)}e^{p}p^{a+b-c-d}& +\dfrac{\Gamma(c) \Gamma(d)\Gamma(b-a)}{\Gamma(b) \Gamma(c-a)\Gamma(d-a)}(-p)^{-a} \nonumber\\
            %     & + \dfrac{\Gamma(c) \Gamma(d)\Gamma(a-b)}{\Gamma(a) \Gamma(c-b)\Gamma(d-b)}(-p)^{-b}\,.
            % \end{align}
    \item Hankel function at $z \lesssim \sqrt{\mu}$ (late-time expansion):
            \begin{align}
                H^{(1)}_{i\mu}(z) &\simeq~  {1+ \coth(\pi \mu) \over \Gamma(1 + i \mu) } \left( {z \over 2} \right)^{i \mu} - i {\Gamma(i \mu) \over \pi} \left( {z \over 2} \right)^{-i \mu}\nonumber\\
                &\xrightarrow[]{\mu \gg 1} e^{-i \pi/4} \sqrt{{2 \over \pi \mu}} e^{\pi\mu/2} e^{i\mu(1-\log \mu)} \left( {z \over 2} \right)^{i \mu}, \label{eq:late_time_expansion_H1}\\
                H^{(2)}_{i\mu}(z) &\simeq~   {1 - \coth(\pi \mu) \over \Gamma(1 + i \mu)} \left( {z \over 2} \right)^{i \mu} + i {\Gamma(i \mu) \over \pi} \left( {z \over 2} \right)^{-i \mu}\nonumber\\ &\xrightarrow[]{\mu \gg 1} e^{i \pi/4} \sqrt{{2 \over \pi \mu}} e^{-\pi\mu/2} e^{-i \mu (1-\log \mu)} \left( {z \over 2} \right)^{-i \mu}\,. \label{eq:late_time_expansion_H2}
            \end{align}
    \item Hankel function at $z \gtrsim \mu^2$ (early-time expansion):
            \begin{align}
                H^{(1)}_{i\mu}(z) \simeq&~ e^{-i \pi/4} \sqrt{{2 \over \pi z}}  e^{\pi \mu /2} \, e^{i z}~,\\ H^{(2)}_{i\mu}(z) \simeq&~ e^{i \pi/4} \sqrt{{2 \over \pi z}}  e^{-\pi \mu /2} \, e^{-i z}\,.
            \end{align}
\end{itemize}

\paragraph{Hypergeometric integrals:}
\begin{align}
	& e^{-\pi \mu /2} \int_{0}^{\infty}\d z \, z^{n} H^{(1)}_{i\mu}(z)\,e^{+ipz} = \nonumber\\
	&\frac{(i/2)^n}{\sqrt{\pi}\,\Gamma(n+3/2)}\Gamma(n+1+i\mu)\Gamma(n+1-i\mu) _{2}F_{1}(n+1-i\mu,n+1+i\mu,n+3/2,(1-p)/2) \label{eq:H1_int} \;,\\
	& e^{\pi \mu /2} \int_{0}^{\infty}\d z \, z^{n} H^{(2)}_{i\mu}(z) \,e^{-ipz}=  \nonumber\\
	& \frac{(-i/2)^n}{\sqrt{\pi}\,\Gamma(n+3/2)}\Gamma(n+1+i\mu)\Gamma(n+1-i\mu) _{2}F_{1}(n+1+i\mu,n+1-i\mu,n+3/2,(1-p)/2) \label{eq:H2_int} \;.
\end{align}

\bibliographystyle{utphys}
\bibliography{references}
\end{document}

%% file: rezadefs.tex
\def\d{{\rm d}}

\def\c{{\rm c}}
\def\s{{\rm s}}

\def\res{{\rm res}}

\def\interaction{{\rm int}}
\def\CA{{\cal A}}

\def\CL{{\cal L}}
\def\CI{{\cal I}}

\def\CS{{\cal S}}

\def\CO{{\cal O}}

\def\CX{{\cal X}}

\def\kk{{\bf k}}

\def\du{{\dot{u}}}

\def\dphi{\dot{\phi}}

%% file: main.bbl
\providecommand{\href}[2]{#2}\begingroup\raggedright\begin{thebibliography}{100}

\bibitem{Planck:2018jri}
{\bfseries Planck} Collaboration, Y.~Akrami {\em et~al.}, ``{Planck 2018
  results. X. Constraints on inflation},''
  \href{http://dx.doi.org/10.1051/0004-6361/201833887}{{\em Astron. Astrophys.}
  {\bfseries 641} (2020) A10},
  \href{http://arxiv.org/abs/1807.06211}{{\ttfamily arXiv:1807.06211
  [astro-ph.CO]}}.

\bibitem{Bartolo:2004if}
N.~Bartolo, E.~Komatsu, S.~Matarrese, and A.~Riotto, ``{Non-Gaussianity from
  inflation: Theory and observations},''
  \href{http://dx.doi.org/10.1016/j.physrep.2004.08.022}{{\em Phys. Rept.}
  {\bfseries 402} (2004) 103--266},
  \href{http://arxiv.org/abs/astro-ph/0406398}{{\ttfamily
  arXiv:astro-ph/0406398}}.

\bibitem{Liguori:2010hx}
M.~Liguori, E.~Sefusatti, J.~R. Fergusson, and E.~P.~S. Shellard, ``{Primordial
  non-Gaussianity and Bispectrum Measurements in the Cosmic Microwave
  Background and Large-Scale Structure},''
  \href{http://dx.doi.org/10.1155/2010/980523}{{\em Adv. Astron.} {\bfseries
  2010} (2010) 980523}, \href{http://arxiv.org/abs/1001.4707}{{\ttfamily
  arXiv:1001.4707 [astro-ph.CO]}}.

\bibitem{Chen:2010xka}
X.~Chen, ``{Primordial Non-Gaussianities from Inflation Models},''
  \href{http://dx.doi.org/10.1155/2010/638979}{{\em Adv. Astron.} {\bfseries
  2010} (2010) 638979}, \href{http://arxiv.org/abs/1002.1416}{{\ttfamily
  arXiv:1002.1416 [astro-ph.CO]}}.

\bibitem{Wang:2013zva}
Y.~Wang, ``{Inflation, Cosmic Perturbations and Non-Gaussianities},''
  \href{http://dx.doi.org/10.1088/0253-6102/62/1/19}{{\em Commun. Theor. Phys.}
  {\bfseries 62} (2014) 109--166},
  \href{http://arxiv.org/abs/1303.1523}{{\ttfamily arXiv:1303.1523 [hep-th]}}.

\bibitem{Chen:2009zp}
X.~Chen and Y.~Wang, ``{Quasi-Single Field Inflation and Non-Gaussianities},''
  \href{http://dx.doi.org/10.1088/1475-7516/2010/04/027}{{\em JCAP} {\bfseries
  04} (2010) 027}, \href{http://arxiv.org/abs/0911.3380}{{\ttfamily
  arXiv:0911.3380 [hep-th]}}.

\bibitem{Arkani-Hamed:2015bza}
N.~Arkani-Hamed and J.~Maldacena, ``{Cosmological Collider Physics},''
  \href{http://arxiv.org/abs/1503.08043}{{\ttfamily arXiv:1503.08043
  [hep-th]}}.

\bibitem{Chen:2009we}
X.~Chen and Y.~Wang, ``{Large non-Gaussianities with Intermediate Shapes from
  Quasi-Single Field Inflation},''
  \href{http://dx.doi.org/10.1103/PhysRevD.81.063511}{{\em Phys. Rev. D}
  {\bfseries 81} (2010) 063511},
  \href{http://arxiv.org/abs/0909.0496}{{\ttfamily arXiv:0909.0496
  [astro-ph.CO]}}.

\bibitem{Baumann:2011nk}
D.~Baumann and D.~Green, ``{Signatures of Supersymmetry from the Early
  Universe},'' \href{http://dx.doi.org/10.1103/PhysRevD.85.103520}{{\em Phys.
  Rev. D} {\bfseries 85} (2012) 103520},
  \href{http://arxiv.org/abs/1109.0292}{{\ttfamily arXiv:1109.0292 [hep-th]}}.

\bibitem{Assassi:2012zq}
V.~Assassi, D.~Baumann, and D.~Green, ``{On Soft Limits of Inflationary
  Correlation Functions},''
  \href{http://dx.doi.org/10.1088/1475-7516/2012/11/047}{{\em JCAP} {\bfseries
  11} (2012) 047}, \href{http://arxiv.org/abs/1204.4207}{{\ttfamily
  arXiv:1204.4207 [hep-th]}}.

\bibitem{Chen:2012ge}
X.~Chen and Y.~Wang, ``{Quasi-Single Field Inflation with Large Mass},''
  \href{http://dx.doi.org/10.1088/1475-7516/2012/09/021}{{\em JCAP} {\bfseries
  09} (2012) 021}, \href{http://arxiv.org/abs/1205.0160}{{\ttfamily
  arXiv:1205.0160 [hep-th]}}.

\bibitem{Sefusatti:2012ye}
E.~Sefusatti, J.~R. Fergusson, X.~Chen, and E.~P.~S. Shellard, ``{Effects and
  Detectability of Quasi-Single Field Inflation in the Large-Scale Structure
  and Cosmic Microwave Background},''
  \href{http://dx.doi.org/10.1088/1475-7516/2012/08/033}{{\em JCAP} {\bfseries
  08} (2012) 033}, \href{http://arxiv.org/abs/1204.6318}{{\ttfamily
  arXiv:1204.6318 [astro-ph.CO]}}.

\bibitem{Norena:2012yi}
J.~Norena, L.~Verde, G.~Barenboim, and C.~Bosch, ``{Prospects for constraining
  the shape of non-Gaussianity with the scale-dependent bias},''
  \href{http://dx.doi.org/10.1088/1475-7516/2012/08/019}{{\em JCAP} {\bfseries
  08} (2012) 019}, \href{http://arxiv.org/abs/1204.6324}{{\ttfamily
  arXiv:1204.6324 [astro-ph.CO]}}.

\bibitem{Pi:2012gf}
S.~Pi and M.~Sasaki, ``{Curvature Perturbation Spectrum in Two-field Inflation
  with a Turning Trajectory},''
  \href{http://dx.doi.org/10.1088/1475-7516/2012/10/051}{{\em JCAP} {\bfseries
  10} (2012) 051}, \href{http://arxiv.org/abs/1205.0161}{{\ttfamily
  arXiv:1205.0161 [hep-th]}}.

\bibitem{Noumi:2012vr}
T.~Noumi, M.~Yamaguchi, and D.~Yokoyama, ``{Effective field theory approach to
  quasi-single field inflation and effects of heavy fields},''
  \href{http://dx.doi.org/10.1007/JHEP06(2013)051}{{\em JHEP} {\bfseries 06}
  (2013) 051}, \href{http://arxiv.org/abs/1211.1624}{{\ttfamily arXiv:1211.1624
  [hep-th]}}.

\bibitem{Gong:2013sma}
J.-O. Gong, S.~Pi, and M.~Sasaki, ``{Equilateral non-Gaussianity from heavy
  fields},'' \href{http://dx.doi.org/10.1088/1475-7516/2013/11/043}{{\em JCAP}
  {\bfseries 11} (2013) 043}, \href{http://arxiv.org/abs/1306.3691}{{\ttfamily
  arXiv:1306.3691 [hep-th]}}.

\bibitem{Emami:2013lma}
R.~Emami, ``{Spectroscopy of Masses and Couplings during Inflation},''
  \href{http://dx.doi.org/10.1088/1475-7516/2014/04/031}{{\em JCAP} {\bfseries
  04} (2014) 031}, \href{http://arxiv.org/abs/1311.0184}{{\ttfamily
  arXiv:1311.0184 [hep-th]}}.

\bibitem{Chen:2015lza}
X.~Chen, M.~H. Namjoo, and Y.~Wang, ``{Quantum Primordial Standard Clocks},''
  \href{http://dx.doi.org/10.1088/1475-7516/2016/02/013}{{\em JCAP} {\bfseries
  02} (2016) 013}, \href{http://arxiv.org/abs/1509.03930}{{\ttfamily
  arXiv:1509.03930 [astro-ph.CO]}}.

\bibitem{Dimastrogiovanni:2015pla}
E.~Dimastrogiovanni, M.~Fasiello, and M.~Kamionkowski, ``{Imprints of Massive
  Primordial Fields on Large-Scale Structure},''
  \href{http://dx.doi.org/10.1088/1475-7516/2016/02/017}{{\em JCAP} {\bfseries
  02} (2016) 017}, \href{http://arxiv.org/abs/1504.05993}{{\ttfamily
  arXiv:1504.05993 [astro-ph.CO]}}.

\bibitem{Kehagias:2015jha}
A.~Kehagias and A.~Riotto, ``{High Energy Physics Signatures from Inflation and
  Conformal Symmetry of de Sitter},''
  \href{http://dx.doi.org/10.1002/prop.201500025}{{\em Fortsch. Phys.}
  {\bfseries 63} (2015) 531--542},
  \href{http://arxiv.org/abs/1501.03515}{{\ttfamily arXiv:1501.03515
  [hep-th]}}.

\bibitem{Chen:2016nrs}
X.~Chen, Y.~Wang, and Z.-Z. Xianyu, ``{Loop Corrections to Standard Model
  Fields in Inflation},'' \href{http://dx.doi.org/10.1007/JHEP08(2016)051}{{\em
  JHEP} {\bfseries 08} (2016) 051},
  \href{http://arxiv.org/abs/1604.07841}{{\ttfamily arXiv:1604.07841
  [hep-th]}}.

\bibitem{Lee:2016vti}
H.~Lee, D.~Baumann, and G.~L. Pimentel, ``{Non-Gaussianity as a Particle
  Detector},'' \href{http://dx.doi.org/10.1007/JHEP12(2016)040}{{\em JHEP}
  {\bfseries 12} (2016) 040}, \href{http://arxiv.org/abs/1607.03735}{{\ttfamily
  arXiv:1607.03735 [hep-th]}}.

\bibitem{Meerburg:2016zdz}
P.~D. Meerburg, M.~M\"unchmeyer, J.~B. Mu\~noz, and X.~Chen, ``{Prospects for
  Cosmological Collider Physics},''
  \href{http://dx.doi.org/10.1088/1475-7516/2017/03/050}{{\em JCAP} {\bfseries
  03} (2017) 050}, \href{http://arxiv.org/abs/1610.06559}{{\ttfamily
  arXiv:1610.06559 [astro-ph.CO]}}.

\bibitem{Chen:2016uwp}
X.~Chen, Y.~Wang, and Z.-Z. Xianyu, ``{Standard Model Background of the
  Cosmological Collider},''
  \href{http://dx.doi.org/10.1103/PhysRevLett.118.261302}{{\em Phys. Rev.
  Lett.} {\bfseries 118} no.~26, (2017) 261302},
  \href{http://arxiv.org/abs/1610.06597}{{\ttfamily arXiv:1610.06597
  [hep-th]}}.

\bibitem{Chen:2016hrz}
X.~Chen, Y.~Wang, and Z.-Z. Xianyu, ``{Standard Model Mass Spectrum in
  Inflationary Universe},''
  \href{http://dx.doi.org/10.1007/JHEP04(2017)058}{{\em JHEP} {\bfseries 04}
  (2017) 058}, \href{http://arxiv.org/abs/1612.08122}{{\ttfamily
  arXiv:1612.08122 [hep-th]}}.

\bibitem{Chen:2017ryl}
X.~Chen, Y.~Wang, and Z.-Z. Xianyu, ``{Schwinger-Keldysh Diagrammatics for
  Primordial Perturbations},''
  \href{http://dx.doi.org/10.1088/1475-7516/2017/12/006}{{\em JCAP} {\bfseries
  12} (2017) 006}, \href{http://arxiv.org/abs/1703.10166}{{\ttfamily
  arXiv:1703.10166 [hep-th]}}.

\bibitem{Kehagias:2017cym}
A.~Kehagias and A.~Riotto, ``{On the Inflationary Perturbations of Massive
  Higher-Spin Fields},''
  \href{http://dx.doi.org/10.1088/1475-7516/2017/07/046}{{\em JCAP} {\bfseries
  07} (2017) 046}, \href{http://arxiv.org/abs/1705.05834}{{\ttfamily
  arXiv:1705.05834 [hep-th]}}.

\bibitem{An:2017hlx}
H.~An, M.~McAneny, A.~K. Ridgway, and M.~B. Wise, ``{Quasi Single Field
  Inflation in the non-perturbative regime},''
  \href{http://dx.doi.org/10.1007/JHEP06(2018)105}{{\em JHEP} {\bfseries 06}
  (2018) 105}, \href{http://arxiv.org/abs/1706.09971}{{\ttfamily
  arXiv:1706.09971 [hep-ph]}}.

\bibitem{An:2017rwo}
H.~An, M.~McAneny, A.~K. Ridgway, and M.~B. Wise, ``{Non-Gaussian Enhancements
  of Galactic Halo Correlations in Quasi-Single Field Inflation},''
  \href{http://dx.doi.org/10.1103/PhysRevD.97.123528}{{\em Phys. Rev. D}
  {\bfseries 97} no.~12, (2018) 123528},
  \href{http://arxiv.org/abs/1711.02667}{{\ttfamily arXiv:1711.02667
  [hep-ph]}}.

\bibitem{Iyer:2017qzw}
A.~V. Iyer, S.~Pi, Y.~Wang, Z.~Wang, and S.~Zhou, ``{Strongly Coupled
  Quasi-Single Field Inflation},''
  \href{http://dx.doi.org/10.1088/1475-7516/2018/01/041}{{\em JCAP} {\bfseries
  01} (2018) 041}, \href{http://arxiv.org/abs/1710.03054}{{\ttfamily
  arXiv:1710.03054 [hep-th]}}.

\bibitem{Kumar:2017ecc}
S.~Kumar and R.~Sundrum, ``{Heavy-Lifting of Gauge Theories By Cosmic
  Inflation},'' \href{http://dx.doi.org/10.1007/JHEP05(2018)011}{{\em JHEP}
  {\bfseries 05} (2018) 011}, \href{http://arxiv.org/abs/1711.03988}{{\ttfamily
  arXiv:1711.03988 [hep-ph]}}.

\bibitem{Chen:2018sce}
X.~Chen, W.~Z. Chua, Y.~Guo, Y.~Wang, Z.-Z. Xianyu, and T.~Xie, ``{Quantum
  Standard Clocks in the Primordial Trispectrum},''
  \href{http://dx.doi.org/10.1088/1475-7516/2018/05/049}{{\em JCAP} {\bfseries
  05} (2018) 049}, \href{http://arxiv.org/abs/1803.04412}{{\ttfamily
  arXiv:1803.04412 [hep-th]}}.

\bibitem{Chen:2018xck}
X.~Chen, Y.~Wang, and Z.-Z. Xianyu, ``{Neutrino Signatures in Primordial
  Non-Gaussianities},'' \href{http://dx.doi.org/10.1007/JHEP09(2018)022}{{\em
  JHEP} {\bfseries 09} (2018) 022},
  \href{http://arxiv.org/abs/1805.02656}{{\ttfamily arXiv:1805.02656
  [hep-ph]}}.

\bibitem{Chua:2018dqh}
W.~Z. Chua, Q.~Ding, Y.~Wang, and S.~Zhou, ``{Imprints of Schwinger Effect on
  Primordial Spectra},'' \href{http://dx.doi.org/10.1007/JHEP04(2019)066}{{\em
  JHEP} {\bfseries 04} (2019) 066},
  \href{http://arxiv.org/abs/1810.09815}{{\ttfamily arXiv:1810.09815
  [hep-th]}}.

\bibitem{Kumar:2018jxz}
S.~Kumar and R.~Sundrum, ``{Seeing Higher-Dimensional Grand Unification In
  Primordial Non-Gaussianities},''
  \href{http://dx.doi.org/10.1007/JHEP04(2019)120}{{\em JHEP} {\bfseries 04}
  (2019) 120}, \href{http://arxiv.org/abs/1811.11200}{{\ttfamily
  arXiv:1811.11200 [hep-ph]}}.

\bibitem{Wu:2018lmx}
Y.-P. Wu, ``{Higgs as heavy-lifted physics during inflation},''
  \href{http://dx.doi.org/10.1007/JHEP04(2019)125}{{\em JHEP} {\bfseries 04}
  (2019) 125}, \href{http://arxiv.org/abs/1812.10654}{{\ttfamily
  arXiv:1812.10654 [hep-ph]}}.

\bibitem{MoradinezhadDizgah:2018ssw}
A.~Moradinezhad~Dizgah, H.~Lee, J.~B. Mu\~noz, and C.~Dvorkin, ``{Galaxy
  Bispectrum from Massive Spinning Particles},''
  \href{http://dx.doi.org/10.1088/1475-7516/2018/05/013}{{\em JCAP} {\bfseries
  05} (2018) 013}, \href{http://arxiv.org/abs/1801.07265}{{\ttfamily
  arXiv:1801.07265 [astro-ph.CO]}}.

\bibitem{Saito:2018omt}
R.~Saito and T.~Kubota, ``{Heavy Particle Signatures in Cosmological
  Correlation Functions with Tensor Modes},''
  \href{http://dx.doi.org/10.1088/1475-7516/2018/06/009}{{\em JCAP} {\bfseries
  06} (2018) 009}, \href{http://arxiv.org/abs/1804.06974}{{\ttfamily
  arXiv:1804.06974 [hep-th]}}.

\bibitem{Tong:2018tqf}
X.~Tong, Y.~Wang, and S.~Zhou, ``{Unsuppressed primordial standard clocks in
  warm quasi-single field inflation},''
  \href{http://dx.doi.org/10.1088/1475-7516/2018/06/013}{{\em JCAP} {\bfseries
  06} (2018) 013}, \href{http://arxiv.org/abs/1801.05688}{{\ttfamily
  arXiv:1801.05688 [hep-th]}}.

\bibitem{Alexander:2019vtb}
S.~Alexander, S.~J. Gates, L.~Jenks, K.~Koutrolikos, and E.~McDonough,
  ``{Higher Spin Supersymmetry at the Cosmological Collider: Sculpting SUSY
  Rilles in the CMB},'' \href{http://dx.doi.org/10.1007/JHEP10(2019)156}{{\em
  JHEP} {\bfseries 10} (2019) 156},
  \href{http://arxiv.org/abs/1907.05829}{{\ttfamily arXiv:1907.05829
  [hep-th]}}.

\bibitem{Lu:2019tjj}
S.~Lu, Y.~Wang, and Z.-Z. Xianyu, ``{A Cosmological Higgs Collider},''
  \href{http://dx.doi.org/10.1007/JHEP02(2020)011}{{\em JHEP} {\bfseries 02}
  (2020) 011}, \href{http://arxiv.org/abs/1907.07390}{{\ttfamily
  arXiv:1907.07390 [hep-th]}}.

\bibitem{Hook:2019zxa}
A.~Hook, J.~Huang, and D.~Racco, ``{Searches for other vacua. Part II. A new
  Higgstory at the cosmological collider},''
  \href{http://dx.doi.org/10.1007/JHEP01(2020)105}{{\em JHEP} {\bfseries 01}
  (2020) 105}, \href{http://arxiv.org/abs/1907.10624}{{\ttfamily
  arXiv:1907.10624 [hep-ph]}}.

\bibitem{Hook:2019vcn}
A.~Hook, J.~Huang, and D.~Racco, ``{Minimal signatures of the Standard Model in
  non-Gaussianities},''
  \href{http://dx.doi.org/10.1103/PhysRevD.101.023519}{{\em Phys. Rev. D}
  {\bfseries 101} no.~2, (2020) 023519},
  \href{http://arxiv.org/abs/1908.00019}{{\ttfamily arXiv:1908.00019
  [hep-ph]}}.

\bibitem{Kumar:2019ebj}
S.~Kumar and R.~Sundrum, ``{Cosmological Collider Physics and the Curvaton},''
  \href{http://dx.doi.org/10.1007/JHEP04(2020)077}{{\em JHEP} {\bfseries 04}
  (2020) 077}, \href{http://arxiv.org/abs/1908.11378}{{\ttfamily
  arXiv:1908.11378 [hep-ph]}}.

\bibitem{Wang:2019gbi}
L.-T. Wang and Z.-Z. Xianyu, ``{In Search of Large Signals at the Cosmological
  Collider},'' \href{http://dx.doi.org/10.1007/JHEP02(2020)044}{{\em JHEP}
  {\bfseries 02} (2020) 044}, \href{http://arxiv.org/abs/1910.12876}{{\ttfamily
  arXiv:1910.12876 [hep-ph]}}.

\bibitem{Liu:2019fag}
T.~Liu, X.~Tong, Y.~Wang, and Z.-Z. Xianyu, ``{Probing P and CP Violations on
  the Cosmological Collider},''
  \href{http://dx.doi.org/10.1007/JHEP04(2020)189}{{\em JHEP} {\bfseries 04}
  (2020) 189}, \href{http://arxiv.org/abs/1909.01819}{{\ttfamily
  arXiv:1909.01819 [hep-ph]}}.

\bibitem{Wang:2019gok}
D.-G. Wang, ``{On the inflationary massive field with a curved field
  manifold},'' \href{http://dx.doi.org/10.1088/1475-7516/2020/01/046}{{\em
  JCAP} {\bfseries 01} (2020) 046},
  \href{http://arxiv.org/abs/1911.04459}{{\ttfamily arXiv:1911.04459
  [astro-ph.CO]}}.

\bibitem{Wang:2020uic}
Y.~Wang and Y.~Zhu, ``{Cosmological Collider Signatures of Massive Vectors from
  Non-Gaussian Gravitational Waves},''
  \href{http://dx.doi.org/10.1088/1475-7516/2020/04/049}{{\em JCAP} {\bfseries
  04} (2020) 049}, \href{http://arxiv.org/abs/2001.03879}{{\ttfamily
  arXiv:2001.03879 [astro-ph.CO]}}.

\bibitem{Li:2020xwr}
L.~Li, S.~Lu, Y.~Wang, and S.~Zhou, ``{Cosmological Signatures of Superheavy
  Dark Matter},'' \href{http://dx.doi.org/10.1007/JHEP07(2020)231}{{\em JHEP}
  {\bfseries 07} (2020) 231}, \href{http://arxiv.org/abs/2002.01131}{{\ttfamily
  arXiv:2002.01131 [hep-ph]}}.

\bibitem{Wang:2020ioa}
L.-T. Wang and Z.-Z. Xianyu, ``{Gauge Boson Signals at the Cosmological
  Collider},'' \href{http://dx.doi.org/10.1007/JHEP11(2020)082}{{\em JHEP}
  {\bfseries 11} (2020) 082}, \href{http://arxiv.org/abs/2004.02887}{{\ttfamily
  arXiv:2004.02887 [hep-ph]}}.

\bibitem{Fan:2020xgh}
J.~Fan and Z.-Z. Xianyu, ``{A Cosmic Microscope for the Preheating Era},''
  \href{http://dx.doi.org/10.1007/JHEP01(2021)021}{{\em JHEP} {\bfseries 01}
  (2021) 021}, \href{http://arxiv.org/abs/2005.12278}{{\ttfamily
  arXiv:2005.12278 [hep-ph]}}.

\bibitem{Kogai:2020vzz}
K.~Kogai, K.~Akitsu, F.~Schmidt, and Y.~Urakawa, ``{Galaxy imaging surveys as
  spin-sensitive detector for cosmological colliders},''
  \href{http://dx.doi.org/10.1088/1475-7516/2021/03/060}{{\em JCAP} {\bfseries
  03} (2021) 060}, \href{http://arxiv.org/abs/2009.05517}{{\ttfamily
  arXiv:2009.05517 [astro-ph.CO]}}.

\bibitem{Bodas:2020yho}
A.~Bodas, S.~Kumar, and R.~Sundrum, ``{The Scalar Chemical Potential in
  Cosmological Collider Physics},''
  \href{http://dx.doi.org/10.1007/JHEP02(2021)079}{{\em JHEP} {\bfseries 02}
  (2021) 079}, \href{http://arxiv.org/abs/2010.04727}{{\ttfamily
  arXiv:2010.04727 [hep-ph]}}.

\bibitem{Aoki:2020zbj}
S.~Aoki and M.~Yamaguchi, ``{Disentangling mass spectra of multiple fields in
  cosmological collider},''
  \href{http://dx.doi.org/10.1007/JHEP04(2021)127}{{\em JHEP} {\bfseries 04}
  (2021) 127}, \href{http://arxiv.org/abs/2012.13667}{{\ttfamily
  arXiv:2012.13667 [hep-th]}}.

\bibitem{Arkani-Hamed:2018kmz}
N.~Arkani-Hamed, D.~Baumann, H.~Lee, and G.~L. Pimentel, ``{The Cosmological
  Bootstrap: Inflationary Correlators from Symmetries and Singularities},''
  \href{http://dx.doi.org/10.1007/JHEP04(2020)105}{{\em JHEP} {\bfseries 04}
  (2020) 105}, \href{http://arxiv.org/abs/1811.00024}{{\ttfamily
  arXiv:1811.00024 [hep-th]}}.

\bibitem{Baumann:2019oyu}
D.~Baumann, C.~Duaso~Pueyo, A.~Joyce, H.~Lee, and G.~L. Pimentel, ``{The
  cosmological bootstrap: weight-shifting operators and scalar seeds},''
  \href{http://dx.doi.org/10.1007/JHEP12(2020)204}{{\em JHEP} {\bfseries 12}
  (2020) 204}, \href{http://arxiv.org/abs/1910.14051}{{\ttfamily
  arXiv:1910.14051 [hep-th]}}.

\bibitem{Baumann:2020dch}
D.~Baumann, C.~Duaso~Pueyo, A.~Joyce, H.~Lee, and G.~L. Pimentel, ``{The
  Cosmological Bootstrap: Spinning Correlators from Symmetries and
  Factorization},'' \href{http://dx.doi.org/10.21468/SciPostPhys.11.3.071}{{\em
  SciPost Phys.} {\bfseries 11} (2021) 071},
  \href{http://arxiv.org/abs/2005.04234}{{\ttfamily arXiv:2005.04234
  [hep-th]}}.

\bibitem{Maru:2021ezc}
N.~Maru and A.~Okawa, ``{Non-Gaussianity from $X, Y$ gauge bosons in
  Cosmological Collider Physics},''
  \href{http://arxiv.org/abs/2101.10634}{{\ttfamily arXiv:2101.10634
  [hep-ph]}}.

\bibitem{Lu:2021gso}
S.~Lu, ``{Axion isocurvature collider},''
  \href{http://dx.doi.org/10.1007/JHEP04(2022)157}{{\em JHEP} {\bfseries 04}
  (2022) 157}, \href{http://arxiv.org/abs/2103.05958}{{\ttfamily
  arXiv:2103.05958 [hep-th]}}.

\bibitem{Lu:2021wxu}
Q.~Lu, M.~Reece, and Z.-Z. Xianyu, ``{Missing scalars at the cosmological
  collider},'' \href{http://dx.doi.org/10.1007/JHEP12(2021)098}{{\em JHEP}
  {\bfseries 12} (2021) 098}, \href{http://arxiv.org/abs/2108.11385}{{\ttfamily
  arXiv:2108.11385 [hep-ph]}}.

\bibitem{Wang:2021qez}
L.-T. Wang, Z.-Z. Xianyu, and Y.-M. Zhong, ``{Precision calculation of
  inflation correlators at one loop},''
  \href{http://dx.doi.org/10.1007/JHEP02(2022)085}{{\em JHEP} {\bfseries 02}
  (2022) 085}, \href{http://arxiv.org/abs/2109.14635}{{\ttfamily
  arXiv:2109.14635 [hep-ph]}}.

\bibitem{Tong:2021wai}
X.~Tong, Y.~Wang, and Y.~Zhu, ``{Cutting rule for cosmological collider
  signals: a bulk evolution perspective},''
  \href{http://dx.doi.org/10.1007/JHEP03(2022)181}{{\em JHEP} {\bfseries 03}
  (2022) 181}, \href{http://arxiv.org/abs/2112.03448}{{\ttfamily
  arXiv:2112.03448 [hep-th]}}.

\bibitem{Pinol:2021aun}
L.~Pinol, S.~Aoki, S.~Renaux-Petel, and M.~Yamaguchi, ``{Inflationary flavor
  oscillations and the cosmic spectroscopy},''
  \href{http://arxiv.org/abs/2112.05710}{{\ttfamily arXiv:2112.05710
  [hep-th]}}.

\bibitem{Cui:2021iie}
Y.~Cui and Z.-Z. Xianyu, ``{Probing Leptogenesis with the Cosmological
  Collider},'' \href{http://arxiv.org/abs/2112.10793}{{\ttfamily
  arXiv:2112.10793 [hep-ph]}}.

\bibitem{Tong:2022cdz}
X.~Tong and Z.-Z. Xianyu, ``{Large Spin-2 Signals at the Cosmological
  Collider},'' \href{http://arxiv.org/abs/2203.06349}{{\ttfamily
  arXiv:2203.06349 [hep-ph]}}.

\bibitem{Reece:2022soh}
M.~Reece, L.-T. Wang, and Z.-Z. Xianyu, ``{Large-Field Inflation and the
  Cosmological Collider},'' \href{http://arxiv.org/abs/2204.11869}{{\ttfamily
  arXiv:2204.11869 [hep-ph]}}.

\bibitem{Pimentel:2022fsc}
G.~L. Pimentel and D.-G. Wang, ``{Boostless Cosmological Collider Bootstrap},''
  \href{http://arxiv.org/abs/2205.00013}{{\ttfamily arXiv:2205.00013
  [hep-th]}}.

\bibitem{Qin:2022lva}
Z.~Qin and Z.-Z. Xianyu, ``{Phase Information in Cosmological Collider
  Signals},'' \href{http://arxiv.org/abs/2205.01692}{{\ttfamily
  arXiv:2205.01692 [hep-th]}}.

\bibitem{Birrell:1982ix}
N.~D. Birrell and P.~C.~W. Davies,
  \href{http://dx.doi.org/10.1017/CBO9780511622632}{{\em {Quantum Fields in
  Curved Space}}}.
\newblock Cambridge Monographs on Mathematical Physics. Cambridge Univ. Press,
  Cambridge, UK, 2, 1984.

\bibitem{ParticleDataGroup:2020ssz}
{\bfseries Particle Data Group} Collaboration, P.~A. Zyla {\em et~al.},
  ``{Review of Particle Physics},''
  \href{http://dx.doi.org/10.1093/ptep/ptaa104}{{\em PTEP} {\bfseries 2020}
  no.~8, (2020) 083C01}.

\bibitem{Sou:2021juh}
C.~M. Sou, X.~Tong, and Y.~Wang, ``{Chemical-potential-assisted particle
  production in FRW spacetimes},''
  \href{http://dx.doi.org/10.1007/JHEP06(2021)129}{{\em JHEP} {\bfseries 06}
  (2021) 129}, \href{http://arxiv.org/abs/2104.08772}{{\ttfamily
  arXiv:2104.08772 [hep-th]}}.

\bibitem{Creminelli:2003iq}
P.~Creminelli, ``{On non-Gaussianities in single-field inflation},''
  \href{http://dx.doi.org/10.1088/1475-7516/2003/10/003}{{\em JCAP} {\bfseries
  10} (2003) 003}, \href{http://arxiv.org/abs/astro-ph/0306122}{{\ttfamily
  arXiv:astro-ph/0306122}}.

\bibitem{Chluba:2015bqa}
J.~Chluba, J.~Hamann, and S.~P. Patil, ``{Features and New Physical Scales in
  Primordial Observables: Theory and Observation},''
  \href{http://dx.doi.org/10.1142/S0218271815300232}{{\em Int. J. Mod. Phys. D}
  {\bfseries 24} no.~10, (2015) 1530023},
  \href{http://arxiv.org/abs/1505.01834}{{\ttfamily arXiv:1505.01834
  [astro-ph.CO]}}.

\bibitem{Slosar:2019gvt}
A.~Slosar {\em et~al.}, ``{Scratches from the Past: Inflationary Archaeology
  through Features in the Power Spectrum of Primordial Fluctuations},'' {\em
  Bull. Am. Astron. Soc.} {\bfseries 51} no.~3, (2019) 98,
  \href{http://arxiv.org/abs/1903.09883}{{\ttfamily arXiv:1903.09883
  [astro-ph.CO]}}.

\bibitem{WMAP:2003syu}
{\bfseries WMAP} Collaboration, H.~V. Peiris {\em et~al.}, ``{First year
  Wilkinson Microwave Anisotropy Probe (WMAP) observations: Implications for
  inflation},'' \href{http://dx.doi.org/10.1086/377228}{{\em Astrophys. J.
  Suppl.} {\bfseries 148} (2003) 213--231},
  \href{http://arxiv.org/abs/astro-ph/0302225}{{\ttfamily
  arXiv:astro-ph/0302225}}.

\bibitem{Chen:2011zf}
X.~Chen, ``{Primordial Features as Evidence for Inflation},''
  \href{http://dx.doi.org/10.1088/1475-7516/2012/01/038}{{\em JCAP} {\bfseries
  01} (2012) 038}, \href{http://arxiv.org/abs/1104.1323}{{\ttfamily
  arXiv:1104.1323 [hep-th]}}.

\bibitem{Chen:2011tu}
X.~Chen, ``{Fingerprints of Primordial Universe Paradigms as Features in
  Density Perturbations},''
  \href{http://dx.doi.org/10.1016/j.physletb.2011.11.009}{{\em Phys. Lett. B}
  {\bfseries 706} (2011) 111--115},
  \href{http://arxiv.org/abs/1106.1635}{{\ttfamily arXiv:1106.1635
  [astro-ph.CO]}}.

\bibitem{Chen:2012ja}
X.~Chen and C.~Ringeval, ``{Searching for Standard Clocks in the Primordial
  Universe},'' \href{http://dx.doi.org/10.1088/1475-7516/2012/08/014}{{\em
  JCAP} {\bfseries 08} (2012) 014},
  \href{http://arxiv.org/abs/1205.6085}{{\ttfamily arXiv:1205.6085
  [astro-ph.CO]}}.

\bibitem{Battefeld:2013xka}
T.~Battefeld, J.~C. Niemeyer, and D.~Vlaykov, ``{Probing Two-Field Open
  Inflation by Resonant Signals in Correlation Functions},''
  \href{http://dx.doi.org/10.1088/1475-7516/2013/05/006}{{\em JCAP} {\bfseries
  05} (2013) 006}, \href{http://arxiv.org/abs/1302.3877}{{\ttfamily
  arXiv:1302.3877 [astro-ph.CO]}}.

\bibitem{Gao:2013ota}
X.~Gao, D.~Langlois, and S.~Mizuno, ``{Oscillatory features in the curvature
  power spectrum after a sudden turn of the inflationary trajectory},''
  \href{http://dx.doi.org/10.1088/1475-7516/2013/10/023}{{\em JCAP} {\bfseries
  10} (2013) 023}, \href{http://arxiv.org/abs/1306.5680}{{\ttfamily
  arXiv:1306.5680 [hep-th]}}.

\bibitem{Noumi:2013cfa}
T.~Noumi and M.~Yamaguchi, ``{Primordial spectra from sudden turning
  trajectory},'' \href{http://dx.doi.org/10.1088/1475-7516/2013/12/038}{{\em
  JCAP} {\bfseries 12} (2013) 038},
  \href{http://arxiv.org/abs/1307.7110}{{\ttfamily arXiv:1307.7110 [hep-th]}}.

\bibitem{Saito:2012pd}
R.~Saito, M.~Nakashima, Y.-i. Takamizu, and J.~Yokoyama, ``{Resonant Signatures
  of Heavy Scalar Fields in the Cosmic Microwave Background},''
  \href{http://dx.doi.org/10.1088/1475-7516/2012/11/036}{{\em JCAP} {\bfseries
  11} (2012) 036}, \href{http://arxiv.org/abs/1206.2164}{{\ttfamily
  arXiv:1206.2164 [astro-ph.CO]}}.

\bibitem{Saito:2013aqa}
R.~Saito and Y.-i. Takamizu, ``{Localized Features in Non-Gaussianity from
  Heavy Physics},'' \href{http://dx.doi.org/10.1088/1475-7516/2013/06/031}{{\em
  JCAP} {\bfseries 06} (2013) 031},
  \href{http://arxiv.org/abs/1303.3839}{{\ttfamily arXiv:1303.3839
  [astro-ph.CO]}}.

\bibitem{Chen:2014joa}
X.~Chen and M.~H. Namjoo, ``{Standard Clock in Primordial Density Perturbations
  and Cosmic Microwave Background},''
  \href{http://dx.doi.org/10.1016/j.physletb.2014.11.002}{{\em Phys. Lett. B}
  {\bfseries 739} (2014) 285--292},
  \href{http://arxiv.org/abs/1404.1536}{{\ttfamily arXiv:1404.1536
  [astro-ph.CO]}}.

\bibitem{Chen:2014cwa}
X.~Chen, M.~H. Namjoo, and Y.~Wang, ``{Models of the Primordial Standard
  Clock},'' \href{http://dx.doi.org/10.1088/1475-7516/2015/02/027}{{\em JCAP}
  {\bfseries 02} (2015) 027}, \href{http://arxiv.org/abs/1411.2349}{{\ttfamily
  arXiv:1411.2349 [astro-ph.CO]}}.

\bibitem{Huang:2016quc}
Q.-G. Huang and S.~Pi, ``{Power-law modulation of the scalar power spectrum
  from a heavy field with a monomial potential},''
  \href{http://dx.doi.org/10.1088/1475-7516/2018/04/001}{{\em JCAP} {\bfseries
  04} (2018) 001}, \href{http://arxiv.org/abs/1610.00115}{{\ttfamily
  arXiv:1610.00115 [hep-th]}}.

\bibitem{Domenech:2018bnf}
G.~Dom\`enech, J.~Rubio, and J.~Wons, ``{Mimicking features in alternatives to
  inflation with interacting spectator fields},''
  \href{http://dx.doi.org/10.1016/j.physletb.2019.01.039}{{\em Phys. Lett. B}
  {\bfseries 790} (2019) 263--269},
  \href{http://arxiv.org/abs/1811.08224}{{\ttfamily arXiv:1811.08224
  [astro-ph.CO]}}.

\bibitem{Braglia:2021ckn}
M.~Braglia, X.~Chen, and D.~K. Hazra, ``{Comparing multi-field primordial
  feature models with the Planck data},''
  \href{http://dx.doi.org/10.1088/1475-7516/2021/06/005}{{\em JCAP} {\bfseries
  06} (2021) 005}, \href{http://arxiv.org/abs/2103.03025}{{\ttfamily
  arXiv:2103.03025 [astro-ph.CO]}}.

\bibitem{Braglia:2021sun}
M.~Braglia, X.~Chen, and D.~K. Hazra, ``{Uncovering the History of Cosmic
  Inflation from Anomalies in Cosmic Microwave Background Spectra},''
  \href{http://arxiv.org/abs/2106.07546}{{\ttfamily arXiv:2106.07546
  [astro-ph.CO]}}.

\bibitem{Braglia:2021rej}
M.~Braglia, X.~Chen, and D.~K. Hazra, ``{Primordial Standard Clock Models and
  CMB Residual Anomalies},'' \href{http://arxiv.org/abs/2108.10110}{{\ttfamily
  arXiv:2108.10110 [astro-ph.CO]}}.

\bibitem{Chen:2008wn}
X.~Chen, R.~Easther, and E.~A. Lim, ``{Generation and Characterization of Large
  Non-Gaussianities in Single Field Inflation},''
  \href{http://dx.doi.org/10.1088/1475-7516/2008/04/010}{{\em JCAP} {\bfseries
  04} (2008) 010}, \href{http://arxiv.org/abs/0801.3295}{{\ttfamily
  arXiv:0801.3295 [astro-ph]}}.

\bibitem{Flauger:2009ab}
R.~Flauger, L.~McAllister, E.~Pajer, A.~Westphal, and G.~Xu, ``{Oscillations in
  the CMB from Axion Monodromy Inflation},''
  \href{http://dx.doi.org/10.1088/1475-7516/2010/06/009}{{\em JCAP} {\bfseries
  06} (2010) 009}, \href{http://arxiv.org/abs/0907.2916}{{\ttfamily
  arXiv:0907.2916 [hep-th]}}.

\bibitem{Flauger:2010ja}
R.~Flauger and E.~Pajer, ``{Resonant Non-Gaussianity},''
  \href{http://dx.doi.org/10.1088/1475-7516/2011/01/017}{{\em JCAP} {\bfseries
  01} (2011) 017}, \href{http://arxiv.org/abs/1002.0833}{{\ttfamily
  arXiv:1002.0833 [hep-th]}}.

\bibitem{Chen:2010bka}
X.~Chen, ``{Folded Resonant Non-Gaussianity in General Single Field
  Inflation},'' \href{http://dx.doi.org/10.1088/1475-7516/2010/12/003}{{\em
  JCAP} {\bfseries 12} (2010) 003},
  \href{http://arxiv.org/abs/1008.2485}{{\ttfamily arXiv:1008.2485 [hep-th]}}.

\bibitem{Starobinsky:1992ts}
A.~A. Starobinsky, ``{Spectrum of adiabatic perturbations in the universe when
  there are singularities in the inflation potential},'' {\em JETP Lett.}
  {\bfseries 55} (1992) 489--494.

\bibitem{Adams:2001vc}
J.~A. Adams, B.~Cresswell, and R.~Easther, ``{Inflationary perturbations from a
  potential with a step},''
  \href{http://dx.doi.org/10.1103/PhysRevD.64.123514}{{\em Phys. Rev. D}
  {\bfseries 64} (2001) 123514},
  \href{http://arxiv.org/abs/astro-ph/0102236}{{\ttfamily
  arXiv:astro-ph/0102236}}.

\bibitem{Bean:2008na}
R.~Bean, X.~Chen, G.~Hailu, S.~H.~H. Tye, and J.~Xu, ``{Duality Cascade in
  Brane Inflation},''
  \href{http://dx.doi.org/10.1088/1475-7516/2008/03/026}{{\em JCAP} {\bfseries
  03} (2008) 026}, \href{http://arxiv.org/abs/0802.0491}{{\ttfamily
  arXiv:0802.0491 [hep-th]}}.

\bibitem{Achucarro:2010da}
A.~Achucarro, J.-O. Gong, S.~Hardeman, G.~A. Palma, and S.~P. Patil,
  ``{Features of heavy physics in the CMB power spectrum},''
  \href{http://dx.doi.org/10.1088/1475-7516/2011/01/030}{{\em JCAP} {\bfseries
  01} (2011) 030}, \href{http://arxiv.org/abs/1010.3693}{{\ttfamily
  arXiv:1010.3693 [hep-ph]}}.

\bibitem{Bartolo:2013exa}
N.~Bartolo, D.~Cannone, and S.~Matarrese, ``{The Effective Field Theory of
  Inflation Models with Sharp Features},''
  \href{http://dx.doi.org/10.1088/1475-7516/2013/10/038}{{\em JCAP} {\bfseries
  10} (2013) 038}, \href{http://arxiv.org/abs/1307.3483}{{\ttfamily
  arXiv:1307.3483 [astro-ph.CO]}}.

\bibitem{Kofman:1997yn}
L.~Kofman, A.~D. Linde, and A.~A. Starobinsky, ``{Towards the theory of
  reheating after inflation},''
  \href{http://dx.doi.org/10.1103/PhysRevD.56.3258}{{\em Phys. Rev. D}
  {\bfseries 56} (1997) 3258--3295},
  \href{http://arxiv.org/abs/hep-ph/9704452}{{\ttfamily arXiv:hep-ph/9704452}}.

\bibitem{Chung:1999ve}
D.~J.~H. Chung, E.~W. Kolb, A.~Riotto, and I.~I. Tkachev, ``{Probing Planckian
  physics: Resonant production of particles during inflation and features in
  the primordial power spectrum},''
  \href{http://dx.doi.org/10.1103/PhysRevD.62.043508}{{\em Phys. Rev. D}
  {\bfseries 62} (2000) 043508},
  \href{http://arxiv.org/abs/hep-ph/9910437}{{\ttfamily arXiv:hep-ph/9910437}}.

\bibitem{Kofman:2004yc}
L.~Kofman, A.~D. Linde, X.~Liu, A.~Maloney, L.~McAllister, and E.~Silverstein,
  ``{Beauty is attractive: Moduli trapping at enhanced symmetry points},''
  \href{http://dx.doi.org/10.1088/1126-6708/2004/05/030}{{\em JHEP} {\bfseries
  05} (2004) 030}, \href{http://arxiv.org/abs/hep-th/0403001}{{\ttfamily
  arXiv:hep-th/0403001}}.

\bibitem{Barnaby:2009mc}
N.~Barnaby, Z.~Huang, L.~Kofman, and D.~Pogosyan, ``{Cosmological Fluctuations
  from Infra-Red Cascading During Inflation},''
  \href{http://dx.doi.org/10.1103/PhysRevD.80.043501}{{\em Phys. Rev. D}
  {\bfseries 80} (2009) 043501},
  \href{http://arxiv.org/abs/0902.0615}{{\ttfamily arXiv:0902.0615 [hep-th]}}.

\bibitem{Flauger:2016idt}
R.~Flauger, M.~Mirbabayi, L.~Senatore, and E.~Silverstein, ``{Productive
  Interactions: heavy particles and non-Gaussianity},''
  \href{http://dx.doi.org/10.1088/1475-7516/2017/10/058}{{\em JCAP} {\bfseries
  10} (2017) 058}, \href{http://arxiv.org/abs/1606.00513}{{\ttfamily
  arXiv:1606.00513 [hep-th]}}.

\bibitem{Kim:2021ida}
J.~H. Kim, S.~Kumar, A.~Martin, and Y.~Tsai, ``{Cosmological particle
  production and pairwise hotspots on the CMB},''
  \href{http://dx.doi.org/10.1007/JHEP11(2021)158}{{\em JHEP} {\bfseries 11}
  (2021) 158}, \href{http://arxiv.org/abs/2107.09061}{{\ttfamily
  arXiv:2107.09061 [hep-ph]}}.

\bibitem{Maldacena:2002vr}
J.~M. Maldacena, ``{Non-Gaussian features of primordial fluctuations in single
  field inflationary models},''
  \href{http://dx.doi.org/10.1088/1126-6708/2003/05/013}{{\em JHEP} {\bfseries
  05} (2003) 013}, \href{http://arxiv.org/abs/astro-ph/0210603}{{\ttfamily
  arXiv:astro-ph/0210603}}.

\bibitem{Weinberg:2005vy}
S.~Weinberg, ``{Quantum contributions to cosmological correlations},''
  \href{http://dx.doi.org/10.1103/PhysRevD.72.043514}{{\em Phys. Rev. D}
  {\bfseries 72} (2005) 043514},
  \href{http://arxiv.org/abs/hep-th/0506236}{{\ttfamily arXiv:hep-th/0506236}}.

\bibitem{Achucarro:2022qrl}
A.~Ach\'ucarro {\em et~al.}, ``{Inflation: Theory and Observations},''
  \href{http://arxiv.org/abs/2203.08128}{{\ttfamily arXiv:2203.08128
  [astro-ph.CO]}}.

\end{thebibliography}\endgroup
